\documentclass[aps,prd,reprint,nofootinbib,twocolumn,superscriptaddress,preprintnumbers]{revtex4-2}

\usepackage{amsthm}
\usepackage{amsmath}
\usepackage{graphicx}
\usepackage{slashed}
\usepackage{amssymb}
\usepackage[colorlinks=True, citecolor=blue, urlcolor=blue, linkcolor=blue]{hyperref}
\usepackage{multirow}
\usepackage{orcidlink}
\usepackage{qcircuit}
\usepackage{physics}
\usepackage{wasysym}
\usepackage[caption=false]{subfig}

\newcommand{\nn}{\nonumber}
\newcommand{\bea}{\begin{align}}
\newcommand{\eea}{\end{align}}

\begin{document}
\title{Tame the Umklapp Processes in Real-Time Lattice Simulation for Hydrodynamics:
An Ising Field Theory Study}

\author{Xiaojun Yao}
\email{xjyao@uw.edu}
\affiliation{InQubator for Quantum Simulation, Department of Physics, University of Washington, Seattle, Washington 98195, USA}

%\date{\today}
\preprint{IQuS@UW-21-128}

\begin{abstract}
We calculate the real-time symmetric correlation function of the stress-energy tensor for a near-integrable Ising field theory consisting of three stable scalar particles via lattice Hamiltonian simulation. Using classical exact diagonalization and the matrix product state tensor network methods, we find that in the scaling region of the lattice theory, Umklapp processes that break continuum momentum conservation are suppressed and the sound modes of relativistic hydrodynamics emerge at long wavelength and late time in non-equilibrium dynamics. The extracted ratio of bulk viscosity to entropy density is $\zeta/s=3.06\pm 0.68$ and the speed of sound is $c_s/c=0.76 \pm 0.02$ at the temperature $T\approx 7.14$ in units of the lowest stable particle's mass. Our study demonstrates the utility of real-time lattice Hamiltonian simulation for describing hydrodynamization and calculating transport coefficients nonperturbatively. 
\end{abstract}

\maketitle

\section{Introduction}
Hydrodynamics is one of the oldest effective theories discovered. It describes low-frequency and long-wavelength dynamics of interacting systems that are typically many-body in nature. Its applicability ranges from early Universe expansion and galaxy formation to transport of fundamental particles such as electrons, quarks and gluons at the atomic and subatomic levels.

In the field of heavy ion collisions, relativistic hydrodynamics with a low shear viscosity has been successfully applied to describe the azimuthal distribution of particles produced~\cite{Song:2010mg,Schenke:2010rr,Bernhard:2019bmu,Nijs:2020ors}. The specific shear viscosity, ratio of the shear viscosity to the entropy density, extracted from the experimental data is consistent at low temperature with a strong-coupling result obtained for supersymmetric Yang-Mills theory, which is $1/(4\pi)$~\cite{Policastro:2001yc}. However, a precision calculation of this ratio in QCD is still missing~\cite{Moore:2020pfu}, due to the poor convergence of perturbation theory at finite temperature~\cite{Jeon:1994if,Jeon:1995zm,Arnold:2000dr,Arnold:2003zc,Ghiglieri:2018dib} and the challenging spectral reconstruction task in the Euclidean lattice QCD approach~\cite{Meyer:2007ic,Mages:2015rea,Itou:2020azb,Itou:2021hsj,Altenkort:2022yhb}. This, together with the development of quantum computing, motivates one to consider calculating transport coefficients of gauge theories via the lattice Hamiltonian approach~\cite{Cohen:2021imf,Turro:2024pxu}, which is more natural for real-time observables and avoids the spectral reconstruction. In addition to calculating transport coefficients, real-time Hamiltonian simulation enables studying thermalization~\cite{Mueller:2021gxd, Yao:2023pht,Ebner:2023ixq,Lee:2023urk,Ebner:2024mee,Lin:2024eiz,Ebner:2024qtu,Mueller:2024mmk,Florio:2024aix,Florio:2025hoc,Halimeh:2025vvp,Das:2025utp,Li:2025sgo,Ebner:2025pdm,Than:2025gso,Gupta:2026tcg,Hayata:2026rmv,Chen:2026tvd} and the onset of hydrodynamics in out-of-equilibrium systems~\cite{Turro:2025sec}. Hydrodynamization is also studied in the context of condensed matter physics~\cite{Ye2020Emergent,Zu:2021irm}.

A crucial lattice artifact that needs removing in order to study physical observables via the lattice Hamiltonian simulation is the Umklapp process~\cite{Pitaevskii1981Physical}. It breaks the continuum momentum conservation, which originates from bounces off the lattice grid and turns the propagating sound modes into a diffusive mode~\cite{Arnold:1997gh}. It is expected that the Umklapp process will be suppressed when taking the continuum limit~\cite{Turro:2025sec}. This is a critical difference between lattice simulations of a quantum field theory and a generic lattice model. 

In this work, we take a simple field theory, i.e., a near-integrable $1+1$-dimensional ($1+1$D) Ising field theory as an example, perform real-time lattice Hamiltonian simulation, and study the continuum limit. By computing real-time symmetric correlation functions of stress-energy tensors using exact diagonalization and a tensor network method known as the matrix product state (MPS)~\cite{Vidal2003,PerezGarcia2007}, we explicitly demonstrate the suppression of the Umklapp processes in the scaling region of the lattice theory on a lattice of size 20. Furthermore, we show numerical evidence for emergent hydrodynamic sound modes on a lattice of size 64. We also
discuss the renormalization flow of the bare coupling and extrapolate the specific bulk viscosity and the speed of sound in the continuum limit. A previous work using MPS to study real-time correlators in this field theory can be found in Ref.~\cite{Banuls:2019qrq}, which didn't study correlators of the stress-energy tensors, neither showed the suppression of the Umklapp process, nor extracted transport coefficients.

This paper is organized as follows: In Sec.~\ref{sec:ift}, we will briefly review the $1+1$D Ising field theory and explain its lattice formulation. Then we will introduce the lattice formulation of the stress-energy tensor and the real-time symmetric correlators in Sec.~\ref{sec:Gs}. Their expected hydrodynamic behavior with and without the Umklapp process will be discussed in Sec.~\ref{sec:hydro}, which is followed by Sec.~\ref{sec:umklapp} that identifies the scaling region and shows the suppression of the Umklapp processes therein. Furthermore, we will present MPS calculation results in Sec.~\ref{sec:mps} and analyze the renormalization group equation for the coupling and take the continuum limit for the speed of light. With these preparations, we will analyze the emergent sound modes in Sec.~\ref{sec:bulk} and extract the specific bulk viscosity and the speed of sound in the continuum limit. Finally, conclusions will be drawn in Sec.~\ref{sec:conclusions} with a prospect for future studies.

\section{Lattice Formulation of Ising Field Theory}
\label{sec:ift}
A family of $1+1$D Ising field theories can be obtained by deforming the $1+1$D Ising conformal field theory that describes a free massless Majorana fermion. The action of the theory can be written as~\cite{Delfino:2005bh}
\begin{align}
    S_{\rm IFT} = S_{\rm ICFT} - \tau\int {\rm d}^2x \,\epsilon(x) - h\int {\rm d}^2x \,\sigma(x) \,,
\end{align}
where $\epsilon$ and $\sigma$ are the relevant energy and spin operators, with the conformal scaling dimensions $\Delta_\epsilon=1$ and $\Delta_\sigma=1/8$, respectively. As the conformal scaling dimensions are equal to the mass dimensions, we see the mass dimensions of the coupling parameters are $[\tau]=1$ and $[h]=15/8$. An Ising field theory can be characterized by a dimensionless quantity\footnote{Previous work used $\eta$ for this quantity. We use $\xi$ to avoid confusion with the shear viscosity.}
\begin{align}
    \xi = \frac{\tau}{|h|^{8/15}} \,.
\end{align}
At fixed $\xi$, $\tau^{-1}$ or equivalently $|h|^{-8/15}$ provides a length scale in the theory. For $\tau \neq0$ and $h=0$, corresponding to $\xi=\infty$, the theory describes a free massive Majorana fermion while for $\tau=0$ and $h\neq 0$, corresponding to $\xi=0$, the theory is the integrable $E_8$ theory found by Zamolodchikov~\cite{Zamolodchikov:1989hfa,Zamolodchikov:1989fp}. For $\tau\neq0$ and $h\neq0$, corresponding to finite $\xi\neq 0$, the theory is a generic non-integrable quantum field theory.
As $\xi>0$ increases, the number of stable particles drops from three to two at around $\xi\approx 0.022$ and from two to one at around $\xi\approx 0.333$~\cite{Delfino:2005bh}, see e.g., Fig.~2 of Ref.~\cite{Jha:2024jan}.

This family of Ising field theories can be formulated on a spatial lattice in the scaling region of the Ising model with both transverse and longitudinal fields
\begin{align}
\label{eqn:H}
    H = \sum_{i=0}^{L-1} \left(J \sigma_i^z \sigma_{i+1}^z + h_x \sigma_i^x + h_z \sigma_i^z \right) \,.
\end{align}
For periodic boundary conditions (PBC), we set $\sigma_L^z = \sigma_0^z$ while for open boundary conditions, we choose $\sigma_L^z = 0$. We will use PBC throughout this paper since it gives well-defined lattice momenta and makes finding the scaling region through the mass spectrum easy.
The scaling region is reached by taking $J=-1$, $h_z\to 0^-$ and $|h_z|^{8/15}L\to \infty$ with
\begin{align}
\label{eqn:hx}
    h_x = -1-\xi_{\rm lat} |h_z|^{8/15}\,.
\end{align}
$\xi_{\rm lat}$ is a fixed parameter that determines the field theory obtained in the continuum limit and is closely related to $\xi$ (roughly proportional~\cite{Jha:2024jan}). The parameter $h_z$ in the lattice setup is a unitless version of the parameter $h$ in the continuum setup. Since the mass dimension of $h$ is 15/8, we would expect $|h_z|^{8/15} \propto a|h|^{8/15}$, where $a$ is the lattice spacing. So $|h_z|^{8/15}$ is an effective measure of the lattice spacing (see Sec.~\ref{sec:beta_func} for a numerical demonstration). 
The condition $|h_z|^{8/15}L\to \infty$ corresponds to the infinite volume limit. 
For $\xi_{\rm lat}=0$, the continuum limit corresponds to the integrable $E_8$ theory found by Zamolodchikov~\cite{Zamolodchikov:1989hfa,Zamolodchikov:1989fp}, while for $\xi_{\rm lat}=\infty$, the continuum theory describes the free massive Majorana fermion. For finite $\xi_{\rm lat}\neq0$, the theory is non-integrable and the number of stable particles in the continuum theory drops from three to one as $\xi_{\rm lat}$ increases, see e.g., studies in Refs.~\cite{Coldea2010Quantum,PhysRevD.18.1259,DELFINO1995724,fonseca2001ising,delfino2007particle,Gabai2022On,fitzpatrick2023lightcone}. The mass spectrum can also be reasonably described by using a two-particle approximation and solving a Bethe-Salpeter equation~\cite{Fonseca:2006au,Litvinov:2025geb}. Low-energy scattering in the Ising field theory has been studied on a Hamiltonian lattice by using the MPS~\cite{Jha:2024jan} and quantum computing~\cite{Farrell:2025nkx}. The scaling region for the $E_8$ integrable theory has also been found via the MPS method in a version of truncated SU(2) lattice gauge theory on a plaquette chain~\cite{Siew:2025thj}. 

We close this section by making a remark that a recent work shows 1+1D quantum field theories as generated by perturbing conformal field theories (like the Ising field theories with $0<\xi_{\rm lat}<\infty$ studied here) can hydrodynamize at late time when conformal perturbation theory breaks down~\cite{Davison:2024msq}.

\section{Symmetric Correlation Function of Stress-energy Tensors}
\label{sec:Gs}
In this study we mainly compute the real-time symmetric correlation function of stress energy tensors for a non-integrable Ising field theory at finite temperature
\begin{align}
    G_s^{\mu\nu}(t,x) &= \langle \{T^{\mu\nu}(t,x), T^{\mu\nu}(0,0) \} \rangle_T \nonumber\\
    &= {\rm Tr}[ \{T^{\mu\nu}(t,x), T^{\mu\nu}(0,0) \} \rho_T ]\,,
\end{align}
where $\rho_T$ denotes the density matrix at thermal equilibrium $\rho_T = e^{-\beta H}/Z$. The temperature of the state is given by $T=1/\beta$ and $Z$ denotes the partition function $Z = {\rm Tr}(e^{-\beta H})$.

For this computation on the lattice, we need to write out the expressions of the lattice-regularized stress-energy tensors explicitly. For the energy density, we use
\begin{align}
  T^{00}(0,i)\to H_i = \frac{J}{2}\left( \sigma_{i-1}^z\sigma_i^z + \sigma_i^z\sigma_{i+1}^z \right) + h_x\sigma_i^x + h_z \sigma_i^z \,,
\end{align}
such that $H = \sum_{i=0}^{L-1} H_i$ under PBC. To construct the energy current operator, we utilize the Heisenberg picture and the conservation of energy in $1+1$D,
\begin{align}
    \partial_t H_i(t) &= \partial_t\! \left( e^{iHt} H_i e^{-iHt} \right) = i[H, H_i(t)] \,,\nonumber \\
    & =  e^{iHt}\! \left( i[H_{i+1},H_i] - i[H_{i},H_{i-1}] \right) e^{-iHt} \,,\\
    \partial_t T^{00}(t,x) &= - \partial_x T^{10}(t,x) \,,
\end{align}
which lead us to write
\begin{align}
    T^{10}(0,i) \to J_i &= - i[H_{i+1},H_i] \nonumber\\
    \label{eqn:T10lattice}
    & = Jh_x(\sigma^z_{i+1}\sigma^y_i - \sigma^z_i\sigma^y_{i+1}) \,.
\end{align}
A similar construction has been shown to accurately describe the energy flow and exactly conserve the energy in the calculation of energy-energy correlators for minimally truncated $2+1$D SU(2) lattice gauge theory~\cite{Lee:2024jnt}.

If the theory respects Lorentz invariance, we would expect the momentum operator to be $T^{01} = T^{10}$ and the total momentum to be conserved
\begin{align}
    \frac{\rm d}{{\rm d}t} \int {\rm d}x\, T^{01}(t, x) = 0 \,.
\end{align}
However, an explicit calculation shows that $J_{\rm tot}=\sum_{i=0}^{L-1} J_i$ is not conserved
\begin{align}
    \left[H,\, \sum_{i=0}^{L-1} J_i \right] = -2i Jh_xh_z \sum_{i=0}^{L-1} (\sigma^z_{i+1}\sigma^x_i - \sigma^z_i\sigma^x_{i+1}) \,,
\end{align}
whose 2-norm can be bounded as
\begin{align}
\label{eqn:comm_H_Jtot}
    ||\left[H, J_{\rm tot} \right]||_2 \leq 4|Jh_xh_z| L \,.
\end{align}
As mentioned earlier, the continuum limit is taken by setting $J=-1$, $h_z\to 0^-$ and $|h_z|^{8/15}L\to \infty$ with $h_x$ given in Eq.~\eqref{eqn:hx}. In this limit, there is still room for $|h_z| L\to 0$ simultaneously with $|h_z|^{8/15}L\to \infty$. Specifically, at a fixed volume $|h_z|^{8/15}L={\rm const}$, sending $h_z\to 0^-$ gives $|h_z| L\to 0$. Thus in the continuum limit, we can realize
\begin{align}
    ||\left[H, J_{\rm tot} \right]||_2 \to 0\,,
\end{align}
which means the total momentum becomes conserved. Recovering the continuum momentum conservation is crucial for studying hydrodynamics in lattice simulation. In the following, we will use $H_i$ and $J_i$ as the corresponding components of the stress-energy tensor for the lattice Ising field theory and expect them to produce the correct physics in the continuum limit.

\section{Hydrodynamic Behavior and Umklapp Processes}
\label{sec:hydro}
In this section we provide a brief review of the hydrodynamic behavior of $G_s^{\mu\nu}$ and the effect of Umklapp processes on lattice simulation. More details can be found in Ref.~\cite{Turro:2025sec}.

The symmetric correlator of the stress-energy tensor can be thought of as the expectation value of $T^{\mu\nu}$ as a function of spacetime in quenched dynamics. In particular, the quench is given by $T^{\mu\nu}$ at the spacetime origin
\begin{align}
    \rho_T \rightarrow \rho(t=0) = \rho_T + T^{\mu\nu}(0) \rho_T + \rho_T T^{\mu\nu}(0) \,,
\end{align}
where we have used the Schr\"odinger picture and the hermiticity of $T^{\mu\nu}$. After time evolution, the expectation value of $T^{\mu\nu}$ at position $x$ is given by
\begin{align}
    \langle T^{\mu\nu}(x) \rangle(t) = {\rm Tr}[ T^{\mu\nu}(x) \rho(t)] \,.
\end{align}
Its deviation from the starting thermal expectation value is exactly the symmetric correlation function
\begin{align}
    \delta \langle T^{\mu\nu}(x) \rangle(t) = \langle T^{\mu\nu}(x) \rangle(t) - {\rm Tr}[ T^{\mu\nu}(x) \rho_T] = G_s^{\mu\nu}(t,x) \,.
\end{align}

Since the late-time behavior of non-equilibrium dynamics is governed by hydrodynamics at long wavelength, we expect $G_s^{\mu\nu}(t,x)$ to develop hydrodynamic behavior at late time and long wavelength. Specifically, if we denote the energy density and momentum density perturbations in this quenched dynamics as
\begin{align}
    \delta\varepsilon(t,x) &= \delta \langle T^{00}(x) \rangle(t) \,,\\
    g^x(t,x) &= \delta \langle T^{01}(x) \rangle(t) = \delta \langle T^{10}(x) \rangle(t) \,,
\end{align}
hydrodynamics ($\nabla_\mu T^{\mu\nu}=0$ with first-order gradient expansion) will give
\begin{align}
    \label{eqn:e_gx1}
    &\partial_t \delta\varepsilon + \partial_x g^x = 0 \,, \\
    \label{eqn:e_gx2}
    & \partial_t g^x + c_s^2\partial_x \delta \varepsilon - \gamma_\zeta \partial_x^2 g^x = 0 \,,
\end{align}
where $c_s$ is the speed of sound, connecting the energy density and pressure perturbations via $c_s^2 = \delta P /\delta \varepsilon$ and $\gamma_\zeta$ is the bulk viscous damping
\begin{align}
\label{eqn:gamma_zeta}
    \gamma_\zeta = \frac{\zeta}{\varepsilon_0+P_0} \,,
\end{align}
with $\zeta$, $\varepsilon_0$ and $P_0$ being the bulk viscosity, energy density and pressure at thermal equilibrium. This set of hydrodynamic equations predicts two sound modes as poles of $g^x(\omega, k)$ in the Fourier space
\begin{align}
\label{eqn:sound_mode}
    \omega_{s\pm} = \pm \sqrt{c_s^2k^2 - \frac{\gamma_\zeta^2k^4}{4}} - i\frac{\gamma_\zeta k^2}{2} \,.
\end{align}
Under the validity condition of the gradient expansion for hydrodynamics, i.e., $c_s k \gg \gamma_\zeta k^2/2$, the solution to $g^x(t, k)$ under an initial perturbation can be written as a damped oscillation~\cite{Arnold:1997gh}
\begin{align}
\label{eqn:gx_sol}
g^x(t,k) =&\ g^x(t=0,k) \cos(c_skt) e^{-\frac{\gamma_\zeta k^2t}{2}} \nn\\
&- ic_s \delta\varepsilon(t=0,k) \sin(c_skt) e^{-\frac{\gamma_\zeta k^2t}{2}} \,.
\end{align}

The Umklapp processes on the lattice break the continuum momentum conservation down to the crystal momentum conservation, i.e., conservation modulo $2\pi/a$, where $a$ is the lattice spacing. As a result, Eq.~\eqref{eqn:e_gx2} no longer holds and the above set of hydrodynamic equations becomes an energy diffusion equation under first-order gradient expansion
\begin{align}
    \partial_t \delta \varepsilon -D_\varepsilon \partial_x^2 \delta \varepsilon = 0\,,
\end{align}
where $D_\varepsilon$ denotes the energy diffusion coefficient.
The pole of $\delta \varepsilon(\omega, k)$ in the Fourier space corresponds to a diffusive mode with the featured frequency
\begin{align}
\omega_d = -i D_\varepsilon k^2 \,.
\end{align}

A previous real-time lattice computation for a truncated SU(2) gauge theory has shown that $G_s^{00}$ behaves diffusively at late time when the theory is far away from the continuum limit~\cite{Turro:2025sec}, as generic lattice models do. In the next section, we will show the suppression of the Umklapp processes and the gradual appearance of the sound modes in the scaling region of lattice simulation. 

\section{Suppression of Umklapp Processes in Scaling Region}
\label{sec:umklapp}
A previous study argues that the Umklapp processes will be suppressed in the continuum limit~\cite{Turro:2025sec}. Here we show this explicitly in the lattice simulation of a non-integrable Ising field theory. First we need to identify the scaling region of the lattice theory.

\subsection{Scaling Region}
We take an Ising lattice of size $L=20$ with PBC and $\xi_{\rm lat}=0.01$. By exactly diagonalizing the Hamiltonian in each lattice momentum sector labeled by $k=2\pi n_k/L$, $n_k\in \mathbb{Z}_L$, we obtain all the eigenenergies $E_n(k)$ and eigenstates $|E_n(k)\rangle$. By analyzing ratios of mass gaps, we can identify the scaling region on this $L=20$ lattice. For bigger lattices, we can estimate the corresponding scaling regions by maintaining $|h_z|^{8/15}L$ to be in a similar range.

\begin{figure}[h]
\centering
\includegraphics[width=0.45\textwidth]{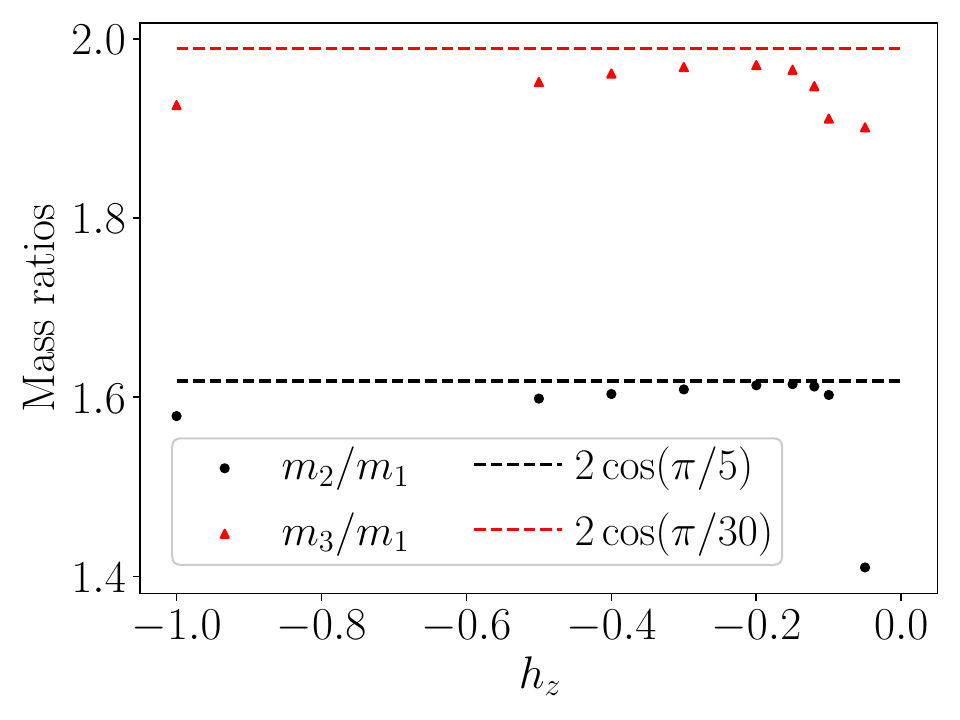}
\caption{Mass ratios of the three stable particles calculated on the periodic $L=20$ lattice with $\xi_{\rm lat}=0.01$ as functions of $h_z$. The dashed lines are predictions from the $E_8$ theory. The lattice results are expected to be close to the $E_8$ predictions but not exactly equal.}
\label{fig:mass_ratio}
\end{figure}

The theory has three stable particles in the continuum. On the lattice, their masses can be obtained as
\begin{align}
    m_n = E_n(k=0) - E_0(k=0) \,,
\end{align}
for $n=1,2,3$.
In Fig.~\ref{fig:mass_ratio}, we show the mass ratios as functions of $h_z$ calculated on this $L=20$ lattice. The dashed lines show the corresponding mass ratios in the integrable $E_8$ theory. We do not expect the lattice results to exactly agree with the $E_8$ predictions since here we study a non-integrable Ising field theory by choosing $\xi_{\rm lat}=0.01$, but we expect the results to be close since $\xi_{\rm lat}$ is very close to zero. From this plot, we conclude that the scaling region of this $L=20$ lattice likely falls within $h_z \in [-0.3,-0.15]$. For $|h_z|<0.1$, the lattice is too small to support the emergence of a quantum field theory as evidenced by the significant deviation of the mass ratios from the $E_8$ predictions. 

\begin{figure}[h]
\centering
\includegraphics[width=0.45\textwidth]{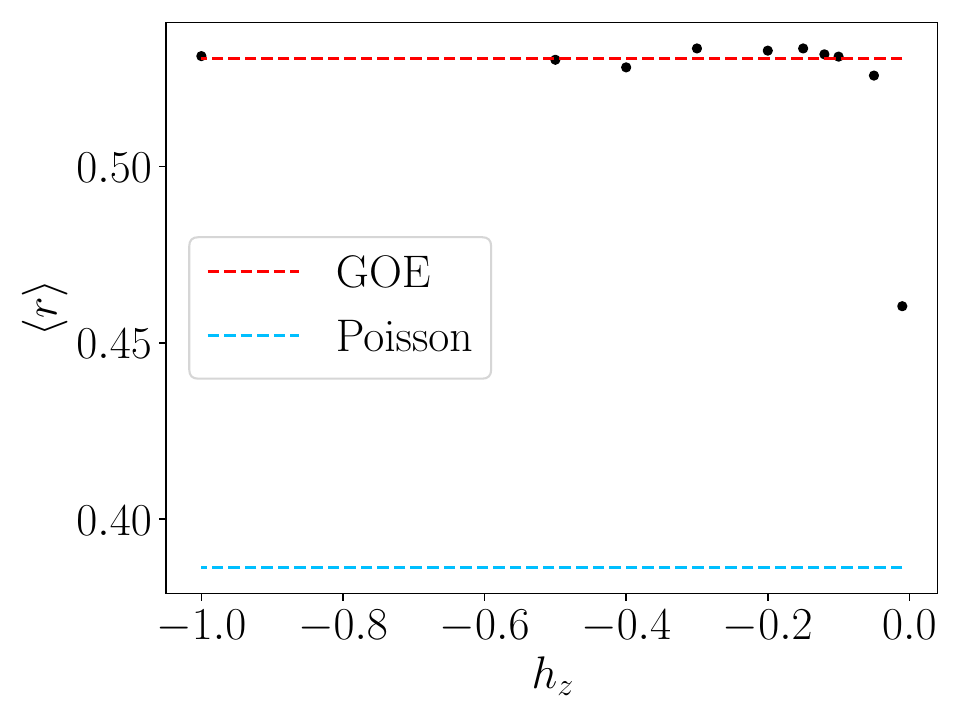}
\caption{Expected values of restricted gap ratios calculated on the periodic $L=20$ lattice with $\xi_{\rm lat}=0.01$ as a function of $h_z$. The middle $70\%$ eigenenergies in the spectrum of the $n_k=1$ sector are used. The red dashed line is the GOE prediction $0.5307$ while the blue dashed line is the value obtained from the Poisson distribution $0.3863$.}
\label{fig:r}
\end{figure}

We also test whether the theory is non-integrable in the scaling region $h_z \in [-0.3,-0.15]$ on this $L=20$ lattice. We use the expected value of the restricted gap ratio $\langle r \rangle$ as a measure of non-integrability. The restricted gap ratio is defined in terms of the gaps between neighboring eigenenergies $\delta_n = E_n - E_{n-1}$ 
\begin{align}
    r_n = \begin{cases}
        \delta_{n+1}/\delta_n \,,\quad {\rm if\ } \delta_{n+1} < \delta_n \\
        \delta_n/\delta_{n+1} \,,\quad {\rm otherwise\ } 
    \end{cases} \,.
\end{align}
We choose the $n_k=1$ sector where there is no more symmetry and thus no more degeneracy in the eigenspectrum. We take the average of $r_n$ for the middle $70\%$ eigenenergies in the spectrum in order to avoid edge effects~\cite{Ebner:2023ixq,Ebner:2024qtu,Das:2025utp}. The calculated $\langle r \rangle$ on this $L=20$ lattice is shown in Fig.~\ref{fig:r}, together with the predictions from the Gaussian orthogonal ensemble (GOE) of random matrices and the Poisson distribution. Agreement with the GOE prediction indicates the theory is non-integrable while agreement with the Poisson one implies integrability. We conclude that the $L=20$ lattice theory in the scaling region $h_z \in [-0.3,-0.15]$ is non-integrable, as expected.

\subsection{Suppression of Umklapp Process}

\begin{figure*}[t]
\centering
\subfloat[$h_z=-1$.\label{fig:Gs_E_-1}]{%
  \includegraphics[width=0.33\linewidth]{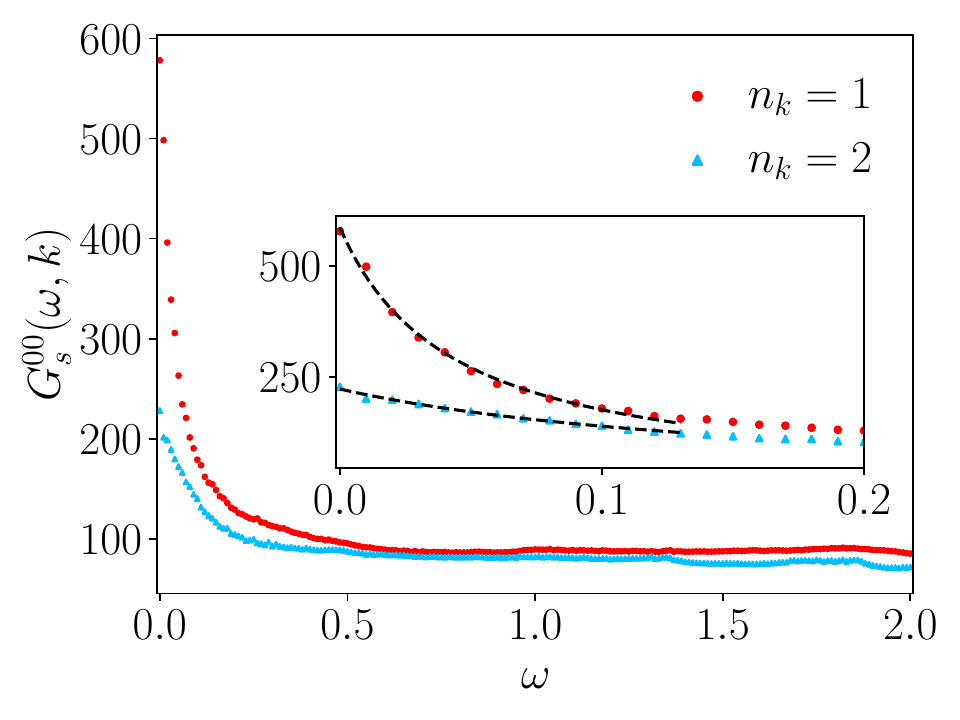}%
}\hfill
\subfloat[$h_z=-0.3$.\label{fig:Gs_E_-0.3}]{%
  \includegraphics[width=0.33\linewidth]{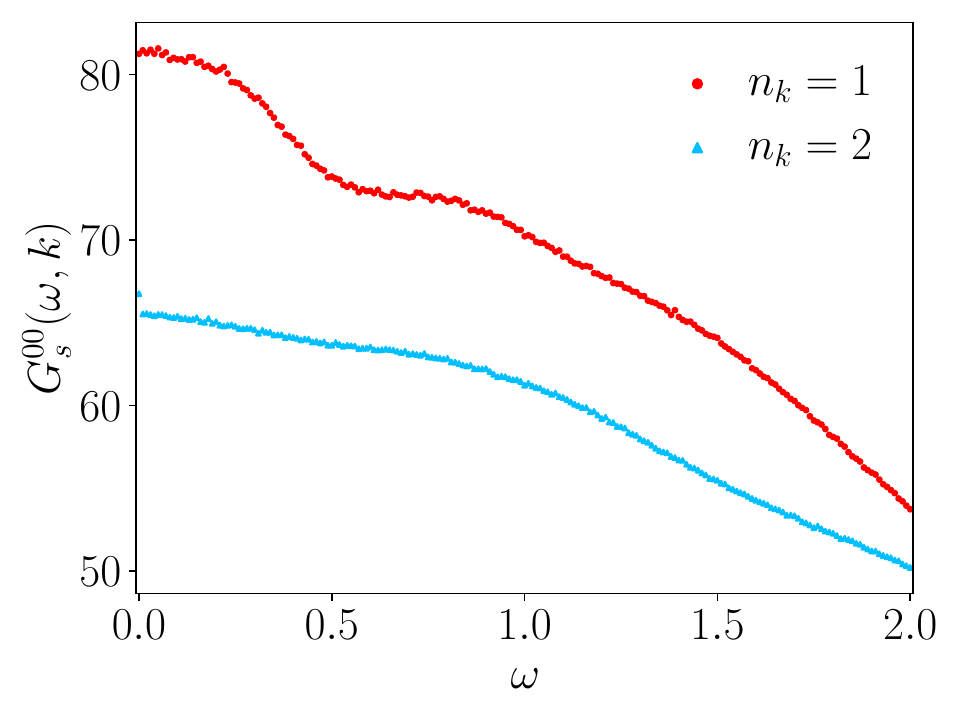}%
}\hfill
\subfloat[$h_z=-0.15$.\label{fig:Gs_E_-0.15}]{%
  \includegraphics[width=0.33\linewidth]{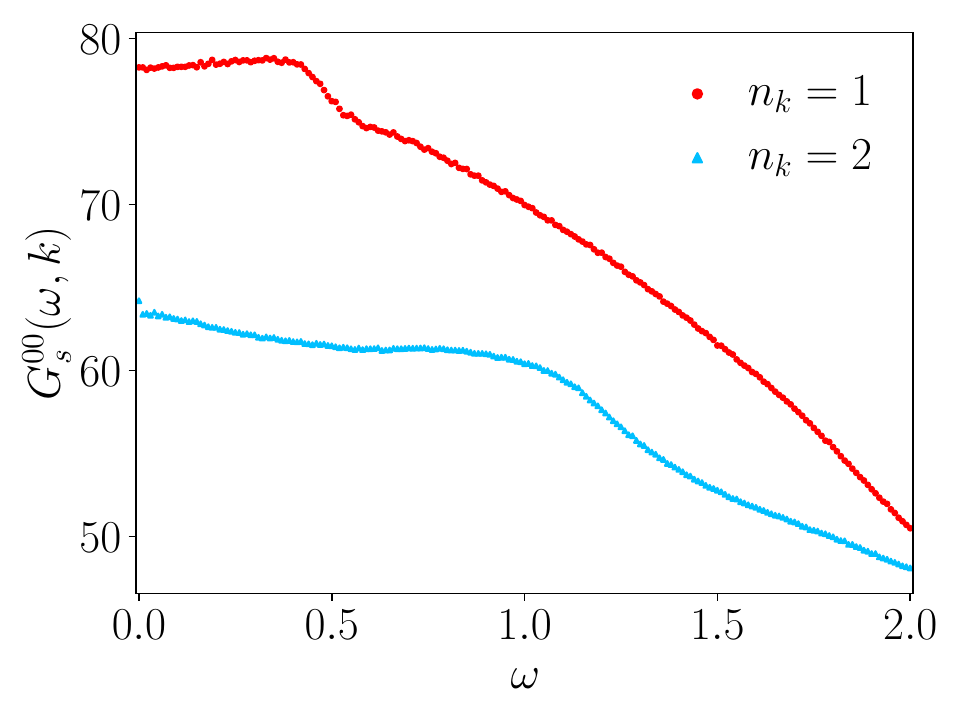}%
}
\caption{Real-time symmetric correlation functions of energy densities in frequency space with two different crystal momenta $k={2\pi n_k}/L$ on a periodic $L=20$ Ising lattice with $\xi_{\rm lat}=0.01$ at $T=10$ for three different longitudinal fields. (a) Outside the scaling region, the symmetric correlator shows a diffusive peak. The black dashed lines in the inset are fits of the form $\frac{b}{|\omega|+4D_\varepsilon \sin^2(k/2)}$ in the range $\omega\in[0,0.13]$. The fitted parameter values are listed in Table~\ref{tab:diffusive_fit}. (b,\,c) Inside the scaling region, the diffusive peak disappears.}
\label{fig:Gs_E_20}
\end{figure*}

\begin{figure*}[t]
\centering
\subfloat[$h_z=-0.3$.\label{fig:Gs_Px_-0.3}]{%
  \includegraphics[width=0.45\linewidth]{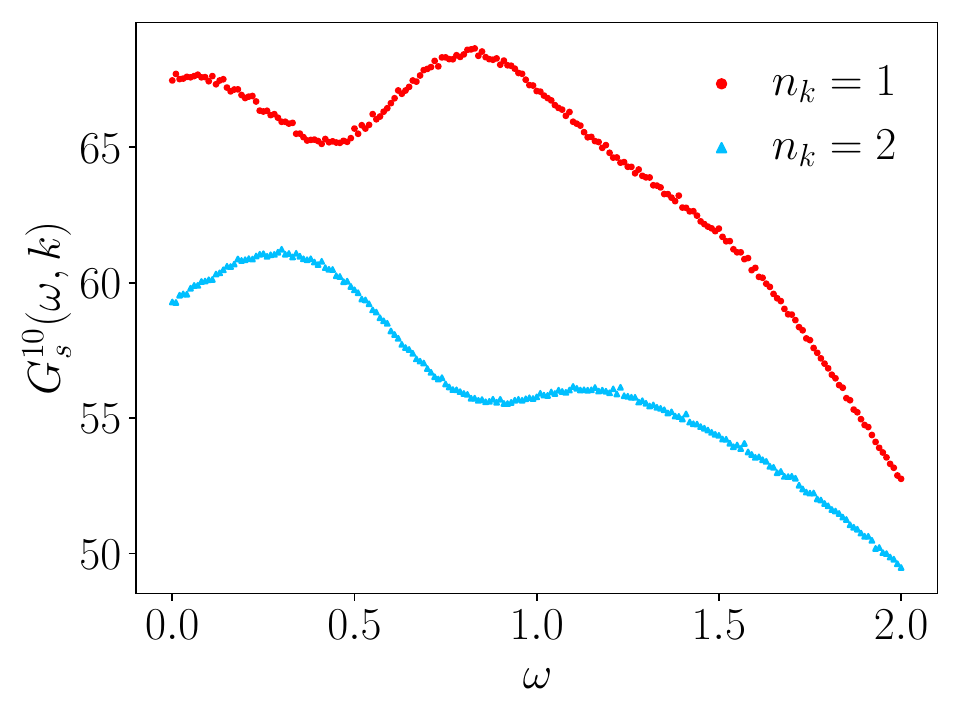}%
}\hfill
\subfloat[$h_z=-0.15$.\label{fig:Gs_Px_-0.15}]{%
  \includegraphics[width=0.45\linewidth]{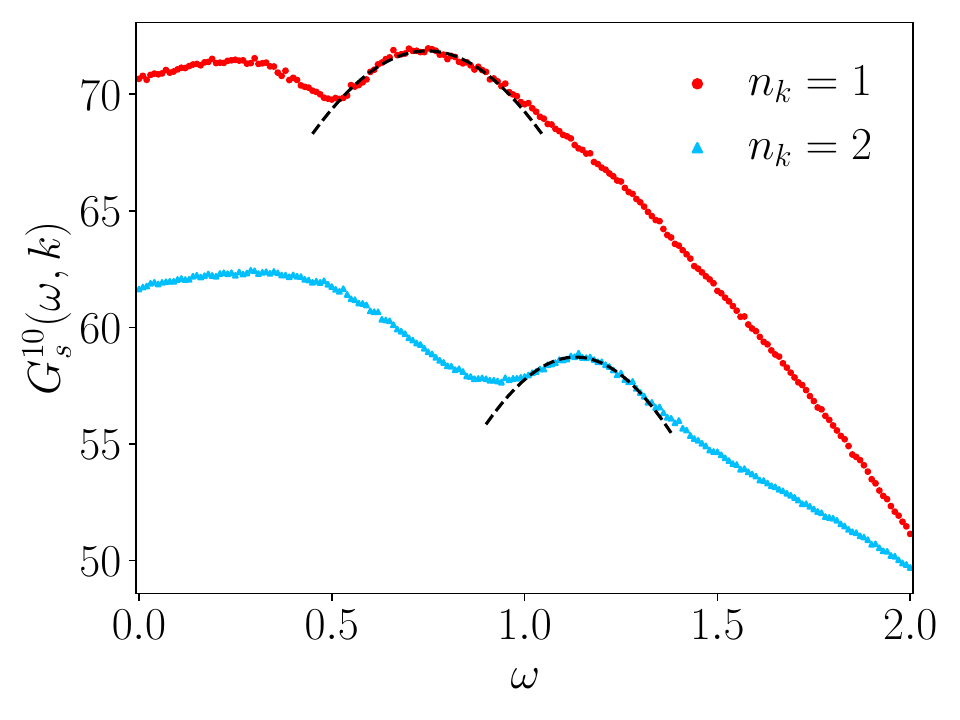}%
}
\caption{Real-time symmetric correlation functions of momentum densities in frequency space with two different crystal momenta $k={2\pi n_k}/L$ on a periodic $L=20$ Ising lattice with $\xi_{\rm lat}=0.01$ at $T=10$ for two different longitudinal fields inside the scaling region. Deep in the scaling region $h_z=-0.15$, two peaks emerge that can be well described by a Lorentzian shape $\frac{b_0}{(\omega-b_1)^2+b_2}$, as indicated by the black dashed lines in (b). The fitting is performed in the interval $\omega\in[0.54,0.95]$ for $n_k=1$ and $\omega\in[1,1.27]$ for $n_k=2$. The fitted parameter values are listed in Table~\ref{tab:lorentz_fit}. These emergent peaks hint the possibility of finding the sound modes if one can take $h_z$ even closer to $0$ while still maintaining the scaling region, which can be done on bigger lattices.}
\label{fig:Gs_Px_20}
\end{figure*}

Using all the eigenstates, we can compute the symmetric correlator in the Fourier space as~\cite{Turro:2025sec}
\begin{align}
\label{eqn:Gs_eigen}
&G_s^{\mu\nu}(\omega, k) \nn\\
=& \sum_p \sum_{E_n(p)} \sum_{E_m(p+k)} \frac{2\pi L} {\Delta\omega Z} |\langle E_n(p) | T^{\mu\nu} | E_m(p+k) \rangle |^2 \nn\\
&\times  \Big[ e^{-\beta E_n(p)} + e^{-\beta E_m(p+k)} \Big]
\bigg|_{ |\omega + E_n(p)- E_m(p+k)| < \frac{\Delta\omega}{2}} \,\nn\\
\approx&\ \sum_p \sum_{E_n(p)} \frac{2\pi L}{Z} \overline{f^2_{\mu\nu}(E_n,p,\omega,k)} \, e^{-\beta E_n(p)} \big( 1 + e^{-\beta \omega} \big)\,,
\end{align}
where we have used $\Delta\omega$ to denote a small frequency window and defined
\begin{align}
\label{eqn:overline_f}
\overline{f^2_{\mu\nu}(E_n,p,\omega,k)} \equiv&\ \frac{1}{\Delta\omega}  \sum_{E_m(p+k)} \bigg|_{ |\omega + E_n(p)- E_m(p+k)| < \frac{\Delta\omega}{2}} \nn\\
&\qquad |\langle E_n(p) | T^{\mu\nu} | E_m(p+k) \rangle |^2 \,.
\end{align}
The momentum summation is over all $L$ momentum sectors and we choose $\Delta \omega = 0.01$. The results are independent of $\Delta \omega$ as long as it is small but the window still contains many states. For $G_s^{00}$, we use $H_i$ for $T^{00}$ while for $G_s^{10}$, we use $J_i$ for $T^{10}$. Using any spatial site $i$ for the $T^{\mu\nu}$ operator in the calculation suffices, due to the lattice-translation invariance.

In Fig.~\ref{fig:Gs_E_20}, we show $G_s^{00}(\omega, k=2\pi n_k/L)$ at $n_k=1$ and $n_k=2$ for three different couplings $h_z=-1$, $-0.3$, and $-0.15$ at the temperature $T=10$. At $h_z=-1$, the lattice theory is outside the scaling region and a diffusive peak located at $\omega=0$ is manifest. The black dashed lines in the inset of Fig.~\ref{fig:Gs_E_-1} are obtained from fits of the form
\begin{align}
    \frac{b}{|\omega| + 4D_\varepsilon \sin^2(\pi n_k/L)} \,.
\end{align}
The fitting is performed in the frequency range $\omega\in[0,0.13]$ and the fitted parameter values are listed in Table~\ref{tab:diffusive_fit}. The closeness of the fitted values of $D_\varepsilon$ for the lowest two nonzero momenta implies that the transport peak at zero frequency is diffusive. On the other hand, as the coupling enters the scaling region, the diffusive transport peak disappears, as shown in Figs.~\ref{fig:Gs_E_-0.3} and~\ref{fig:Gs_E_-0.15}. This shows the suppression of the Umklapp processes in the scaling region of the lattice theory. As the suppression is a consequence of gradually recovering continuum momentum conservation on the lattice, we expect the suppression in the scaling region observed here to be generic, independent of the $1+1$D nature of the theory studied here.

As the Umklapp processes are suppressed in the scaling region, we would expect to see propagating sound modes to appear as transport peaks in $G_s^{10}(\omega,k)$. Figure~\ref{fig:Gs_Px_20} shows $G_s^{10}(\omega,k)$ for two different couplings in the scaling region on the same lattice at the same temperature $T=10$. We find a hint of a sound mode at positive frequency when $h_z=-0.15$. We fit this potential sound mode with a Lorentzian shape 
\begin{align}
    \frac{b_0}{(\omega-b_1)^2+b_2} \,.
\end{align}
The fits for the lowest two nonzero momenta are shown as black dashed lines in Fig.~\ref{fig:Gs_Px_-0.15} and the fitted parameter values are listed in Table~\ref{tab:lorentz_fit}. If this transport peak truly corresponds to a sound mode, we would expect $b_1\propto 2\sin(k/2)$ and $b_2\propto 4\sin^2(k/2)$, where $4\sin^2(k/2)$ is the eigenvalue of the Laplacian operator on periodic lattices. However, the fitted values of $b_1$ and $b_2$ as in Table~\ref{tab:lorentz_fit} do not follow this scaling behavior. We attribute this to the small lattice size that limits us from going into the long wavelength regime, where hydrodynamics applies. In the next section, we will perform the analysis on bigger lattices by using the tensor network method MPS.

\begin{table}[h!]
  \centering
  \begin{tabular}{|c|c|c|}
    \hline
 Parameters & $n_k=1$ & $n_k=2$ \\
 \hline
 $b$ & 25.18 & 37.06 \\
 $D_\varepsilon$ & 0.4384 & 0.4353 \\
 \hline
  \end{tabular}
  \caption{Fitted parameter values for the diffusive peak in Fig.~\ref{fig:Gs_E_-1}, up to four significant numbers. The closeness of the fitted energy diffusion constants for the two momentum transfers supports the diffusive description of the dynamics.}
  \label{tab:diffusive_fit}
\end{table}

\begin{table}[h!]
  \centering
  \begin{tabular}{|c|c|c|}
    \hline
 Parameters & $n_k=1$ & $n_k=2$ \\
 \hline
 $b_0$ & 122.0 & 61.41 \\
 $b_1$ & 0.7474 & 1.132 \\
 $b_2$ & 1.303 & 1.023 \\
 \hline
  \end{tabular}
  \caption{Fitted parameter values for the emergent peaks in Fig.~\ref{fig:Gs_Px_-0.15}, up to four significant numbers. The values of $b_1$ and $b_2$ do not follow the scaling laws expected for a sound mode, i.e., $b_1\propto 2\sin(k/2)$ and $b_2\propto 4\sin^2(k/2)$.}
  \label{tab:lorentz_fit}
\end{table}

\section{MPS Simulation on Larger Lattice}
\label{sec:mps}
In the previous section, we showed the suppression of the Umklapp processes in the scaling region on an $L=20$ lattice. However, to look for the sound modes, we need to use even smaller values of $|h_z|$, which go out of the scaling region on the $L=20$ lattice. We have to use larger lattices that go beyond the capability of the exact diagonalization method due to the exponential growth in the memory usage. Therefore, we use an advanced classical computational method known as the tensor network and in particular, the matrix product state, for simulating the $1+1$D Ising Hamiltonian. The MPS represents a state vector in terms of local tensors on every lattice site that are contracted with nearest neighbors. For tensors with a bond dimension upper limit $D$, the total memory cost is proportional to $LD^2$, rather than exponential in $L$. The bond dimension upper limit caps how much bipartite entanglement the MPS state can contain. Therefore, the MPS method is best for physical systems with low entanglement such as ground states~\cite{PhysRevB.73.094423,Hastings2007,Eisert_2010} and thermal states at high temperature~\cite{PhysRevLett.93.207204,PhysRevLett.100.070502,PhysRevB.98.235154}. As mentioned in the beginning of Sec.~\ref{sec:hydro}, the calculation of the real-time symmetric correlators can be thought of as simulating the dynamics of a thermal state after a local perturbation. Thus, we expect the MPS method to be useful in calculating these correlators.

In this section, we will mainly use a periodic lattice of size $L=32$ with $\xi_{\rm lat}=0.01$ unless specified otherwise. We will study values of $|h_z|$ down to $0.05$, beyond which $|h_z|^{8/15}L$ may be too small to support the scaling region.

\subsection{Mass Gap and Beta Function}
\label{sec:beta_func}
We first study how the mass gap changes with the lattice coupling $h_z$ in order to perform scale fixing and understand the renormalization group equation of the theory, which is crucial for taking the continuum limit. We use the Density Matrix Renormalization Group (DMRG) method implemented in the ITensorMPS package~\cite{itensor,itensor-r0.3} to find the lowest three eigenstates of the Ising lattice Hamiltonian.\footnote{The third eigenstate is not used here, but will be used in the next subsection.} The weight factor for finding excited states is set to be 100, much larger than the mass gap and any momentum excitation energy. We have tested the MPS calculation up to the bond dimension upper limit $D=5000$ and the cutoff $10^{-15}$ and confirmed the convergence.

\begin{figure}[h]
\centering
\includegraphics[width=0.45\textwidth]{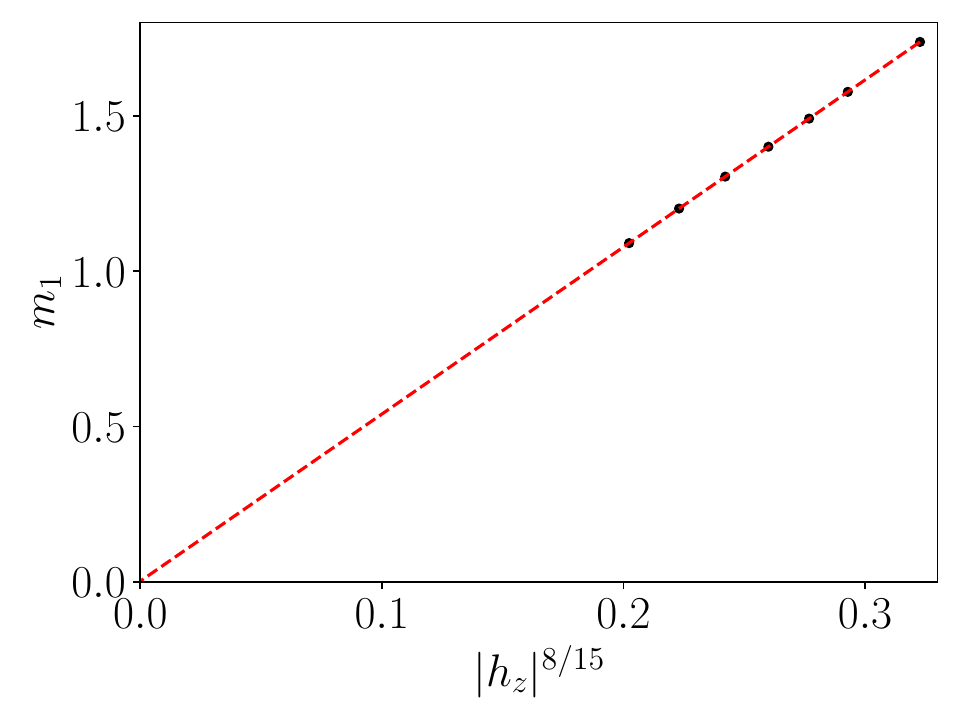}
\caption{Mass gap on a periodic $L=32$ lattice with $\xi_{\rm lat}=0.01$ as a function of the coupling (black points). The red dashed line is a linear fit. The fitted slope is 5.375 and the fitted intercept is $2.790\times10^{-3}$, up to four significant numbers. This linear line establishes the beta function of the theory.}
\label{fig:mass_gap}
\end{figure}

Figure~\ref{fig:mass_gap} shows how the mass gap $m_1$ varies with $|h_z|^{8/15}$ on the $L=32$ lattice with $\xi_{\rm lat}=0.01$ in the coupling range $h_z\in[-0.12,-0.05]$, which exhibits an almost exact linear dependence through the origin. In terms of the physical value of the mass gap $m_1^{\rm phy}$ (Nature does not tell us this value), we know from a dimensional analysis on a line of constant physics
\begin{align}
    m_1 = a\, m_1^{\rm phy} \,,
\end{align}
where $a$ denotes the lattice spacing. The linear dependence then leads to
\begin{align}
    a \propto |h_z|^{8/15} \,,
\end{align}
which gives the beta function
\begin{align}
    \frac{{\rm d}}{{\rm d} \ln a} \ln |h_z|= \frac{15}{8} \,.
\end{align}
This renormalization group equation is also the reason why we stated that $|h_z|^{8/15}L$ effectively measures the physical size of the system earlier.

\subsection{Speed of Light}
In the previous subsection, we showed how the spatial lattice spacing depends on the bare coupling. Now we study the temporal direction. In the lattice Hamiltonian formulation, time is not discretized. But the unit of time is arbitrary since we can multiply the Hamiltonian by an overall constant without changing any physics. In other words, we do not know the speed of light $c$ apriori from the Hamiltonian~\eqref{eqn:H} since it is not derived from a field theory Lagrangian in which $c=1$ is set. We only know $c=2$ in the critical Ising mode, which is defined at $J=h_x=-1$ and $h_z=0$.

To determine the speed of light, we study how the energy of the lowest stable particle changes with its momentum and use the relativistic dispersion relation on the lattice
\begin{align}
\label{eqn:dispersion}
    [E_1(k)]^2 = m_1^2 + 4c^2 \sin^2(k/2) \,,
\end{align}
where $k = 2\pi n_k/L$ for $n_k\in \mathbb{Z}_L$. In the static mass term $c^4$ is not included because both $E_1(k)$ and $m_1$ are obtained from the eigenenergies of the Hamiltonian, i.e., they are measured in the same unit, while the lattice momentum is measured in a different unit. This is why $c$ is not necessarily unity.

\begin{figure}[b]
\centering
\includegraphics[width=0.45\textwidth]{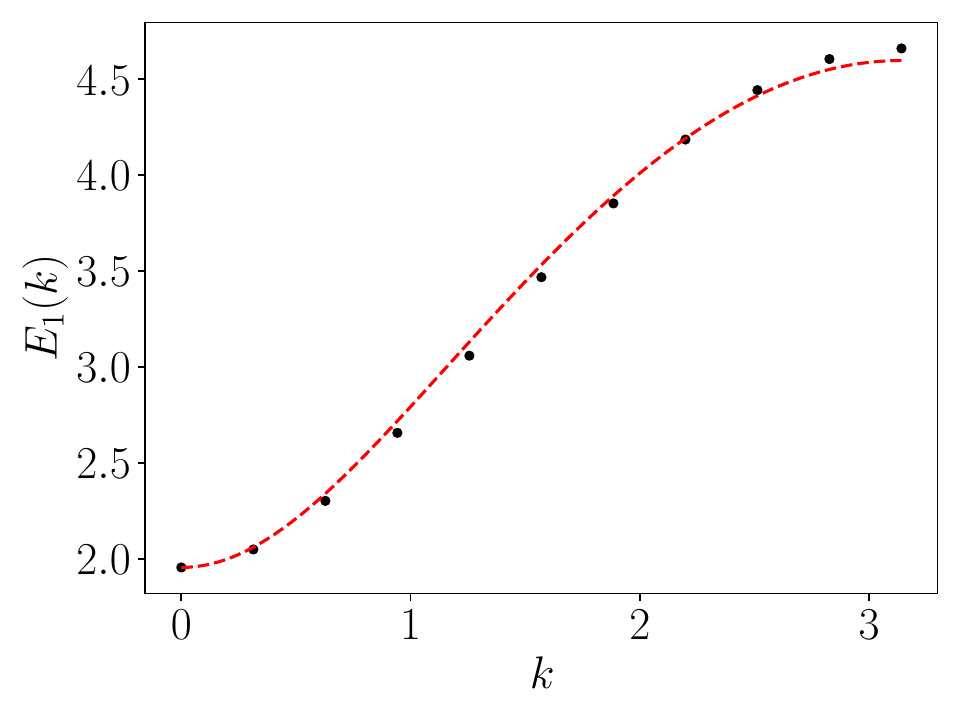}
\caption{Dispersion on a periodic $L=20$ lattice with $\xi_{\rm lat}=0.01$ and $h_z=-0.15$. The black points are obtained from exact diagonalization. The red dashed line is a fit using Eq.~\eqref{eqn:dispersion}. The fitted speed of light is $c=2.080$, up to three digits.}
\label{fig:dispersion}
\end{figure}

If we can calculate the values of $E_1(k)$ for a range of $k$, we can fit $c$ from the calculated dispersion relation. This can be done by exact diagonalization on a periodic $L=20$ lattice. Figure~\ref{fig:dispersion} shows such a calculation for $\xi_{\rm lat}=0.01$ and $h_z=-0.15$. The dispersion relation can be approximately described by Eq.~\eqref{eqn:dispersion}. It provides another evidence for the emergence of a relativistic quantum field theory. We attribute the small deviation of the fit from the numerical results to the opening of the two-particle threshold as energy increases and finite size effect. A similar fit can be found in Fig.~14 of Ref.~\cite{Jha:2024jan}.

\begin{figure}[t]
\centering
\includegraphics[width=0.45\textwidth]{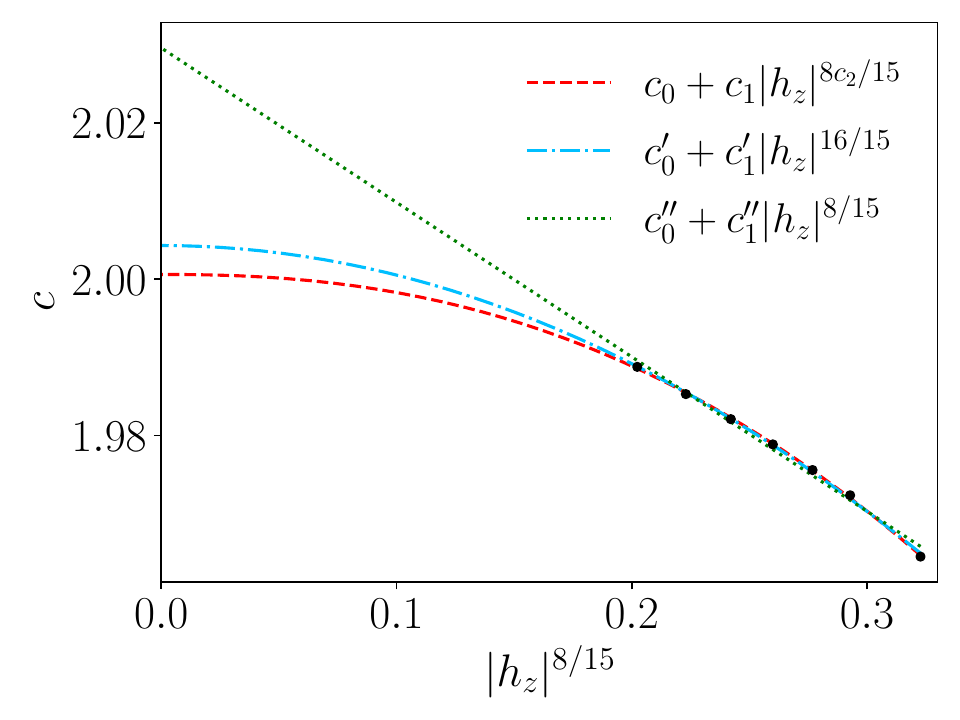}
\caption{Coupling dependence of the speed of light (black points) calculated from the energies $E_1(0)$ and $E_1(2\pi/L)$ on a periodic $L=32$ lattice with $\xi_{\rm lat}=0.01$ by using the relativistic dispersion relation on the lattice~\eqref{eqn:dispersion}. $|h_z|^{8/15}$ is proportional to the lattice spacing $a$. The red dashed line is a fit of the form $c_0 + c_1|h_z|^{8c_2/15}$, the blue dot-dashed line is a fit that is quadratic in $a$, $c_0' + c_1'|h_z|^{16/15}$, and the green dotted line is a fit that is linear in $a$, $c_0'' + c_1''|h_z|^{8/15}$. The fitted parameter values are $c_0=2.0006$, $c_1=-0.50789$, $c_2=2.3416$, $c_0'=2.0043$, $c_1'=-0.37813$, $c_0''=2.0296$, and $c_1''=-0.19777$,
up to five significant numbers.}
\label{fig:c_hz}
\end{figure}

In order to understand how the speed of light changes in the continuum limit on the $L=32$ lattice with $\xi_{\rm lat}=0.01$, we use the MPS results and calculate $c$ by utilizing Eq.~\eqref{eqn:dispersion} and the energies of the first and second excited states of the lattice theory, which are $E_1(0)$ and $E_1(2\pi/L)$, respectively, for the range of couplings studied. The dependence of the calculated $c$ on $|h_z|^{8/15}$ is shown in Fig.~\ref{fig:c_hz}. Since $c$ is computed from the spectral property of the Hamiltonian, we can understand the dependence of $c$ on the lattice spacing $a\propto |h_z|^{8/15}$ in terms of power corrections to the Hamiltonian that vanish in the $a\to0$ limit. Since the Hamiltonian with the sign of $h_z$ flipped can be obtained from the original Hamiltonian by a unitary transformation
\begin{align}
\label{eqn:Qflipspin}
Q = \prod_{i=0}^{L-1}\sigma^x_i \,,
\end{align}
which flips the spin everywhere, the spectral properties of these two Hamiltonians are exactly the same. Thus, we expect $c(h_z) = c(-h_z)$, i.e., $c$ only depends on $|h_z|$. As both $|h_z|^{8/15}$ and $h_z$ appear in Eq.~\eqref{eqn:H}, we expect the corrections to be integer powers of $|h_z|^{8/15}$ and $h_z^2$, which are proportional to $a$ and $a^{15/4}$, respectively.
The dependence of $c$ on $a$ in Fig.~\ref{fig:c_hz} is fitted by three functions: an algebraic function $c_0 + c_1|h_z|^{8c_2/15}$, a function $c_0'+c_1'|h_z|^{16/15}$ quadratic in $a$, and a function $c_0''+c_1''|h_z|^{8/15}$ linear in $a$. We do not consider integer powers of $a^{15/4}$ as it is already a higher-order power compared with $a$ and $a^2$. The extrapolated speeds of light in the continuum limit are $c_0=2.0006$, $c_0'=2.0043$, and $c_0''=2.0296$, up to five significant numbers. The values obtained from the two nonlinear fits are closer to $2$, the speed of light in the critical Ising model, and are close to each other. The fitted exponent in the first algebraic function is also close to $2$. Therefore, we suspect that the dominant power correction to the Hamiltonian is at order $a^2$ and the speed of light in the continuum limit is $c=2$. In other words, if we want to set up a lattice formulation for the Ising field theory that has the unit $c=1$ in the continuum limit, we will just multiply the Hamiltonian in Eq.~\eqref{eqn:H} by a factor of $1/2$. Whenever a physical quantity is obtained on the lattice from the ratio of energy and momentum raised to the $k$th power, we need to multiply it by a factor of $1/2^k$ in order to express it in a unit with $c=1$.

\subsection{MPS Calculation of $G_s^{10}$}

The method we use to calculate a generic real-time symmetric correlator via the MPS consists of three steps: (1) thermal state preparation, (2) applying local perturbation, and (3) time evolution and measurements of local observables. We use PBC throughout. Using open or fixed boundary conditions can enlarge the validity time range of the MPS calculation under a fixed bond dimension upper limit. However, as we aim at demonstrating hydrodynamization and extracting transport coefficients with the boundary effects minimized, we want to focus on the time range before the local perturbation reaches the boundaries, which is more relevant for the physics in heavy ion collision. Thus, the physical impact of the boundary condition choice is small. We use PBC here as it gives well-defined momentum on the lattice.

For the thermal state preparation, we initialize a matrix product operator (MPO) to be the infinite temperature density matrix at $\beta=1/T=0$. Then we apply imaginary time evolution to cool the mixed state to a given temperature $\beta$. This method is known as the purification method~\cite{Feiguin_2005}. We use second-order Trotterization for the imaginary time evolution, which is implemented on the MPO by the Time-Evolving Block Decimation (TEBD) algorithm. The density matrix after the cooling is normalized to have trace 1. We denote the thermal state as $\rho_T$. 

Next we apply a local perturbation at $x=0$ to the thermal state. To minimize the impact of the boundary conditions, we apply the local perturbation at the center of the lattice\footnote{For even-sized lattices, we pick up the left one of the two central sites.} and define this site to be the spatial origin $x=0$. For the example of computing $G_s^{10}$, we take the local perturbation to be $e^{\Delta T^{10}(0)}$, where $\Delta$ is a small parameter that we can choose and $0$ denotes the spatial position. After the perturbation, the state becomes
\begin{align}
    \rho_T &\rightarrow e^{\Delta T^{10}(0)} \rho_T e^{\Delta T^{10}(0)} \,.
\end{align}

\begin{figure}[h]
\centering
\subfloat[$L=32$, $h_z=-0.08$, and $\beta=0.1$.\label{fig:Gs_contour32}]{%
  \includegraphics[width=0.45\textwidth]{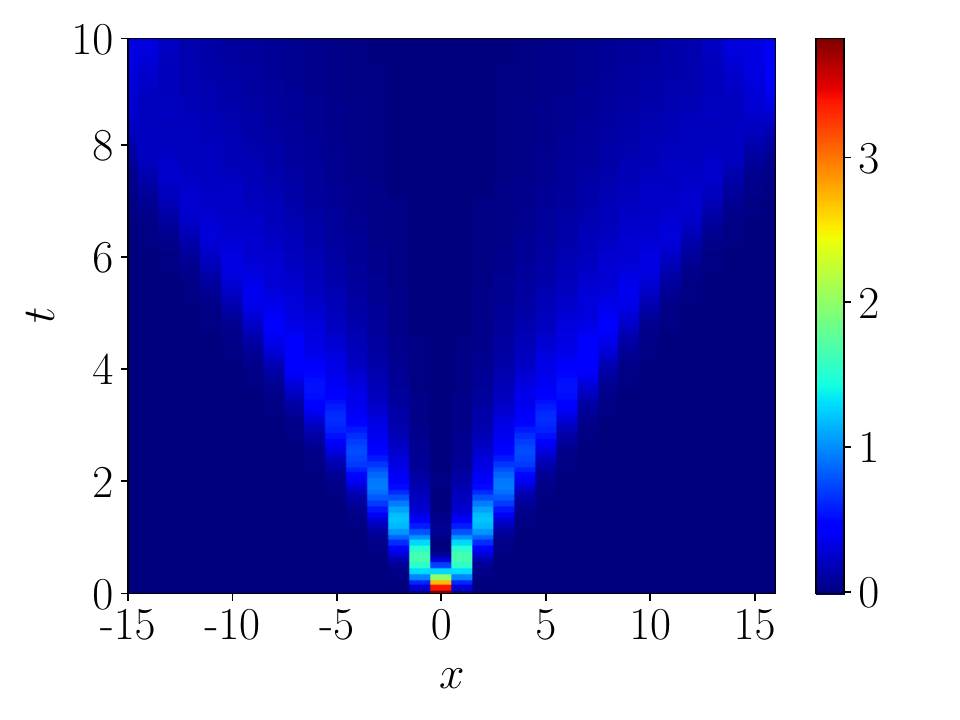}%
}

\subfloat[$L=64$, $h_z=-0.04$, and $\beta=0.1447$.\label{fig:Gs_contour64}]{%
  \includegraphics[width=0.45\textwidth]{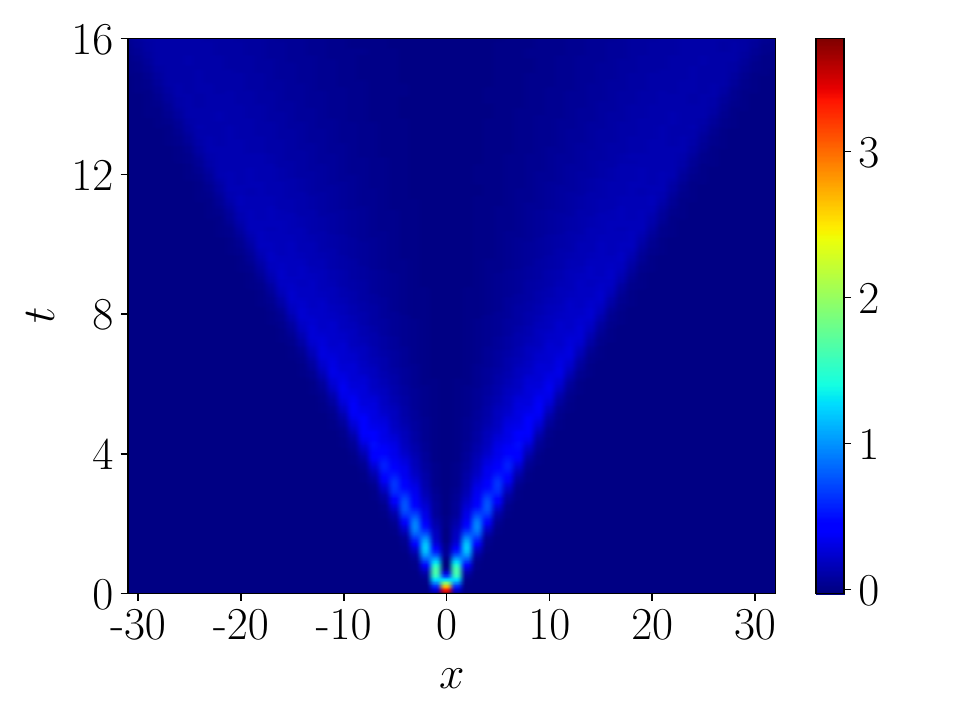}%
}
\caption{Contour plots of MPS-calculated $G_s^{10}(t,x)$ for two lattices with $\xi_{\rm lat}=0.01$ at fixed $T/m_1\approx 7.141$. The result for $L=32$ (a) is obtained with the bond dimension upper limit $D=500$ and the TDVP time step $\Delta t = 0.1$ while that for $L=64$ (b) is with $D=200$ and $\Delta t=0.1$. We use $10$ steps in the second-order Trotterized imaginary time evolution and $\Delta=0.01$ in the local perturbation.}
\label{fig:Gs_contour}
\end{figure}

Finally, we time evolve the perturbed thermal state using the Time-Dependent Variational Principle (TDVP) and measure the local observable $T^{10}$ everywhere on the lattice. We use the ITensor package TensorMixedStates~\cite{houdayer2025tensormixedstates} to implement the TDVP and the measurement. The measurement results can be expressed as
\begin{align}
    \langle T^{10}(x) \rangle_\Delta(t) &= {\rm Tr} [T^{10}(x) e^{-iHt} e^{\Delta T^{10}(0)} \rho_T e^{\Delta T^{10}(0)} e^{iHt} ] \nn\\
    &= {\rm Tr} [T^{10}(t,x) e^{\Delta T^{10}(0)} \rho_T e^{\Delta T^{10}(0)} ]\,,
\end{align}
where we have used the Heisenberg picture for the operator $T^{10}(x)$. When $\Delta$ is small, taking the difference bewteen the cases with $\Delta>0$ and $-\Delta$ leads to
\begin{align}
\label{eqn:G10Delta}
    \langle T^{10}(x) \rangle_\Delta(t) - \langle T^{10}(x) \rangle_{-\Delta}(t) = 2 \Delta G_s^{10}(t,x) + O(\Delta^3) \,,
\end{align}
from which $G_s^{10}(t,x)$ can be obtained up to an accuracy of $O(\Delta^2)$. Contour plots of $G_s^{10}(t,x)$ obtained in this way are shown in Fig.~\ref{fig:Gs_contour} for two different lattice sizes.

\begin{figure*}[t]
\centering
\subfloat[$x=0$.\label{fig:Gs_Px_0_bond_dt}]{%
  \includegraphics[width=0.33\linewidth]{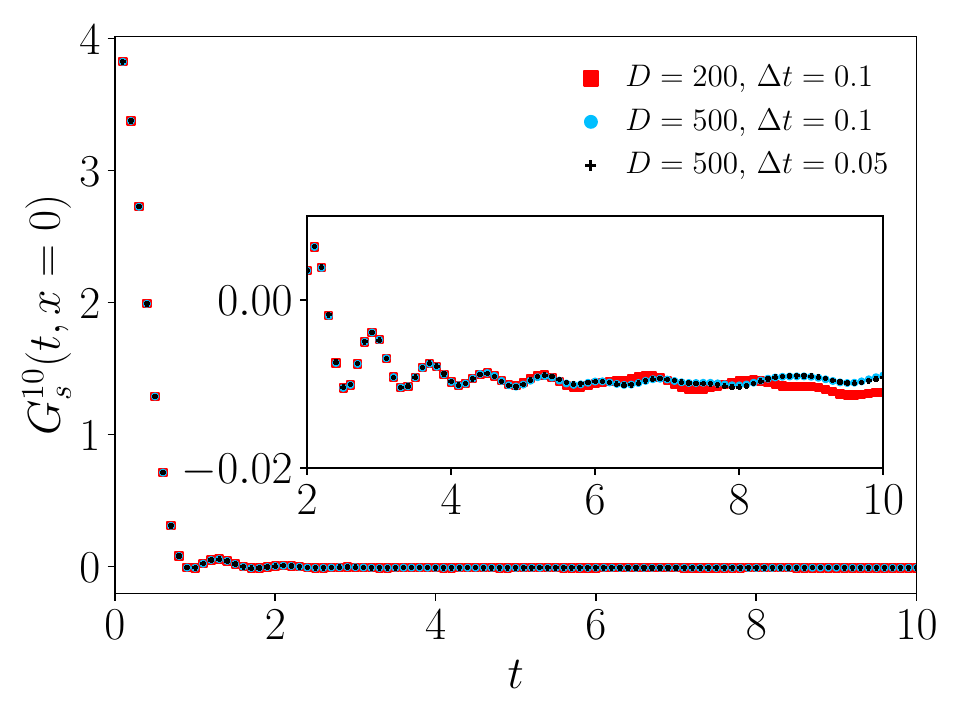}%
}\hfill
\subfloat[$x=2$.\label{fig:Gs_Px_2_bond_dt}]{%
  \includegraphics[width=0.33\linewidth]{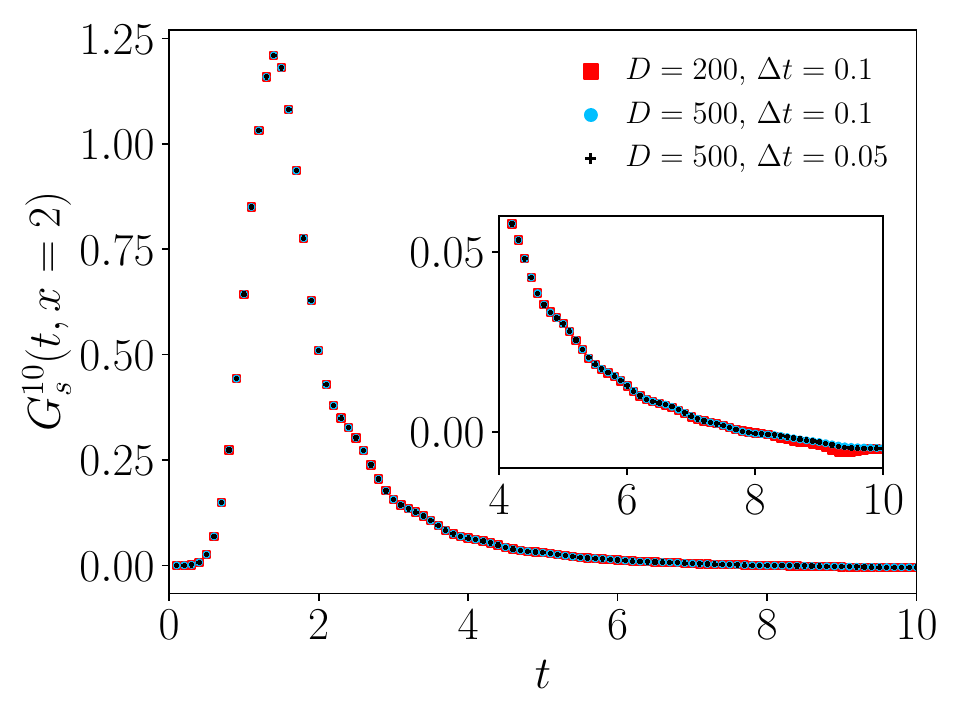}%
}\hfill
\subfloat[$x=4$.\label{fig:Gs_Px_4_bond_dt}]{%
  \includegraphics[width=0.33\linewidth]{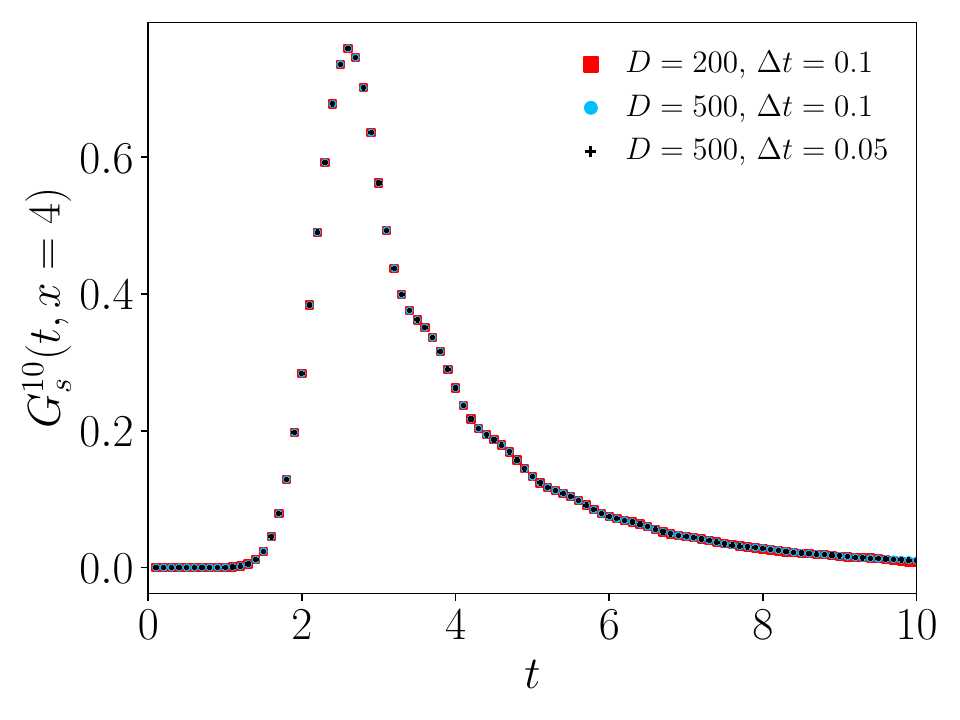}%
}

\subfloat[$x=6$.\label{fig:Gs_Px_6_bond_dt}]{%
  \includegraphics[width=0.33\linewidth]{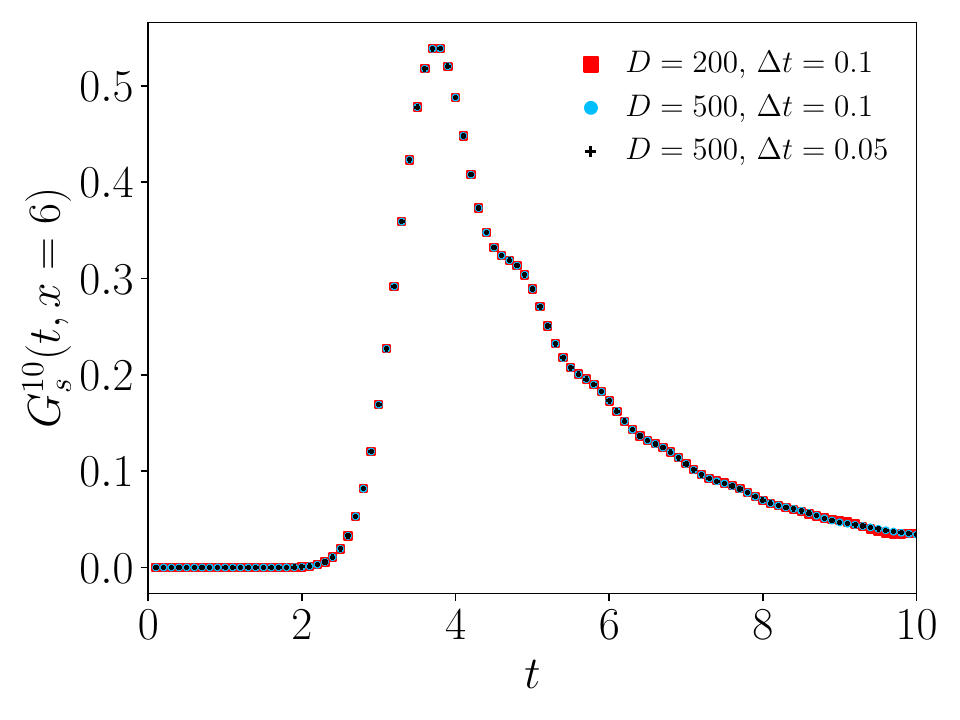}%
}\hfill
\subfloat[$x=8$.\label{fig:Gs_Px_8_bond_dt}]{%
  \includegraphics[width=0.33\linewidth]{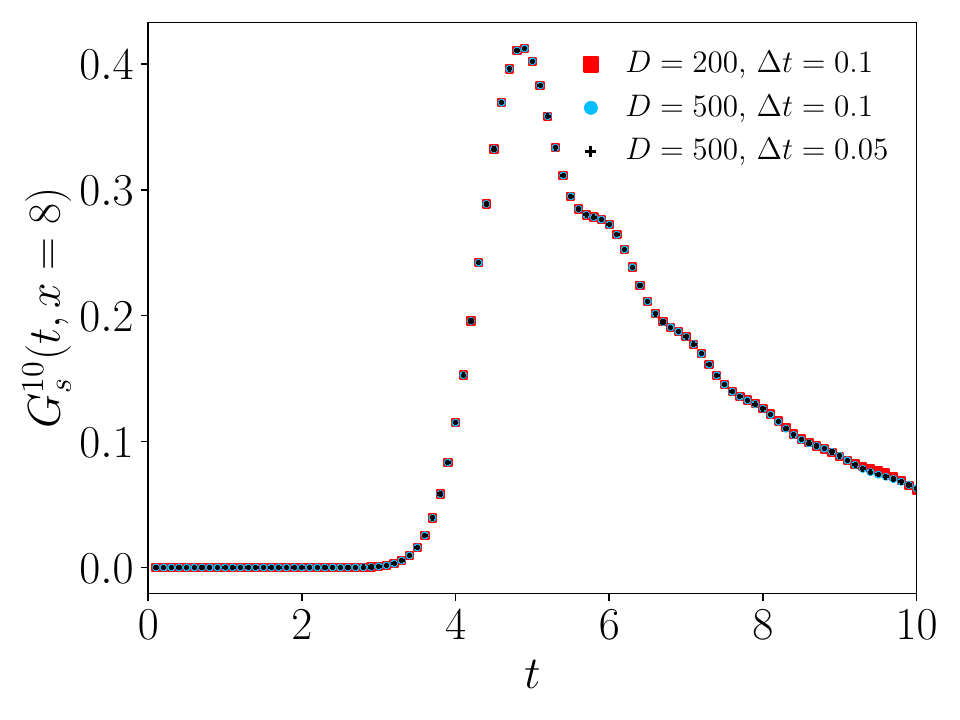}%
}\hfill
\subfloat[$x=10$.\label{fig:Gs_Px_10_bond_dt}]{%
  \includegraphics[width=0.33\linewidth]{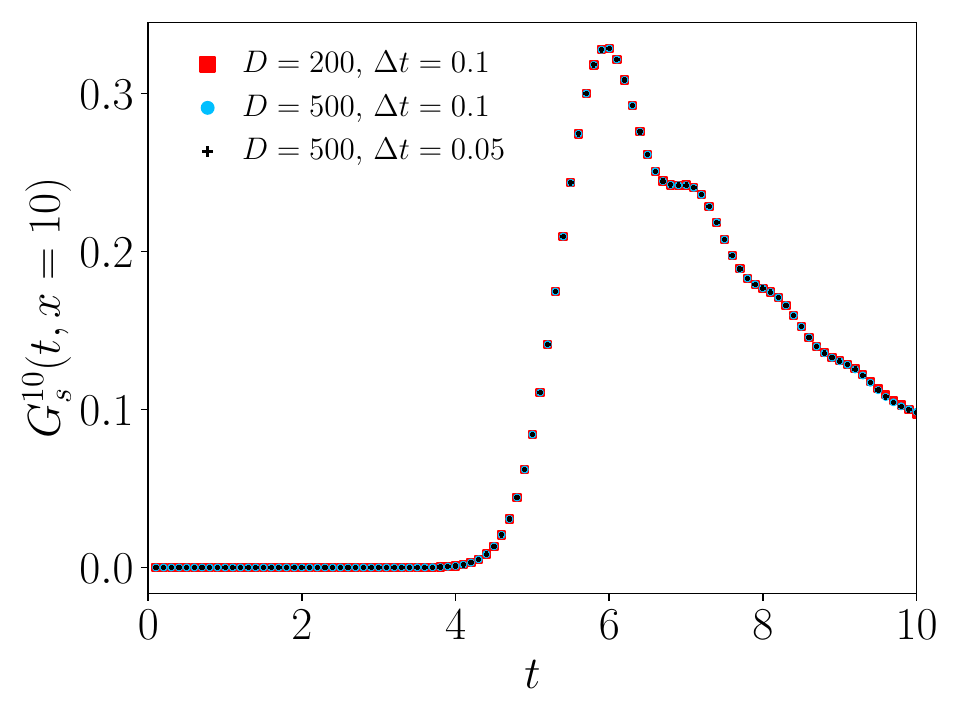}%
}

\subfloat[$x=12$.\label{fig:Gs_Px_12_bond_dt}]{%
  \includegraphics[width=0.33\linewidth]{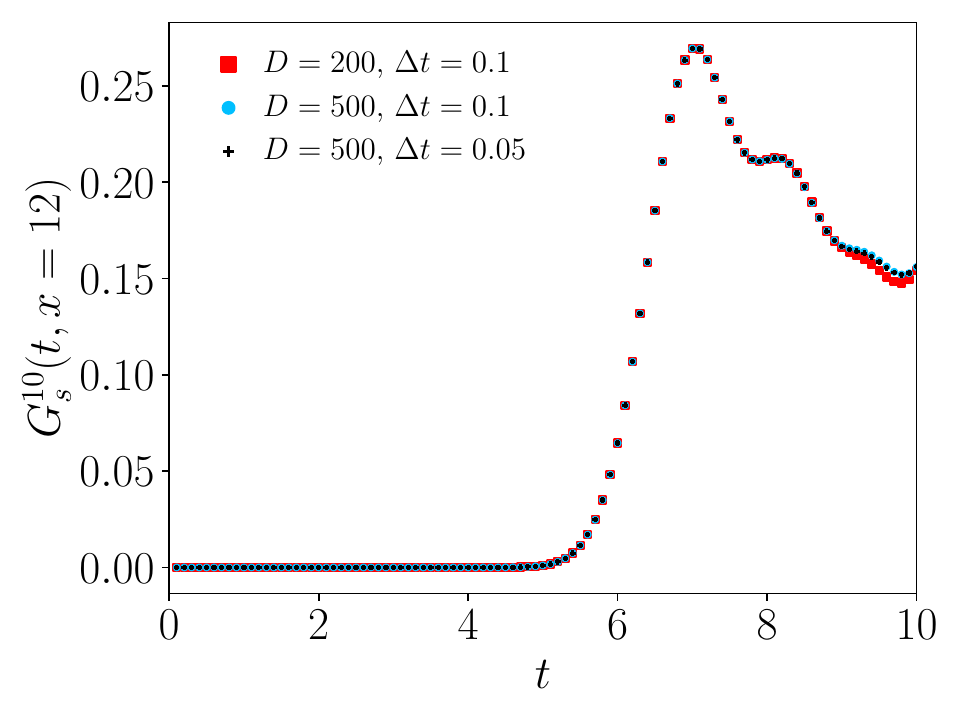}%
}\hfill
\subfloat[$x=14$.\label{fig:Gs_Px_14_bond_dt}]{%
  \includegraphics[width=0.33\linewidth]{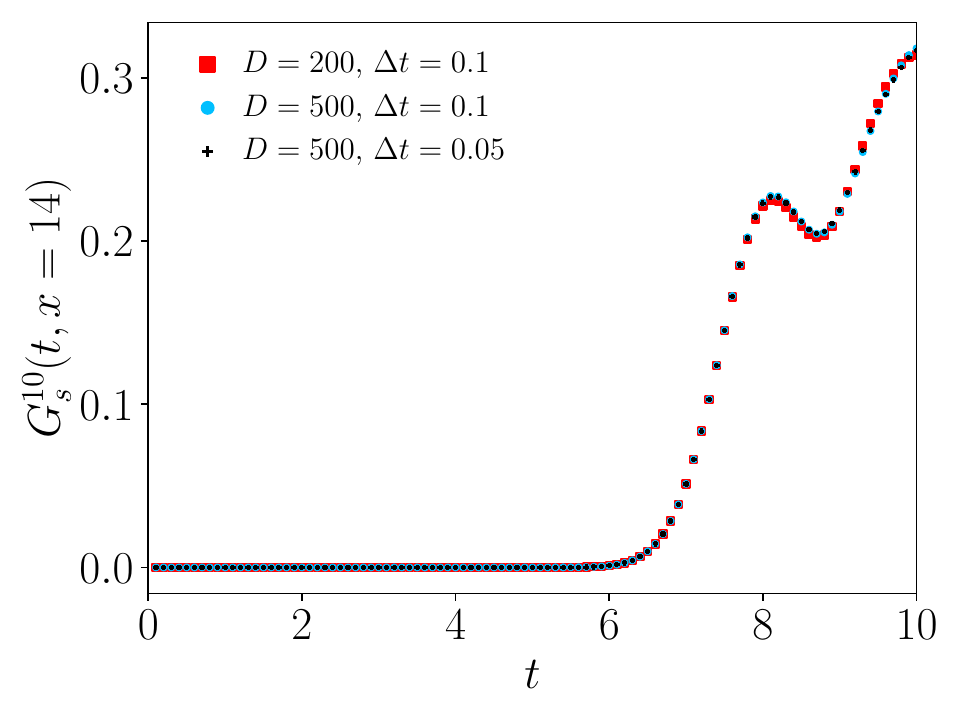}%
}\hfill
\subfloat[$x=16$.\label{fig:Gs_Px_16_bond_dt}]{%
  \includegraphics[width=0.33\linewidth]{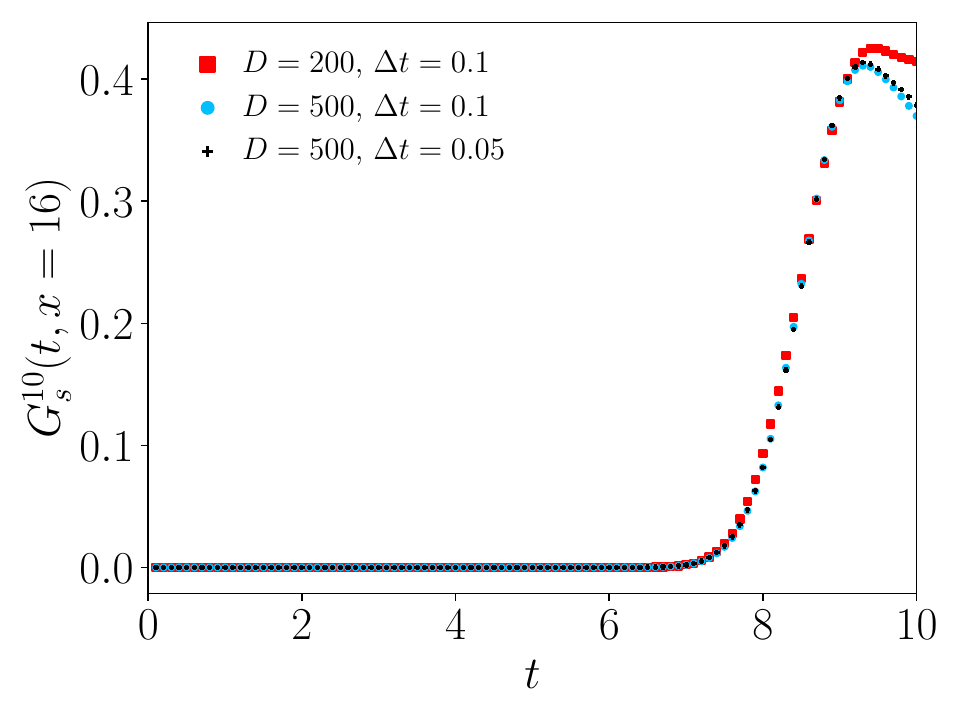}%
}
\caption{MPS calculation results of $G_s^{10}(t,x)$ on a periodic $L=32$ lattice with $\xi_{\rm lat}=0.01$, $h_z=-0.08$, and $T=10$ using two different bond dimension upper limits $D$ and two different TDVP time steps $\Delta t$. We use $10$ steps in the second-order Trotterized imaginary time evolution and $\Delta=0.01$ to apply the local perturbation at $t=0$.}
\label{fig:Gs_Px_32_bond_dt}
\end{figure*}

To understand the systematics of the MPS calculation, we consider a periodic $L=32$ lattice with $\xi_{\rm lat}=0.01$, $h_z=-0.08$, and $\beta=0.1$. We use $N_\tau=10$ steps in the second-order Trotterized imaginary time evolution. The Trotter error of the implemented imaginary time evolution is $O(\beta^3/N^2_\tau) = O(10^{-5})$. We choose $\Delta=0.01$, so the error of reconstructing $G_s^{10}$ via Eq.~\eqref{eqn:G10Delta} is $O(10^{-4})$. The uncertainties associated with the cutoff and bond dimension upper limit in the MPS setup and that originating from the TDVP time step size $\Delta t$ can be studied numerically. Figure~\ref{fig:Gs_Px_32_bond_dt} shows the MPS-calculated $G_s^{10}(t,x)$ for two different bond dimension upper limits and two different TDVP step sizes, where the cutoff of the MPS calculation is chosen to be $10^{-12}$, which will be used in the remaining of the paper unless specified differently. We see that away from the boundaries propagation of the perturbation can already be accurately described by the MPS calculation with the bond dimension upper limit $D=200$ and TDVP step $\Delta t=0.1$. The associated uncertainty is $O(10^{-3})$ as can be seen from the inset of Fig.~\ref{fig:Gs_Px_0_bond_dt}. Next to the right boundary at $x=16$, we observe a large difference between the $D=200$ and $D=500$ calculations in Fig.~\ref{fig:Gs_Px_16_bond_dt}, when the perturbation crosses the boundary around $t=9$. The MPS setup requires a very large bond dimension in order to connect the first and last lattice sites and maintain the periodic boundary conditions. 

We perform similar uncertainty analyses of the MPS calculations on $L=16$ and $L=64$ lattices, for the former of which the MPS results are benchmarked with exact diagonalization results. 
We also study how the maximum bond dimension across the lattice in the MPS calculation grows in the time evolution and how the bond dimension saturation time depends on the cutoff, which determines the accuracy of the MPS calculation at the bond dimension saturation time point. Details can be found in Appendix~\ref{app:mps}. We reach the same conclusion, i.e., to describe the propagation of the perturbation before it crosses the periodic boundary, a MPS calculation with $D=200$ and $\Delta t=0.1$ is good enough.

This has important implications for simulating hydrodynamization on a lattice, which happens rapidly in heavy ion collisions. In the real collision experiment, the system hydrodynamizes without hitting a boundary. This is very different from testing the eigenstate thermalization hypothesis for thermalization. Complete thermalization can only be reached inside a box. If there were no boundary, the system would keep expanding into the vacuum. But hydrodynamization can still happen while the system expands into the vacuum, as long as the system is hot and dense so interactions occur rapidly. As we see above, the MPS calculation only starts to struggle when the perturbation crosses the periodic boundary. So it may be a good tool to simulate hydrodynamization in real time on big lattices, at least in $1+1$D.

\section{Extracting Bulk Viscosity}
\label{sec:bulk}
In the previous section, we discussed the MPS calculation of $G_s^{10}(t,x)$. Now we investigate whether the late-time dynamics is governed by hydrodynamics. If so, we can try to extract the bulk viscosity out of the the symmetric correlator and take the continuum limit.

On a periodic lattice of $L$ sites, we can take a Fourier series of $G_s^{10}(t,x)$ to obtain
\begin{align}
    G_s^{10}(t,k) = \sum_{x=0}^{L-1} e^{ikx} G_s^{10}(t,x) \,,
\end{align}
where $k={2\pi}n_k/{L}$ for $n_k\in\{0,1,\dots,L-1\}$. $G_s^{10}(t,k)$ is real because $G_s^{10}(t,x) = G_s^{10}(t,-x)$ by its symmetric nature and translation invariance on the periodic lattice. 

\subsection{Recover Continuum Momentum Conservation}
For $k=0$, the physical meaning of $G_s^{10}(t,k=0)$ is the total momentum of the perturbed system. As mentioned in Sec.~\ref{sec:Gs}, we expect the total momentum to be conserved in the continuum limit. We can test this by using the numerical results.

\begin{figure}[h]
\centering
\includegraphics[width=0.45\textwidth]{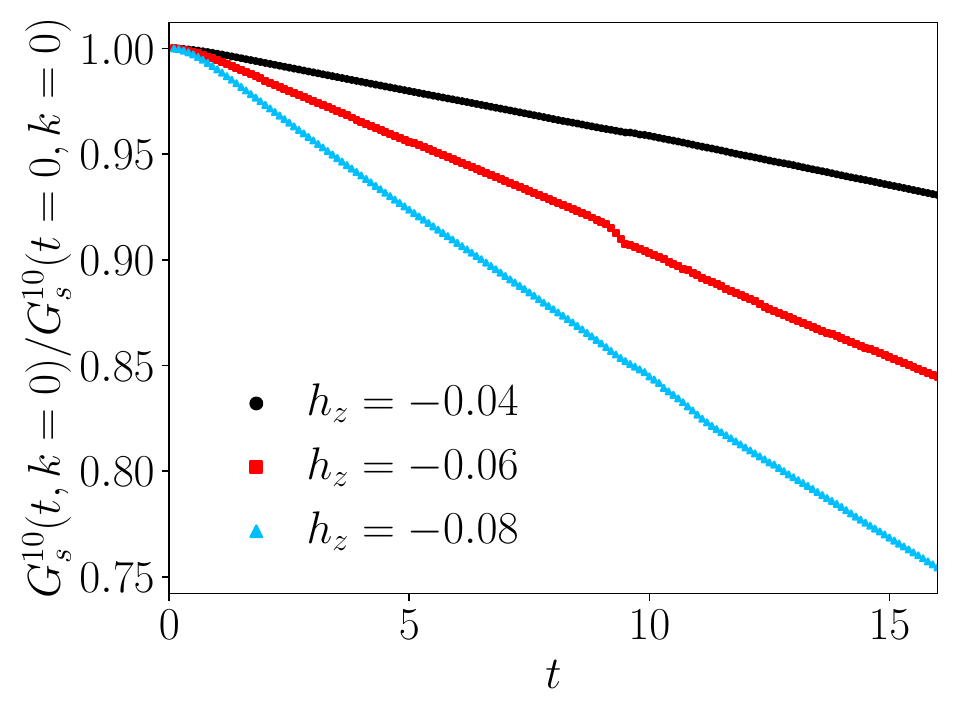}
\caption{Ratio of $G_s^{10}(t,k=0)$ to $G_s^{10}(t=0,k=0)$ as a function of time on a periodic $L=64$ lattice with $\xi_{\rm lat}=0.01$ and $T/m_1\approx7.141$ for three different couplings. The calculation is performed via the MPS method with the bond dimension upper limit $D=200$. The small kink in the $h_z=-0.06$ case disappears when the bond dimension upper limit increases, e.g., when $D=300$.}
\label{fig:p_conserve}
\end{figure}

Figure~\ref{fig:p_conserve} shows the ratio of $G_s^{10}(t,k=0)$ to $G_s^{10}(t=0,k=0)$ as a function of time on a periodic $L=64$ lattice with $\xi_{\rm lat}=0.01$ and $T/m_1\approx 7.141$ for three different couplings ranging from $h_z=-0.08$ to $h_z=-0.04$. If the total momentum is conserved, the ratio will remain unity in the time evolution. We see that the ratio decreases with time but decreases less as the coupling becomes smaller. The momentum conservation is gradually restored when taking the coupling towards zero. We emphasize again that on a lattice of fixed size, one cannot take the coupling to be arbitrarily close to zero as this will go out of the scaling region.

\begin{figure}[h]
\centering
\includegraphics[width=0.45\textwidth]{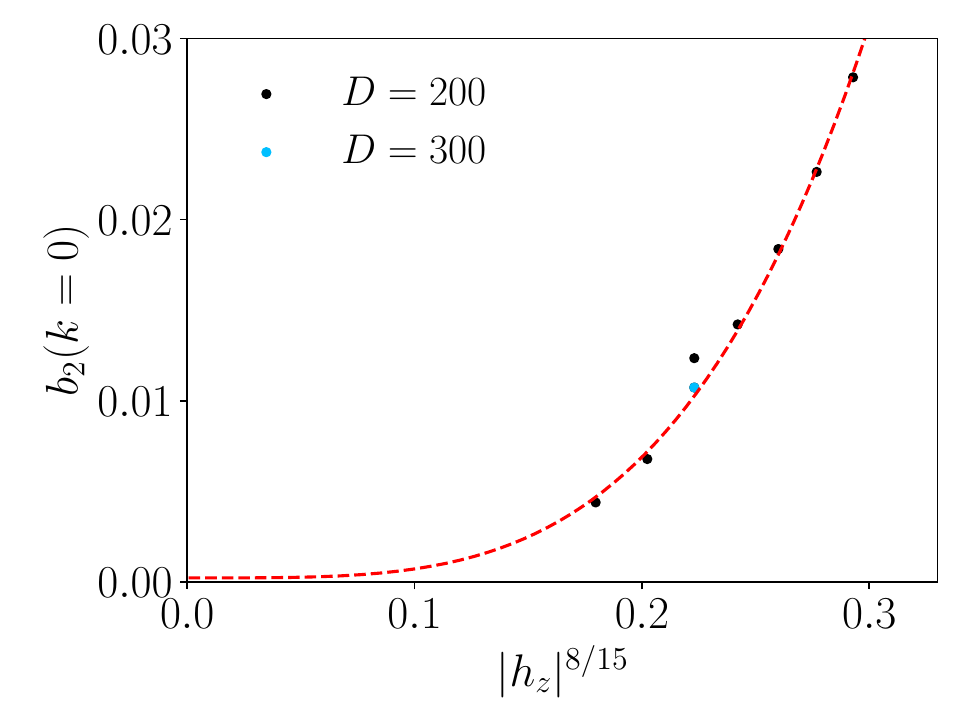}
\caption{Coupling dependence of the decay rate of the total momentum. The black points are obtained from a MPS calculation with the bond dimension upper limit $D=200$ while the blue point is from that with $D=300$. The red dashed line is a fit of the form $b_2^{(0)}+kh_z^2$, corresponding to $b_2^{(0)}+k'a^{15/4}$. The fitted function describes all the black points well except for the point at $h_z=-0.06$. This is due to the small kink observed in Fig.~\ref{fig:p_conserve}, which disappears when the bond dimension upper limit is further increased. At $h_z=-0.06$, the point from $D=300$ is used in the fitting instead of the one from $D=200$.
The fitted parameter values are $b_2^{(0)}=0.0002267$ and $k=2.787$, up to four significant numbers. This fit function gives a value of $b_2^{(0)}$ that is much closer to $0$ than fits using other power laws such as $a$, $a^2$, and $a^{15/8}$ for the $a$ dependence, which give $-0.03455$, $-0.01085$, and $-0.01243$, respectively for the intercepts.}
\label{fig:b2_k0}
\end{figure}

In order to understand how fast the continuum momentum conservation is restored,
we fit to the decay of the total momentum on the same $L=64$ lattice a simple exponential function
\begin{align}
    b_0 e^{-b_2(k=0) t} \,,
\end{align}
in the time domain $t\in[5,16]$ for a range of couplings that will be used in the continuum extrapolation later. 
The fitting parameters $b_0$ and $b_2(k=0)$ have similar physical meanings as those in another function used later for fitting the sound modes. The coupling dependence of $b_2(k=0)$ is shown in Fig.~\ref{fig:b2_k0}. This dependence can be understood from Eq.~\eqref{eqn:comm_H_Jtot}, which vanishes linearly in $h_z$ in the continuum limit. On the other hand, the momentum density operator on the lattice defined in Eq.~\eqref{eqn:T10lattice} is invariant under the overall spin flip given by $Q$ in Eq.~\eqref{eqn:Qflipspin}. Thus, we expect the lattice momentum density operator to converge to the continuum limit with corrections that are even powers of $h_z$. The leading correction is of order $h_z^2 \propto a^{15/4}$. This is confirmed by the red dashed line in Fig.~\ref{fig:b2_k0}, which is a fit of the form $b_2^{(0)}+kh_z^2$. This fit function with the power $a^{15/4}$ gives an intercept that is much closer to $0$ than other power-law functions. We use the proximity of the intercept to zero as the criterion for fitting quality, since the total momentum is conserved in the continuum limit and thus the decay rate vanishes.

\begin{figure*}[t]
\centering
\subfloat[$n_k=1$.\label{fig:Gs_Px_k1}]{%
  \includegraphics[width=0.33\linewidth]{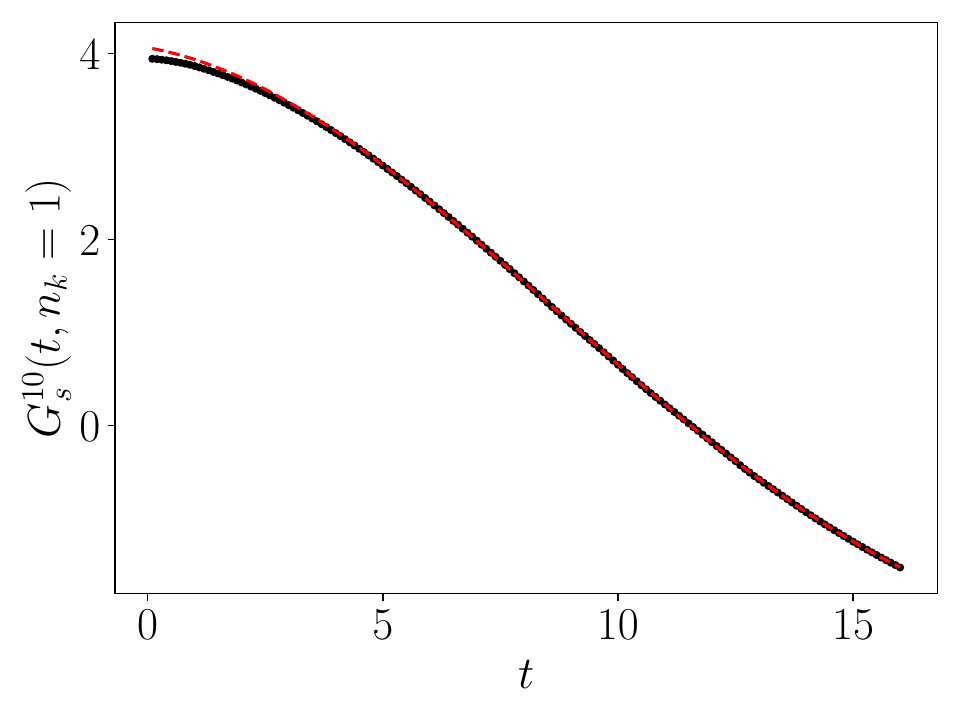}%
}\hfill
\subfloat[$n_k=2$.\label{fig:Gs_Px_k2}]{%
  \includegraphics[width=0.33\linewidth]{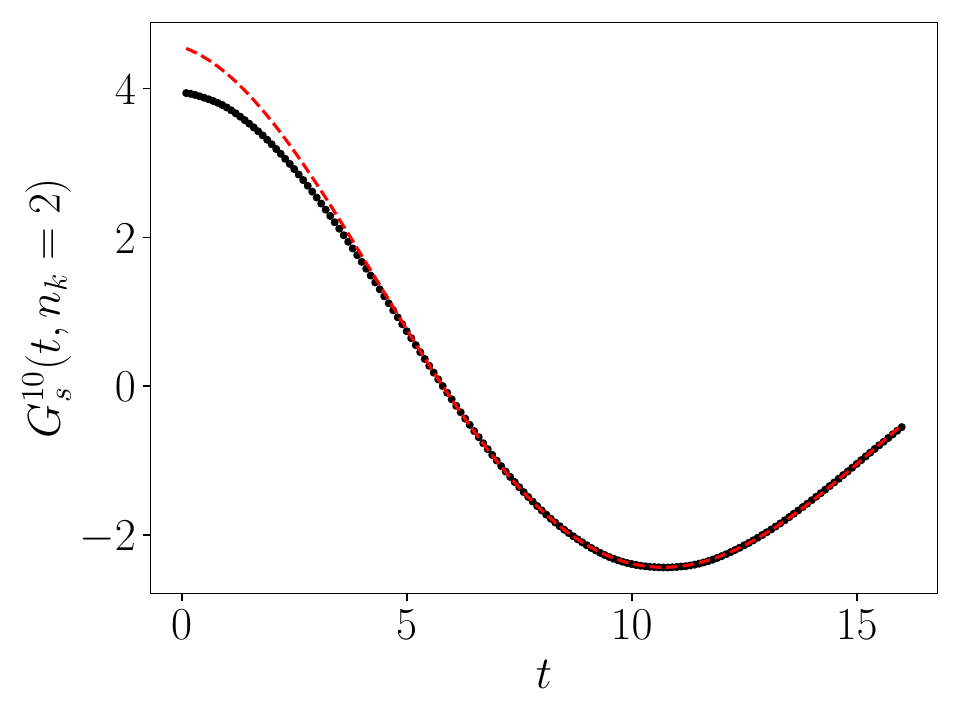}%
}\hfill
\subfloat[$n_k=3$.\label{fig:Gs_Px_k3}]{%
  \includegraphics[width=0.33\linewidth]{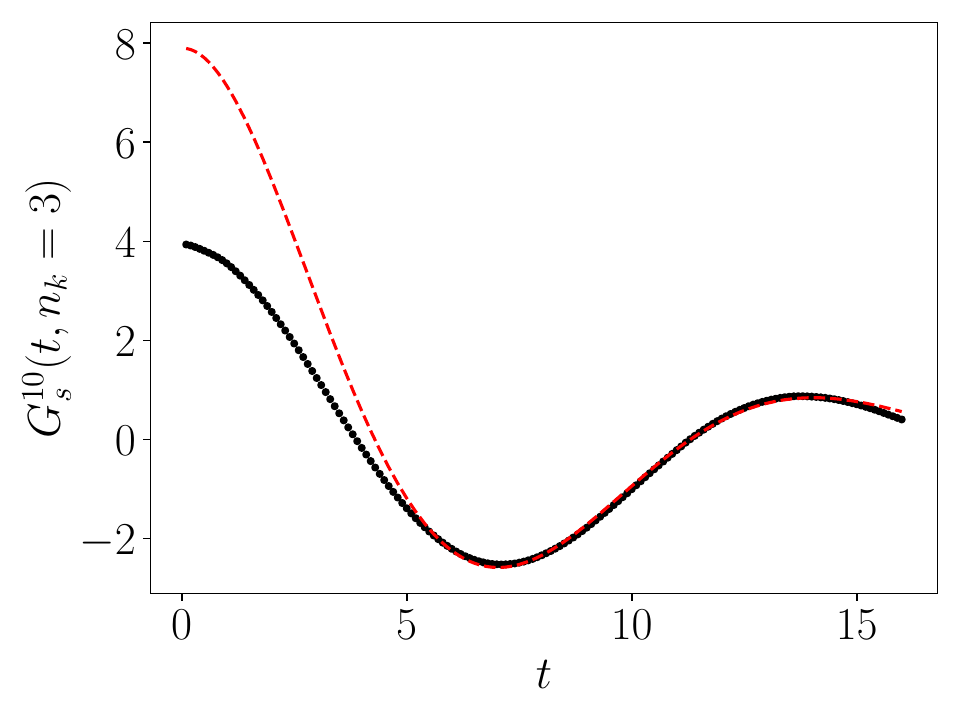}%
}

\subfloat[$n_k=4$.\label{fig:Gs_Px_k4}]{%
  \includegraphics[width=0.33\linewidth]{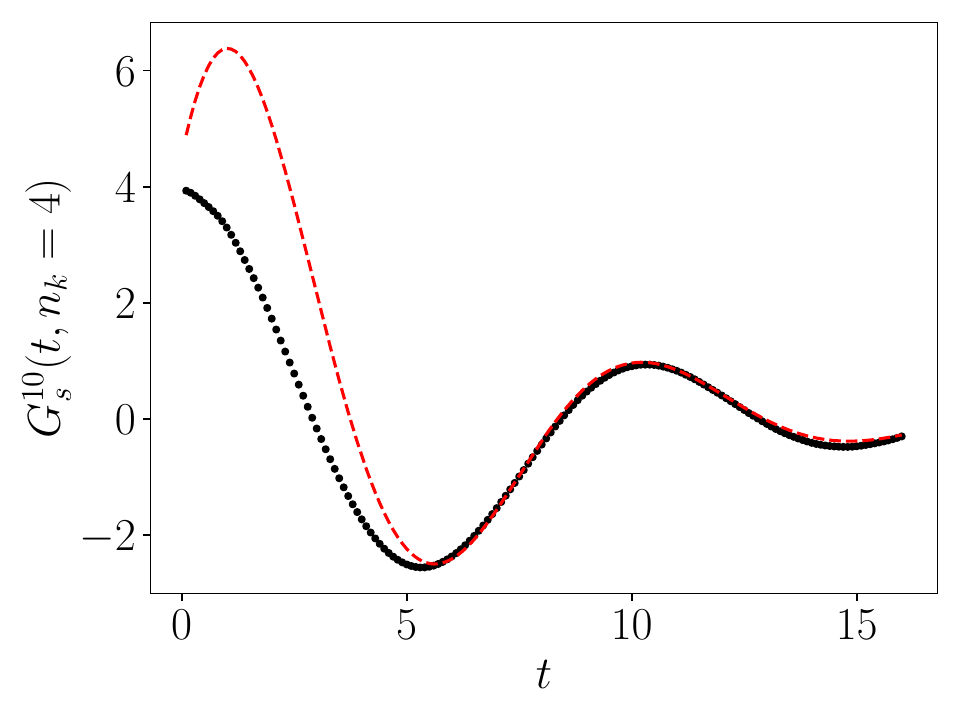}%
}\hfill
\subfloat[$n_k=5$.\label{fig:Gs_Px_k5}]{%
  \includegraphics[width=0.33\linewidth]{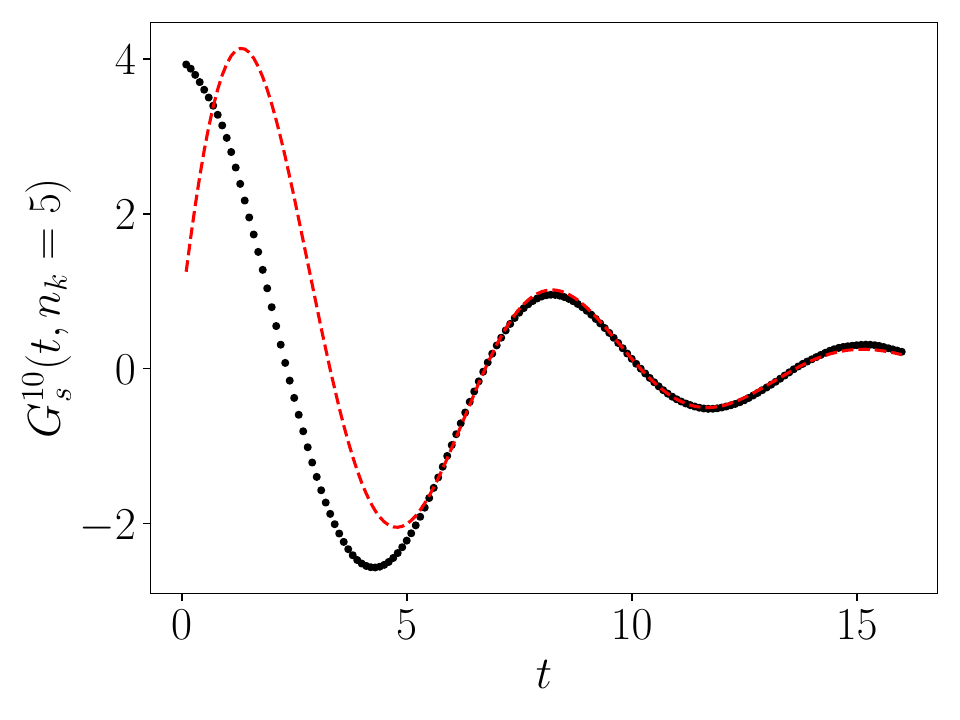}%
}\hfill
\subfloat[$n_k=6$.\label{fig:Gs_Px_k6}]{%
  \includegraphics[width=0.33\linewidth]{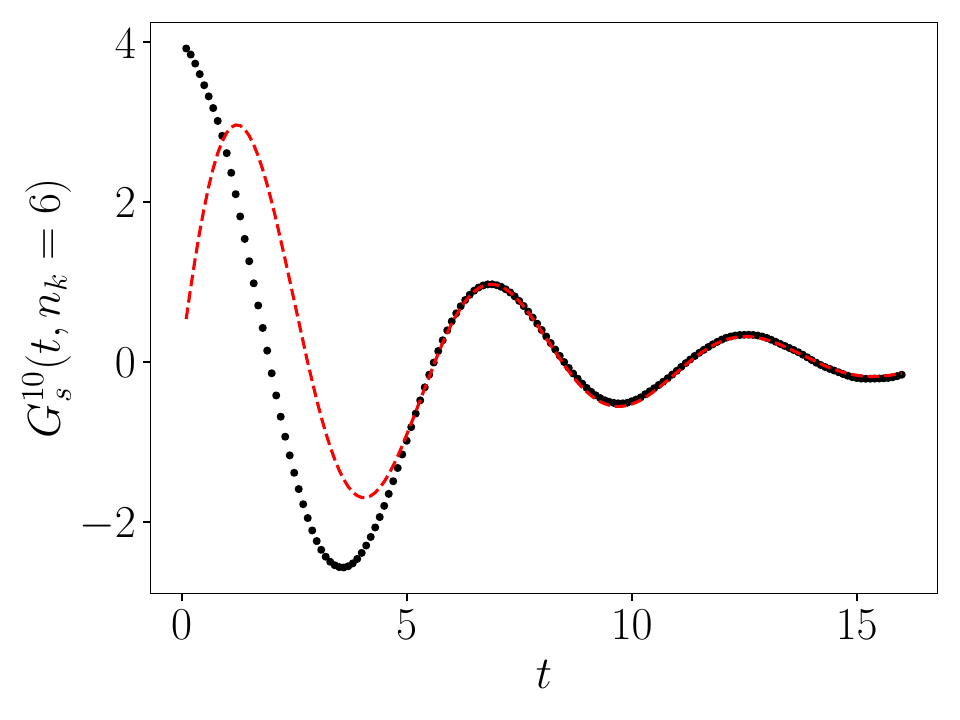}%
}

\subfloat[$n_k=7$.\label{fig:Gs_Px_k7}]{%
  \includegraphics[width=0.33\linewidth]{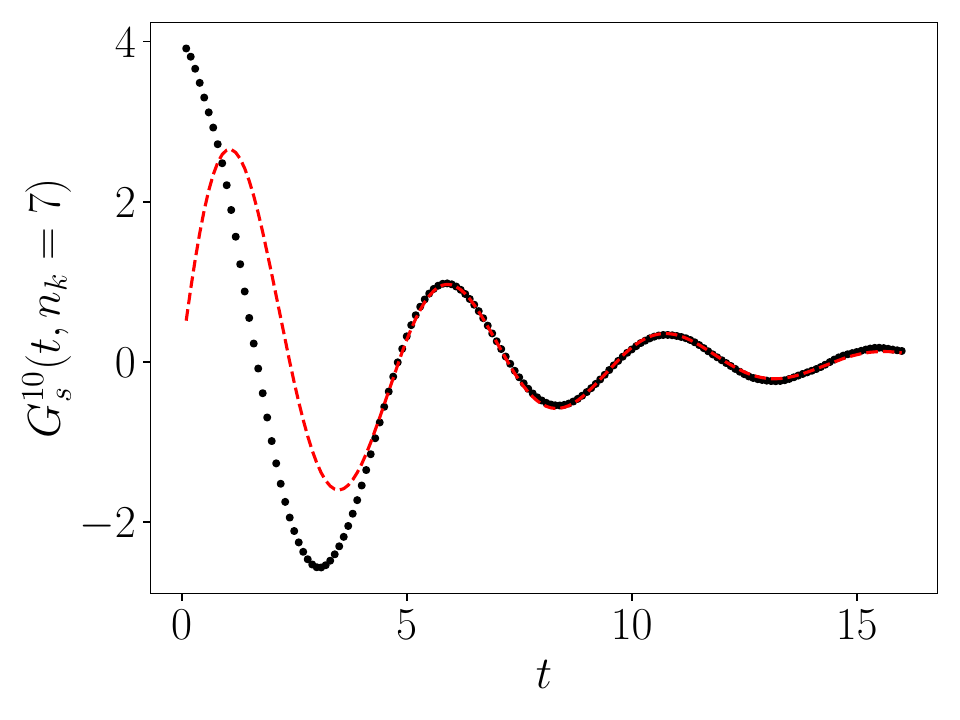}%
}\hfill
\subfloat[$n_k=8$.\label{fig:Gs_Px_k8}]{%
  \includegraphics[width=0.33\linewidth]{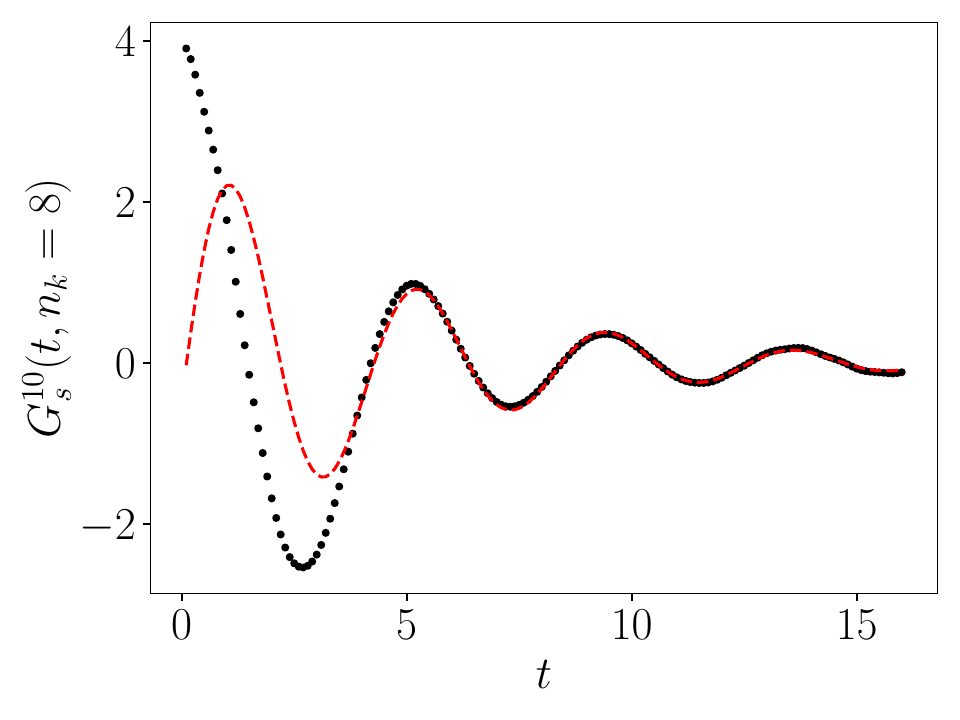}%
}\hfill
\subfloat[$n_k=9$.\label{fig:Gs_Px_k9}]{%
  \includegraphics[width=0.33\linewidth]{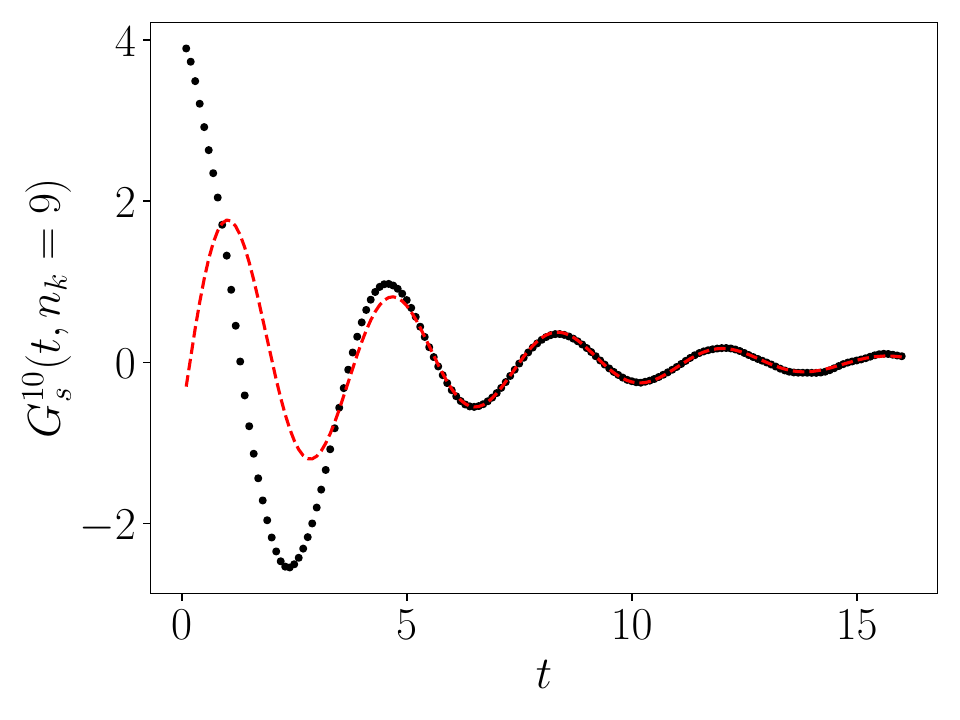}%
}
\caption{$G_s^{10}(t,k=2\pi n_k/L)$ on a periodic $L=64$ lattice with $\xi_{\rm lat}=0.01$, $h_z=-0.08$, and $T=10$. The black points are obtained from a MPS calculation with $D=200$, $\Delta t=0.1$, $N_\tau=10$, and $\Delta=0.01$. The red dashed lines are obtained by fitting the form $b_0 e^{-b_2t} \cos(b_1t + b_3)$ to the black points in the time domain $t\in[5,16]$ and extending the fitted functions to the full time range. The fitted parameter values are listed in Table~\ref{tab:sound_fit}.}
\label{fig:Gs_Px_64_k}
\end{figure*}

\begin{table*}[t]
\centering
\begin{tabular}{|c|c|c|c|c|}
\hline
$n_k$ & $b_0$ & $b_1$ & $b_2$ & $b_3$ \\
\hline
1 & $4.0651 \pm 0.0156$ & $0.1395 \pm 0.0008$ & $0.0276 \pm 0.0002$ & $6.2440 \pm 0.0090$ \\
2 & $4.5668 \pm 0.0243$ & $0.2797 \pm 0.0002$ & $0.0568 \pm 0.0005$ & $6.2319 \pm 0.0014$ \\
3 & $8.4177 \pm 0.1637$ & $0.4483 \pm 0.0020$ & $0.1595 \pm 0.0022$ & $5.9305 \pm 0.0182$ \\
4 & $8.1971 \pm 0.1816$ & $0.6825 \pm 0.0042$ & $0.2039 \pm 0.0030$ & $5.2986 \pm 0.0356$ \\
5 & $5.5538 \pm 0.1156$ & $0.9108 \pm 0.0032$ & $0.2035 \pm 0.0027$ & $4.8532 \pm 0.0260$ \\
6 & $3.8373 \pm 0.0871$ & $1.1116 \pm 0.0023$ & $0.1974 \pm 0.0028$ & $4.7446 \pm 0.0171$ \\
7 & $3.3593 \pm 0.0788$ & $1.2977 \pm 0.0032$ & $0.2085 \pm 0.0031$ & $4.7394 \pm 0.0250$ \\
8 & $2.7932 \pm 0.0632$ & $1.5064 \pm 0.0036$ & $0.2117 \pm 0.0031$ & $4.5508 \pm 0.0288$ \\
9 & $2.2150 \pm 0.0488$ & $1.7158 \pm 0.0026$ & $0.2117 \pm 0.0028$ & $4.4018 \pm 0.0196$ \\
\hline
\end{tabular}
\caption{Fitted values and uncertainties of the parameters $b_0$, $b_1$, $b_2$, $b_3$ for $n_k = 1, \ldots, 9$ on a periodic $L=64$ lattice with $\xi_{\rm lat}=0.01$, $h_z=-0.08$, and $T=10$ as in Fig.~\ref{fig:Gs_Px_64_k}, up to four digits. We constrain $b_0>0$ and $b_3\in[0,2\pi)$.}
\label{tab:sound_fit}
\end{table*}

\subsection{Fit Results to Sound Modes}
If the late-time behavior of the symmetric correlator follows hydrodynamics, it will be described by the sound modes given in Eq.~\eqref{eqn:sound_mode}. In particular, we expect the correlator in the continuum limit to have the form at late time [see Eq.~\eqref{eqn:gx_sol}]
\begin{align}
    G_s^{10}(t,k) = b_0 e^{-\frac{\gamma_\zeta k^2t}{2}} \cos(c_skt + \phi) \,,
\end{align}
where $b_0$ is a constant and $\phi$ denotes a phase shift. On the lattice, we replace the continuum momentum $k$ with $2\sin(k/2)$ as used in the previous sections.

\begin{figure}[hp]
\centering
\subfloat[Pauli-$X$.\label{fig:Gs_xx}]{%
  \includegraphics[width=0.45\textwidth]{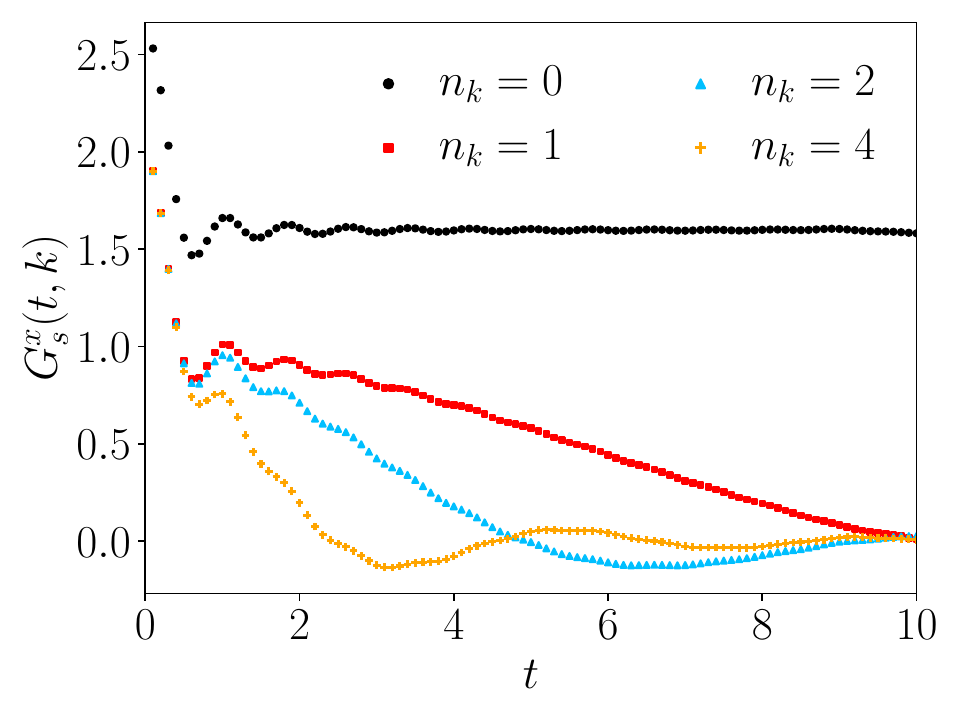}%
}\hfill
\subfloat[Pauli-$Z$.\label{fig:Gs_zz}]{%
  \includegraphics[width=0.45\textwidth]{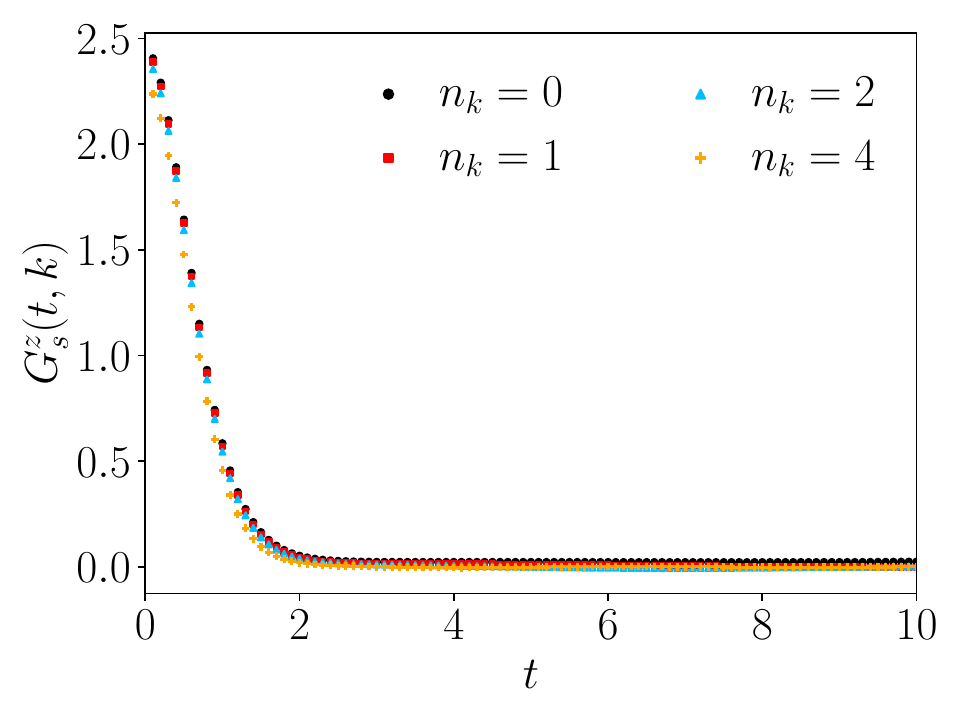}%
}
\caption{Real-time symmetric correlation functions of Pauli operators as defined in Eqs.~\eqref{eqn:Gs_xx} and~\eqref{eqn:Gs_zz} in the time-momentum space on a periodic $L=32$ lattice with $\xi_{\rm lat}=0.01$, $h_z=-0.08$, and $T=10$. The results are obtained via the MPS method with $D=500$, $\Delta t=0.1$, $N_\tau=10$, and $\Delta=0.01$. These correlators reflect genuine spin dynamics on the lattice, rather than the dynamics contained in $G_s^{10}(t,k)$ as shown in Fig.~\ref{fig:Gs_Px_64_k}.}
\label{fig:Gs_xxzz}
\end{figure}

We fit the function
\begin{align}
\label{eqn:sound_fit}
    b_0 \cos(b_1 t + b_3) e^{-b_2 t}
\end{align}
to the MPS calculated $G_s^{10}(t,k)$ at late time ($5\leq t \leq 16$). The fitted results for the lowest nine nonzero momenta are shown in Fig.~\ref{fig:Gs_Px_64_k} for a periodic $L=64$ lattice with $\xi_{\rm lat}=0.01$, $h_z=-0.08$, and $\beta=0.1$. The fitted results are also extended to the full time range for comparison. The MPS results at late time can be well described by the damped oscillating function, which is a feature of the sound mode. The fitted parameter values are listed in Table~\ref{tab:sound_fit}.

To demonstrate that the observed damped oscillation does not originate from some genuine spin dynamics of the lattice theory, we also compute real-time symmetric correlators of Pauli matrices. In particular, we use the MPS method and calculate
\begin{align}
\label{eqn:Gs_xx}
    G_s^x(t,i) = \langle \{\sigma_i^x(t), \sigma_0^x(0) \} \rangle_T \,,\\
\label{eqn:Gs_zz}
    G_s^z(t,i) = \langle \{\sigma_i^z(t), \sigma_0^z(0) \} \rangle_T \,,
\end{align}
on a periodic $L=32$ lattice with $\xi_{\rm lat}=0.01$, $h_z=-0.08$, and $\beta=0.1$. The results Fourier-transformed into the momentum space are shown in Fig.~\ref{fig:Gs_xxzz}. We see that the time and momentum dependence of these symmetric correlators of Pauli matrices is qualitatively very different from that of $G_s^{10}(t,k)$ shown in Fig.~\ref{fig:Gs_Px_64_k}. We conclude that the dynamics of $G_s^{10}(t,k)$ is not some genuine spin dynamics of the lattice theory.

To determine whether the late-time dynamics of $G_s^{10}(t,k)$ is relativistic hydrodynamics, we now analyze if the fitted parameters for different momenta follow the scaling laws given by the sound mode.

\subsection{Speed of Sound}
\begin{figure}[b]
\centering
\includegraphics[width=0.45\textwidth]{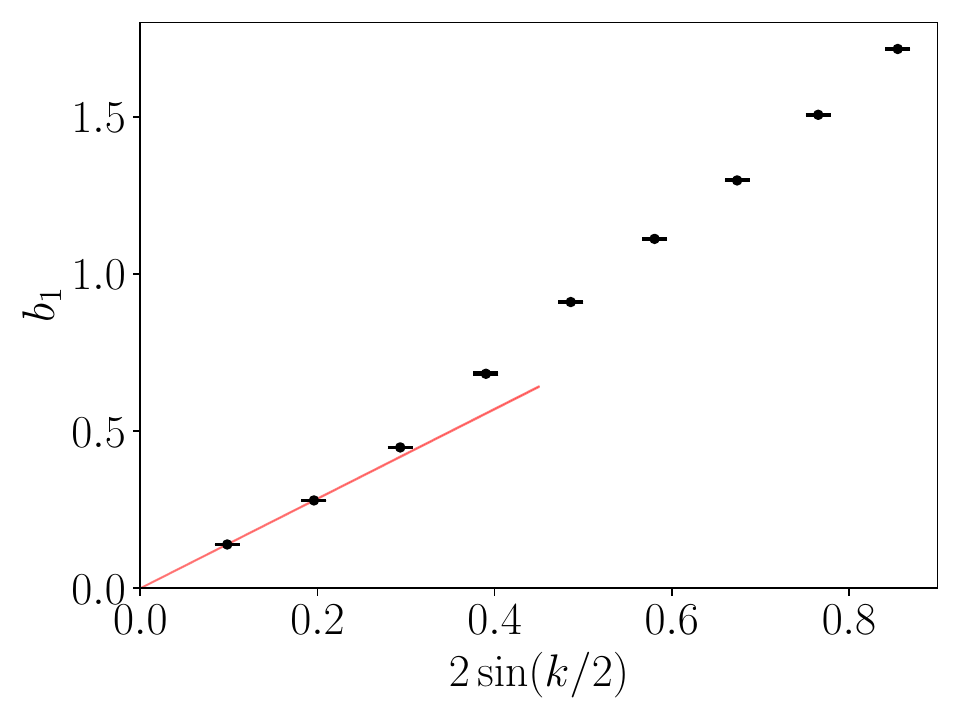}
\caption{Fitted parameter $b_1$ in Table~\ref{tab:sound_fit} as a function of the momentum $2\sin(k/2)$. The horizontal bars indicate the uncertainties of $b_1$. At low momentum, $b_1$ is proportional to the momentum, as predicted from a sound mode. The narrow red band is obtained from a linear fit $2c_s\sin(k/2)$ to the first two data points. The fitted value of the speed of sound is $c_s = 1.4256 \pm 0.0020$, up to four digits.}
\label{fig:cs_k}
\end{figure}
If the damped oscillation originates from the sound mode, we will expect $b_1 \propto 2\sin(k/2)$ and the proportional constant is the speed of sound $c_s$. Figure~\ref{fig:cs_k} shows the fitted parameter $b_1$ as listed in Table~\ref{tab:sound_fit} as a function of the momentum $2\sin(k/2)$. The horizontal bars show the uncertainties of the fitted parameters and are very small here, which are then neglected in the following analysis. We fit a linear function $2c_s\sin(k/2)$ to the first two points. The obtained result is $c_s = 1.4256 \pm 0.0020$ up to four digits (speed of light is $c=2$ in the continuum limit). The narrow red band is obtained by plotting the fitted function with the parameter uncertainty included. The data points at the lowest two nonzero momenta follow the linear scaling closely while the point at the third lowest momentum follows only marginally. The rest of the data points deviate largely from the red band. This is consistent with the validity condition of hydrodynamics that gradient expansion only works at long wavelength. We only expect to see hydrodynamic behavior at small nonzero momentum. We note that the last six data points also look linear but with a different slope, which may originate from some non-hydro mode. 

To further test the hydrodynamic nature of $G_s^{10}(t,k)$ at late time and verify the fitted speed of sound, we estimate $c_s$ at thermal equilibrium. 
Using the thermal state constructed in the MPS calculation, we can calculate the total energy, which is the internal energy $U(L,T)$, as a function of $L$ and $T$. The pressure is given by
\begin{align}
\label{eqn:P_thermodynamics}
    P = - \left( \frac{\partial U}{\partial L}\right)_{S} \,,
\end{align}
where the subscript $S$ means constant entropy. It is not straightforward to maintain the constant entropy as we change the lattice size in our calculation, since the internal energy is obtained as a function of temperature $T$. In order to estimate the speed of sound, we assume the entropy density $s$ is proportional to the temperature.\footnote{For a gas of non-interacting massive scalar particles with mass $m_1$, the mass effect on the relation $s\propto T$ is negligible at $T/m_1\approx 7.141$.} As we increase the lattice size by $\Delta L$, we lower the temperature to $LT/(L+\Delta L)$. In this way, we can estimate the pressure as a function of $L$ and $T$ by using Eq.~\eqref{eqn:P_thermodynamics}. The speed of sound is then obtained as
\begin{align}
    c_s^2 = \frac{{\rm d}P/{\rm d}\beta}{{\rm d}\varepsilon/{\rm d}\beta} \,.
\end{align}
To estimate $c_s$ on the $L=64$ lattice with $\xi_{\rm lat}=0.01$, $h_z=-0.08$, and $\beta = 0.1$, we take $\Delta L=2$ and $\Delta\beta=0.0001$ to calculate the derivatives. We also use $\Delta\beta=0.001$ to confirm that the derivatives are estimated to a good precision. The estimated speed of sound from the thermal state is $1.4039$ up to four digits, in good agreement with that extracted from the damped oscillation. We emphasize that this thermal state estimate assumes the entropy density is proportional to the temperature and thus may introduce unquantified systematic uncertainties. In fact, this thermal state estimate does not change much as the coupling decreases with $T/m_1$ fixed, while the dynamically extracted speed of sound slightly increases, see Sec.~\ref{sec:continuum} below.\footnote{One may try to accurately calculate the pressure from the free energy $F=U-TS$ by calculating the entropy
\begin{align}
    S(\beta) = S(\beta_0) + \int_{\beta_0}^\beta \beta \left(\frac{\partial U}{\partial \beta}\right)_L {\rm d}\beta \nonumber\,.
\end{align}
If one chooses $\beta_0=\infty$, i.e., starts from the ground state, it is expected that the bond dimension needed to construct a thermal state in the MPS setup will increase first and then decrease as temperature grows. The systematic uncertainty associated with the bond dimension may be large this way. So the MPS calculation is under better control if one starts from the infinite temperature $\beta_0=0$, at which $S(0)=L\ln2$. However, at such a high temperature, the lattice setup does not describe the Ising field theory as a low-energy effective theory any more and the obtained entropy probably just describes that of a spin system.}

\subsection{Bulk Viscosity}
\begin{figure}[b]
\centering
\includegraphics[width=0.45\textwidth]{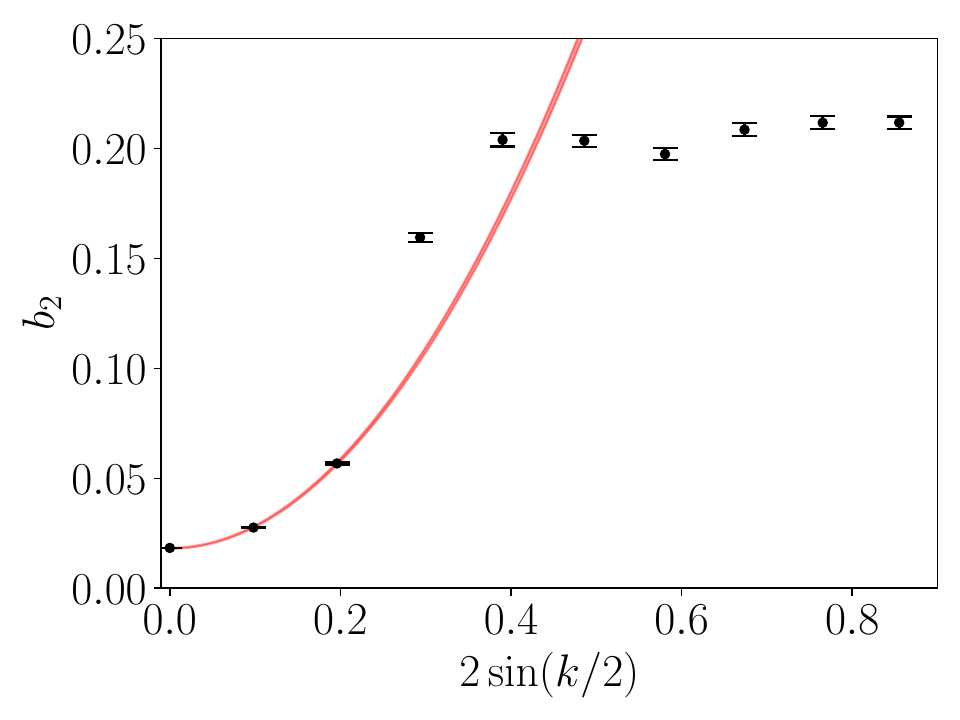}
\caption{Fitted parameter $b_2$ in Table~\ref{tab:sound_fit} as a function of the momentum $2\sin(k/2)$, including the zero momentum. The horizontal bars indicate the uncertainties of $b_2$, which are not used in the fitting. The narrow red band is obtained from fitting the function $\Gamma_0 + 2 \tilde{\gamma}_\zeta \sin^2(k/2)$ to the first three data points including the one at zero momentum. The fitted parameter values are $\Gamma_0=0.0182 \pm 0.0003$ and $\tilde{\gamma}_\zeta=2.0052 \pm 0.0222$, up to four digits.}
\label{fig:gamma}
\end{figure}

We now analyze the dependence of $b_2$ on the momentum and compare with the sound mode prediction $b_2 = 2\gamma_\zeta \sin^2(k/2)$. Since the momentum density operator used in the calculation only exactly conserves momentum in the continuum limit (see Fig.~\ref{fig:p_conserve}), we expect $b_2$ to be nonzero even at zero momentum. In order to reduce systematic uncertainty, we will include the decay rate at $k=0$ in the following analysis, which has been obtained in Fig.~\ref{fig:b2_k0}. The momentum dependence of $b_2$ is shown in Fig.~\ref{fig:gamma}.

From the analysis of the speed of sound in the previous subsection, we know that the damped oscillations observed for the lowest two nonzero momenta probably correspond to the sound mode, and this correspondence becomes marginal for the third lowest nonzero momentum. In order to extract the bulk viscosity, we fit to the momentum dependence of $b_2$ the function
\begin{align}
    \Gamma_0 + 2 \tilde{\gamma}_\zeta \sin^2(k/2) \,,
\end{align}
for the lowest three momenta including zero. The fitting result is plotted as the narrow red band in Fig.~\ref{fig:gamma} with the fitting uncertainty included. The fitted parameter values are $\Gamma_0=0.0182 \pm 0.0003$ and $\tilde{\gamma}_\zeta=2.0052 \pm 0.0222$, up to four digits.

With the fitted values of $c_s$ and $\tilde{\gamma}_\zeta$, we can check whether the validity condition for hydrodynamics ($c_sk\gg \tilde{\gamma}_\zeta k^2/2$) is consistent with the numerical fact that only the lowest two nonzero momenta fall within the hydrodynamic regime. For $n_k=1,2,3$, we find
\begin{align}
    {\rm Kn} \equiv \frac{\tilde{\gamma}_\zeta \sin(k/2)}{c_s} \approx 0.0688,\, 0.138,\,0.206 \,,
\end{align}
respectively, up to three significant numbers. The validity condition for hydrodynamics (${\rm Kn} \ll 1$) works well for $n_k=1$, may be adequate for $n_k=2$, but does not really hold for $n_k=3$. This is consistent with using just the lowest two nonzero momenta for extracting the sound mode properties.

To obtain the ratio of bulk viscosity to entropy density, i.e., the specific bulk viscosity, we use the Euler relation $sT = \varepsilon+P$ and get from Eq.~\eqref{eqn:gamma_zeta}
\begin{align}
\label{eqn:zetaovers}
    \frac{\zeta}{s}  = \gamma_\zeta T = \frac{1}{4}\tilde{\gamma}_\zeta T \,,
\end{align}
where we have added a factor of $1/4$ on the right hand side.
Although the unit of the ratio is $\hbar/k_B$, independent of the speed of light, we still corrected Eq.~\eqref{eqn:zetaovers} by the factor of $1/4$. This is because $b_2$ is obtained by fitting a time dependence so it carries the unit of energy. Then the quantity $\tilde{\gamma}_\zeta T$ has the unit of energy squared divided by momentum squared. Multiplying the right hand side by $1/4$ converts it back to the unit system with $c=1$.
The ratio obtained from the fit is $\zeta/s = 5.013 \pm 0.056$ up to three digits.

\subsection{Continuum Limit}
\label{sec:continuum}
So far, we have only obtained $c_s$ and $\zeta/s$ at one coupling $h_z=-0.08$.
For the continuum limit, we perform similar analyses for a range of couplings while maintaining the ratio $T/m_1$ fixed to be about $7.141$. The couplings and the corresponding temperatures used in taking the continuum limit on the $L=64$ lattice with $\xi_{\rm lat}=0.01$ are listed in Table~\ref{tab:couplings}. For each coupling, we perform similar fits as in Figs.~\ref{fig:cs_k} and~\ref{fig:gamma} to determine $c_s$ and $\tilde{\gamma}_\zeta$. Fitting results are also listed in the table. Plots of the fitting results for couplings other than $h_z=-0.08$ can be found in Appendix~\ref{app:fit}. We note that the fitting uncertainties vary non-uniformly as the coupling changes. They are larger at the two smallest and the largest couplings than the others studied here, e.g. in $\tilde{\gamma}_\zeta$. The uncertainties are closely related to the consistency between the numerical data points and the sound mode predictions. At larger coupling, lattice artifacts from finite spacing are bigger. At smaller coupling, finite size effects can be bigger.

\begin{table}[h]
\centering
\begin{tabular}{| c | c | c | c | c |}
    \hline
    $h_z$ & $\beta$ & $|h_z|^{8/15}L$ & $c_s$ & $\tilde{\gamma}_\zeta$ \\
    \hline
    -0.04 & 0.1447 & 11.50 & $1.4878 \pm 0.0077$ & $2.1484 \pm 0.2409$ \\
    -0.05 & 0.1285 & 12.95 & $1.4889 \pm 0.0391$ & $2.2663 \pm 0.1528$ \\
    -0.06 & 0.1166 & 14.27 & $1.4564 \pm 0.0153$ & $1.8434 \pm 0.0373$ \\
    -0.07 & 0.1074 & 15.50 & $1.4557 \pm 0.0162$ & $2.0243 \pm 0.0562$ \\
    -0.08 & 0.1    & 16.64 & $1.4256 \pm 0.0020$ & $2.0052 \pm 0.0222$ \\
    -0.09 & 0.0939 & 17.72 & $1.4002 \pm 0.0093$ & $2.3693 \pm 0.0472$ \\
    -0.1  & 0.0888 & 18.74 & $1.4152 \pm 0.0303$ & $1.9638 \pm 0.1900$ \\
    \hline
\end{tabular}
\caption{List of couplings used for taking the continuum limit on the periodic $L=64$ lattice with $\xi_{\rm lat}=0.01$. The temperatures change accordingly such that $1/(\beta m_1)\approx 7.141$ is kept fixed. Values of $\beta$ are kept up to four digits. The third column shows the effective physical volume, up to two digits. The last two columns show the fitted values of $c_s$ and $\tilde{\gamma}_\zeta$, up to four digits, obtained in the same way as in Figs.~\ref{fig:cs_k} and~\ref{fig:gamma}. The fitting results are plotted in Appendix~\ref{app:fit} for couplings other than $h_z=-0.08$.}
\label{tab:couplings}
\end{table}

\begin{figure}[t]
\centering
\includegraphics[width=0.45\textwidth]{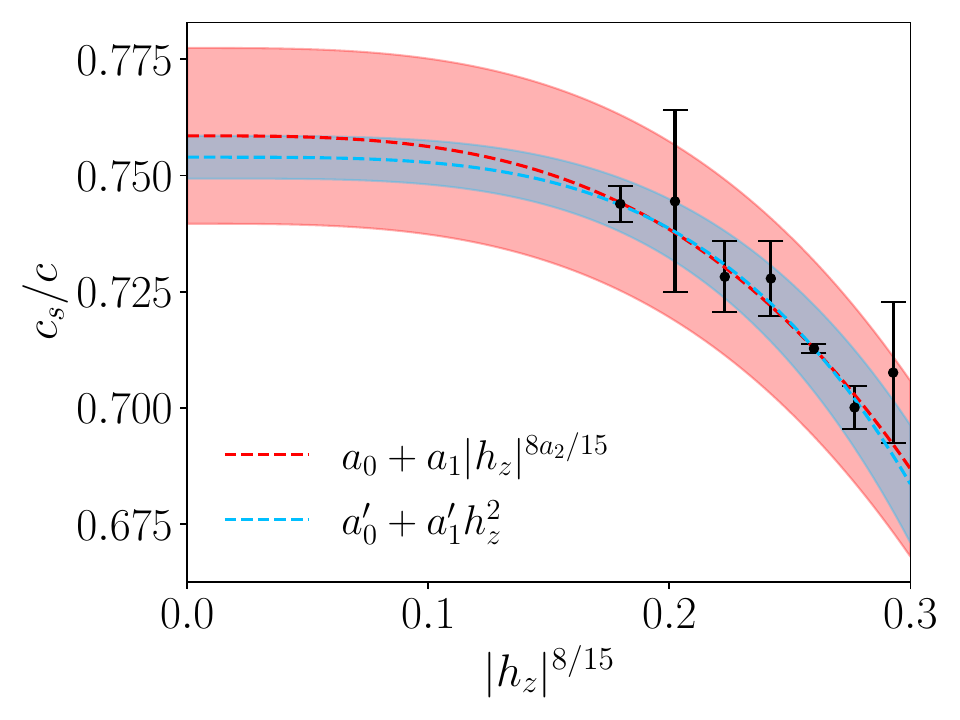}
\caption{Speed of sound in unit of speed of light as a function of the coupling $|h_z|^{8/15}$ on the periodic $L=64$ lattice with $\xi_{\rm lat}=0.01$ and $T/m_1\approx 7.141$. The red dashed line depicts the continuum extrapolation by fitting the function $a_0 + a_1|h_z|^{8a_2/15}$ to the numerical data points with uncertainties included, while the blue one is obtained from fitting the function $a_0'+a_1'h_z^2$. The red band originates from the uncertainty in $a_0$ while the blue band comes from the uncertainties in both $a_0'$ and $a_1'$. The fitted parameter values and their uncertainties are listed in Table~\ref{tab:continuum}.}
\label{fig:cs_continuum}
\end{figure}

Figure.~\ref{fig:cs_continuum} shows the speed of sound extracted from the damped oscillations as a function of $|h_z|^{8/15}$, where we have divided the speed of sound by the continuum speed of light, which is $2$ as shown earlier in Fig.~\ref{fig:c_hz}. 
We take the continuum limit by fitting to the numerical results two algebraic functions
\begin{align}
\label{eqn:continuum1}
    &a_0 + a_1|h_z|^{8a_2/15} \,,\\
\label{eqn:continuum2}
    &a_0' + a_1'h_z^2 \,.
\end{align}
The first function leaves the exponent as a fit parameter while the second one is quadratic in $h_z$, corresponding to the leading power-law correction $a^{15/4}$ found earlier in Fig.~\ref{fig:b2_k0}.
The fitting takes into account the uncertainties associated with $c_s$ at each $h_z$.
The fitted parameter values are listed in Table~\ref{tab:continuum} with uncertainties.
The fitting results with the central values of the parameters are shown as the dashed lines in Fig.~\ref{fig:cs_continuum}. The red band is obtained by considering the uncertainty associated with the $a_0$ parameter, which has the physical meaning of $c_s/c$ in the continuum limit. The blue band describes the fitting uncertainties associated with both $a_0'$ and $a_1'$. The two fitting functions yield consistent continuum extrapolations. But the second fit function~\eqref{eqn:continuum2} is more stable: The fitted parameters have smaller uncertainties. This implies that the leading correction to the symmetric correlation function of $T^{10}$ is at order $a^{15/4}$, consistent with the finding in the analysis of the total momentum conservation restoration.

\begin{figure}[t]
\centering
\includegraphics[width=0.45\textwidth]{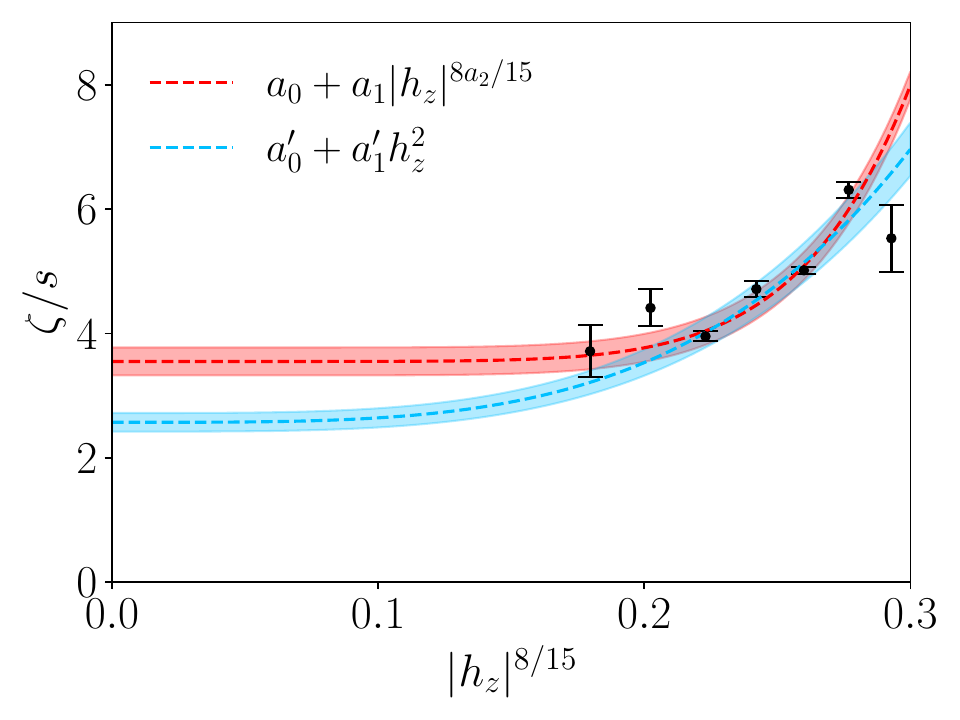}
\caption{Specific bulk viscosity as a function of the coupling $|h_z|^{8/15}$ on the periodic $L=64$ lattice with $\xi_{\rm lat}=0.01$ and $T/m_1\approx 7.141$. The red dashed line depicts the continuum extrapolation by fitting the function $a_0 + a_1|h_z|^{8a_2/15}$ to the numerical data points with uncertainties included, while the blue one is obtained from fitting the function $a_0'+a_1'h_z^2$. The red band originates from the uncertainty in $a_0$ while the blue band comes from the uncertainties in both $a_0'$ and $a_1'$. The fitted parameter values and their uncertainties are listed in Table~\ref{tab:continuum}.}
\label{fig:zeta_s}
\end{figure}

The coupling dependence of $\zeta/s$ is shown in Fig.~\ref{fig:zeta_s}, which is used to fit the same two equations~\eqref{eqn:continuum1} and~\eqref{eqn:continuum2} for the continuum limit. The fitted parameter values are listed in Table~\ref{tab:continuum} with uncertainties. 
The fitting results from the central values of the parameters are shown as the dashed lines. The red band is obtained by considering the uncertainty associated with the $a_0$ parameter, which has the physical meaning of $\zeta/s$ in the continuum limit. 
The blue band describes the fitting uncertainties associated with both $a_0'$ and $a_1'$. The two fitting functions lead to somewhat different continuum extrapolations. Although the second fit function with $h_z^2$ yields much smaller uncertainties in the fitted parameters, and the earlier analyses on the total momentum conservation restoration and speed of sound suggest the leading correction is at order $h_z^2$, we still include the extrapolation result from the first fit function in the final uncertainty estimate. A value of $\zeta/s$ at a smaller coupling magnitude such as $h_z=0.01$ can greatly improve the fitting accuracy, which probably requires a bigger lattice and is left for future work.

We conclude that the Ising field theory defined by $\xi_{\rm lat}=0.01$ has the speed of sound $c_s/c=0.76 \pm 0.02$ and the specific bulk viscosity $\zeta/s=3.06\pm 0.68$ at the temperature $T/m_1\approx 7.141$, up to two digits. For the speed of sound, we use the average of the central values of $a_0$ and $a_0'$ as the final central value and the larger uncertainty in the two fits as the final uncertainty estimate. For $\zeta/s$, we use the average of the central values of $a_0$ and $a_0'$ as the final central value with the final uncertainty estimated as $(3.548-2.569+0.225+0.15)/2$, to reflect the uncertainty in the extrapolation procedure.
If we think of the thermal system as a gas of weakly-coupled massive scalar particles with $\lambda \phi^4$ interaction, we can parametrically interpret the result. The leading perturbative contribution to the bulk viscosity arises from particle number changing processes that are sensitive to soft momenta~\cite{Jeon:1995zm}.  The cross section of $2\to 4$ scattering scales as $\lambda^4$ at leading order. So we expect $\zeta/s \sim 1/\lambda^4$ from the perturbative perspective and thus $\zeta/s$ can be large.\footnote{In $1+1$D, the mass dimension of $\lambda$ is 2. The numerator can be $T^8$ or $m^4T^4$, where $m$ is the scalar particle mass.}

\begin{table}[t]
\centering
\begin{tabular}{| c | c | c |}
    \hline
    Parameters & $c_s/c$ & $\zeta/s$ \\
    \hline
    $a_0$ & $0.759 \pm 0.019$ & $3.548 \pm 0.225$ \\
    $a_1$ & $-3.099 \pm 7.365$ & $(3.361  \pm 5.984 )\times 10^4$ \\
    $a_2$ & $3.130 \pm 2.061$ & $7.419 \pm 1.433$ \\
    $a_0'$ & $0.754 \pm 0.005$  & $2.569 \pm 0.150$ \\
    $a_1'$ & $-6.420 \pm 0.734$ & $401.257 \pm 25.598$ \\
    \hline
\end{tabular}
\caption{Fitted parameter values used to extrapolate $c_s/c$ and $\zeta/s$ to the continuum limit, up to three digits.}
\label{tab:continuum}
\end{table}

\section{Conclusions}
\label{sec:conclusions}
In this paper we studied the real-time symmetric correlation function of stress-energy tensors in a near-integrable Ising field theory, by using classical exact diagonalization and the matrix product state tensor network method. We derived a momentum density operator that respects total momentum conservation in the continuum limit. We found that in the scaling region of the lattice theory, the Umklapp processes are suppressed and the emergent sound modes of relativistic hydrodynamics can be seen on an $L=64$ lattice. Using the MPS method, we calculated the symmetric correlator of the momentum density and observed damped oscillating behavior, from which we extracted the speed of sound and the specific bulk viscosity, ratio of bulk viscosity to entropy density. The extrapolated results in the continuum limit are $c_s/c=0.76 \pm 0.02$ and $\zeta/s=3.06\pm 0.68$ for the Ising field theory defined with $\xi_{\rm lat}=0.01$ at the temperature $T/m_1\approx 7.14$. The precision can be further improved by numerical calculations on bigger lattices and smaller couplings, as well as analytic studies on the parametric scaling of the leading power corrections.

In future work, it would be interesting to study the speed of sound and the specific bulk viscosity as functions of temperature and the $\xi_{\rm lat}$ parameter for different Ising field theories, as well as to understand the impact of the distance to integrability. One may also treat the $1+1$D Ising field thermal system as a gas of weakly coupled scalar particles and perturbatively calculate the bulk viscosity as done in $3+1$D for one scalar particle species~\cite{Jeon:1995zm} and compare with the numerically determined temperature and $\xi_{\rm lat}$ dependence. The temperature dependence of the specific bulk viscosity could shed light on the temperature dependence of the couplings between the scalar particles if they are weakly coupled. As temperature decreases, the entanglement contained in the thermal state increases, which might limit the application of tensor network methods. From an orthogonal perspective, one can perform similar studies for $1+1$D U(1) and non-Abelian gauge theories to investigate if they hydrodynamize and, if so, to calculate their bulk viscosities. For these theories, we know how to perform perturbative calculations, which can be compared with the real-time lattice results. Furthermore, these $1+1$D gauge theories have conserved fermion numbers, which introduce an additional diffusion equation coupled with the hydrodynamic equations. It would be interesting to see both of them emerge in real-time lattice simulations, which also allows the computation of the diffusion constant.

The current work demonstrates the usefulness of real-time lattice Hamiltonian simulation for studying non-equilibrium dynamics. In addition to extracting transport coefficients, the setup also enables nonperturbative simulation of the hydrodynamization process from first principles, since extracting a transport coefficient requires computing the real-time correlator at late time when hydrodynamics becomes applicable and the whole time history is a byproduct. Thus, real-time lattice simulation is also capable of studying non-hydro modes from first principles.

In the long term, one would aim at performing similar calculations for gauge theories in higher dimensions, in particular, QCD. These studies can determine the ratio of shear viscosity to entropy density with uncertainties under control. Furthermore, full QCD real-time simulation can not only benchmark different UV completions of fixed-order hydrodynamic equations~\cite{Muller1967,Israel:1976tn,Israel:1976efz,Israel:1979wp,Baier:2007ix,Denicol:2012cn,Bemfica:2019knx,Kovtun:2019hdm,Armas:2020mpr,Bhambure:2024axa,Heller:2025dxh}, but also benchmark different approaches for hydrodynamization, finding non-hydro modes~\cite{Baier:2000sb, Chesler:2008hg, Chesler:2009cy, Chesler:2010bi, Balasubramanian:2011ur, Balasubramanian:2013oga, vanderSchee:2012qj, Romatschke:2015gic, Heller:2016rtz, Kurkela:2017xis, Brewer:2019oha, Grozdanov:2019uhi, Rajagopal:2025nca}. Finally, real-time lattice simulation is capable of providing first-principle results for jet wake~\cite{Neufeld:2008fi,Neufeld:2009ep,Casalderrey-Solana:2004fdk,Casalderrey-Solana:2006lmc,Wang:2013cia,Casalderrey-Solana:2020rsj,Yang:2022nei}, for which most current studies assumed immediate thermalization of the energy and momentum lost by high energy particles when traversing through a hot medium. As entanglement grows much faster in higher dimensions, tensor network methods may be limited in this regard. Quantum computing may be a promising path towards ultimate simulations for QCD.

\begin{acknowledgments}
We would like to thank Larry Yaffe for inspiring discussions. We would also like to thank Marc Illa, Saurabh Kadam, Joseph Lap, Pavel Meshcheriakov, Berndt M\"uller, Pavel Kovtun, Martin Savage, and Steve Sharpe. 
This work is supported by the U.S. Department of Energy, Office of Science, Office of Nuclear Physics, IQuS under Award Number DOE (NP) Award DE-SC0020970 via the program on Quantum Horizons: QIS Research and Innovation for Nuclear Science.
This research was supported in part by grant NSF PHY-2309135 to the Kavli Institute for Theoretical Physics (KITP).
This research used resources of the National Energy Research Scientific Computing Center (NERSC), a Department of Energy Office of Science User Facility, under NERSC award NP-ERCAP0032083. This work was enabled, in part, by the use of advanced computational, storage and networking infrastructure provided by the Hyak supercomputer system at the University of Washington. All fits are performed via the \verb|curve_fit| method in SciPy.
\end{acknowledgments}

\appendix

\section{MPS Calculations with Different Bond Dimensions for $L=16$ and $L=64$}
\label{app:mps}

\begin{figure*}[p]
\centering
\subfloat[$x=0$.\label{fig:Gs_Px_0_bond}]{%
  \includegraphics[width=0.3\linewidth]{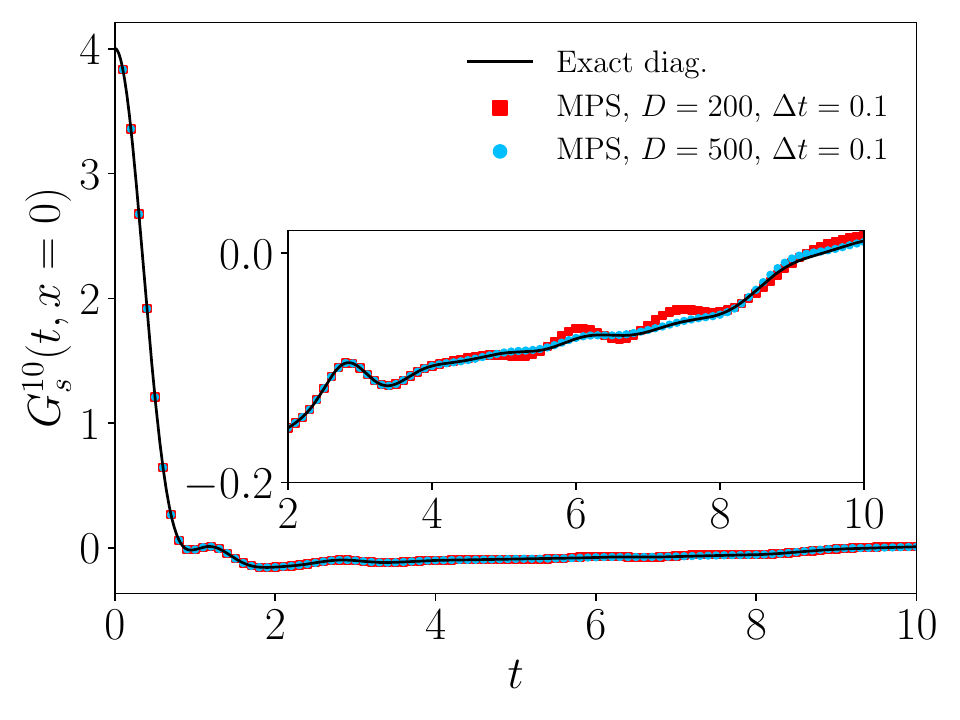}%
}\hfill
\subfloat[$x=2$.\label{fig:Gs_Px_2_bond}]{%
  \includegraphics[width=0.3\linewidth]{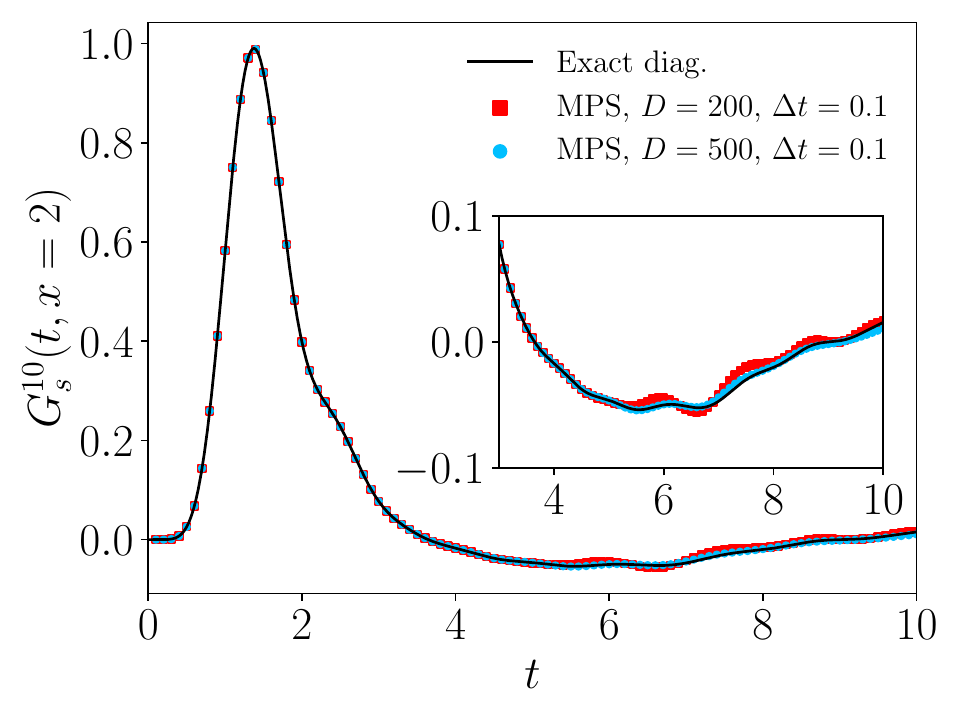}%
}\hfill
\subfloat[$x=4$.\label{fig:Gs_Px_4_bond}]{%
  \includegraphics[width=0.3\linewidth]{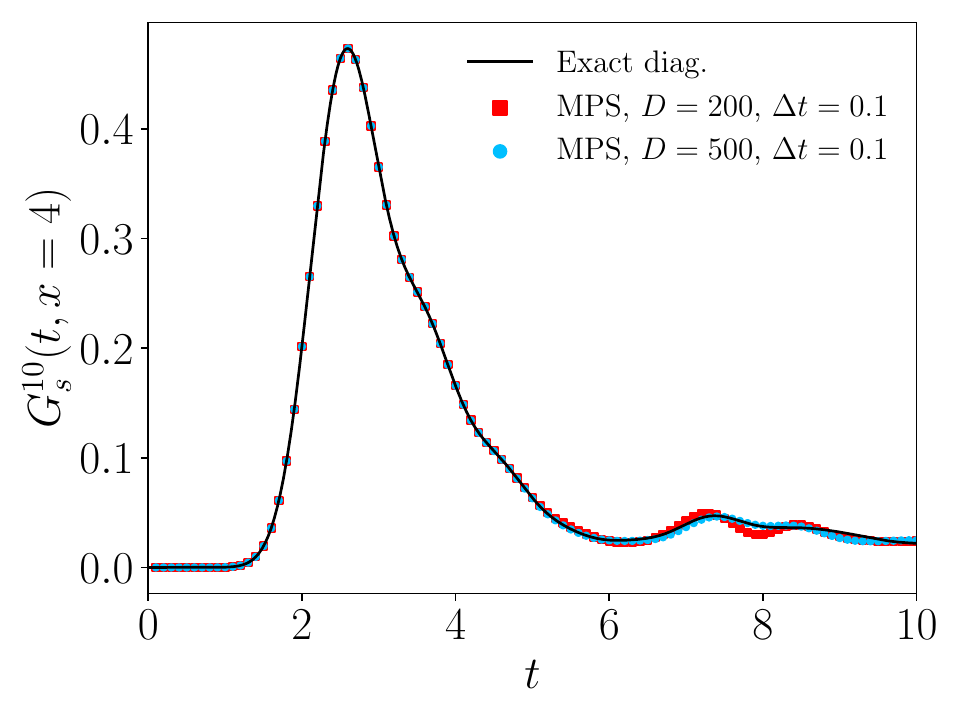}%
}

\subfloat[$x=6$.\label{fig:Gs_Px_6_bond}]{%
  \includegraphics[width=0.3\linewidth]{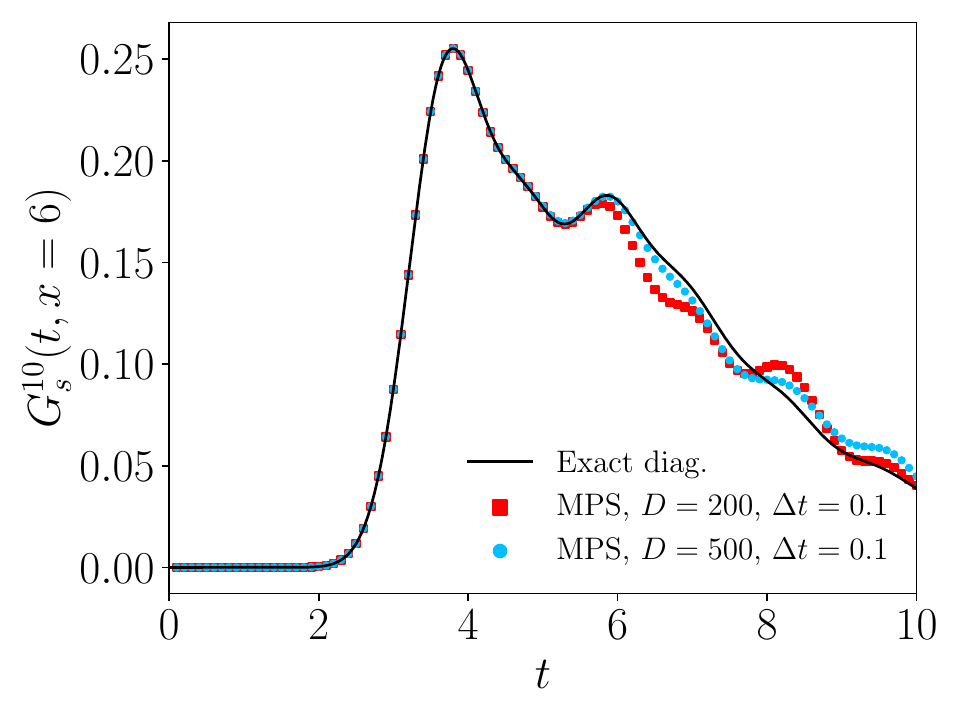}%
}~
\subfloat[$x=8$.\label{fig:Gs_Px_8_bond}]{%
  \includegraphics[width=0.3\linewidth]{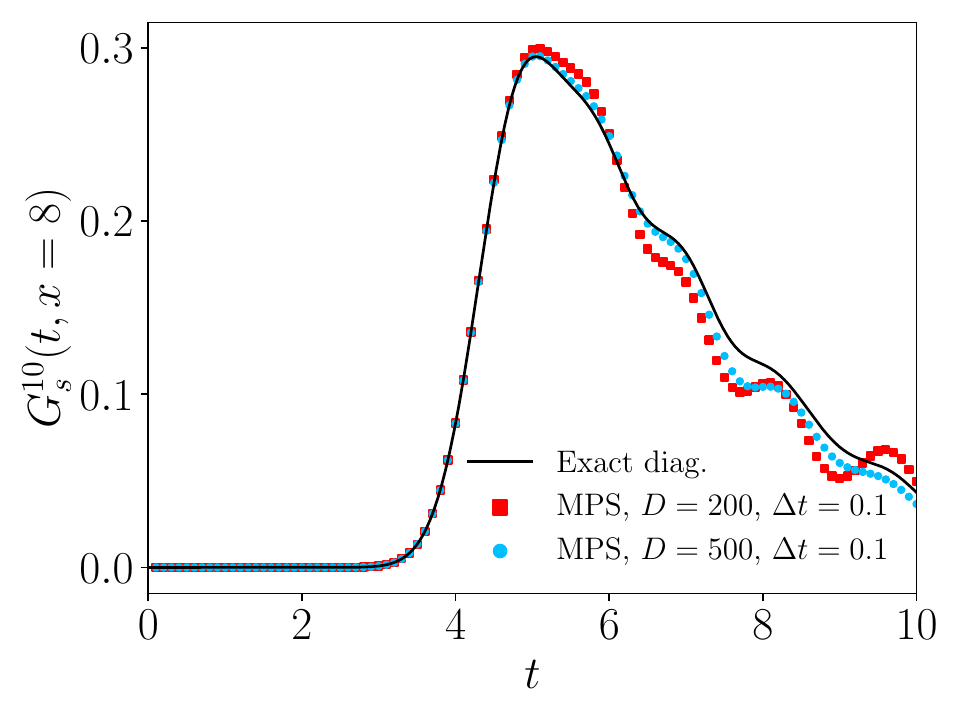}%
}

\subfloat[$x=0$.\label{fig:Delta_Gs_Px_0_bond}]{%
  \includegraphics[width=0.3\linewidth]{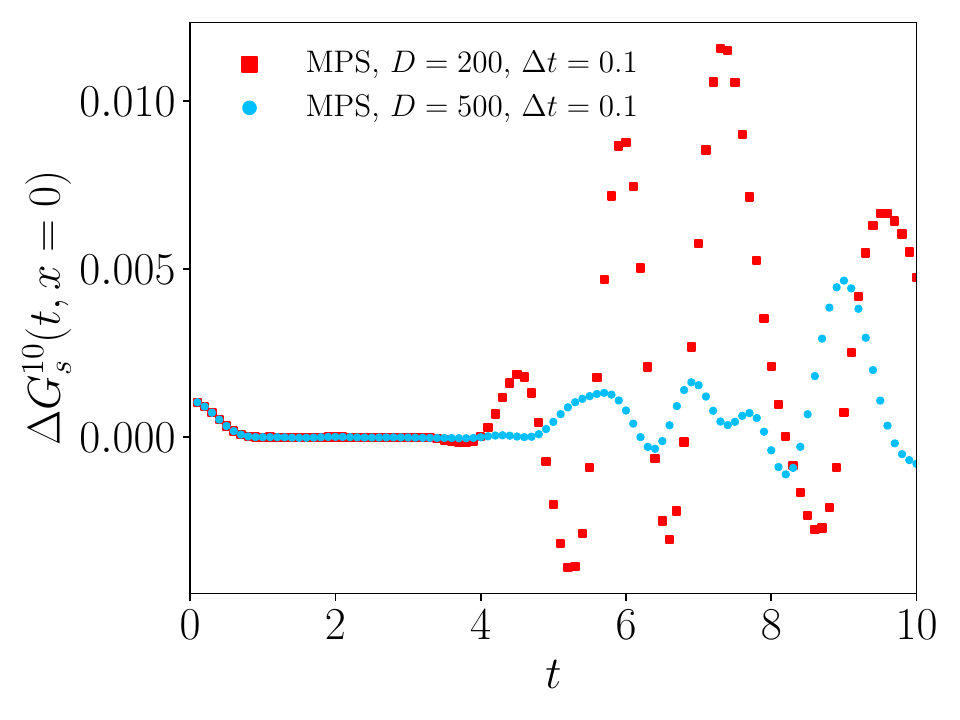}%
}\hfill
\subfloat[$x=2$.\label{fig:Delta_Gs_Px_2_bond}]{%
  \includegraphics[width=0.3\linewidth]{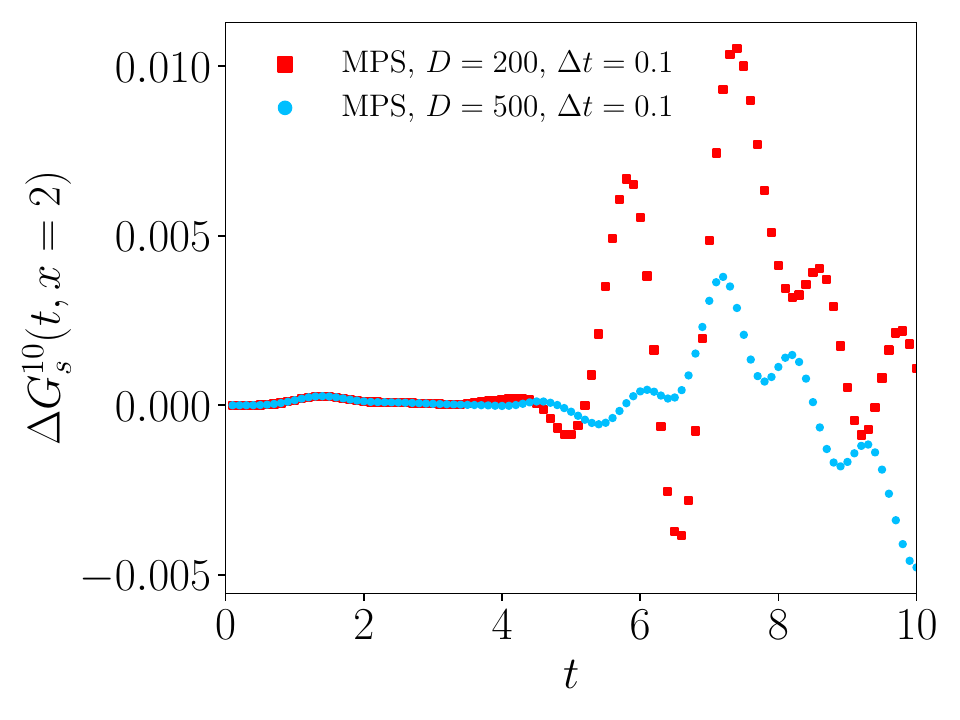}%
}\hfill
\subfloat[$x=4$.\label{fig:Delta_Gs_Px_4_bond}]{%
  \includegraphics[width=0.3\linewidth]{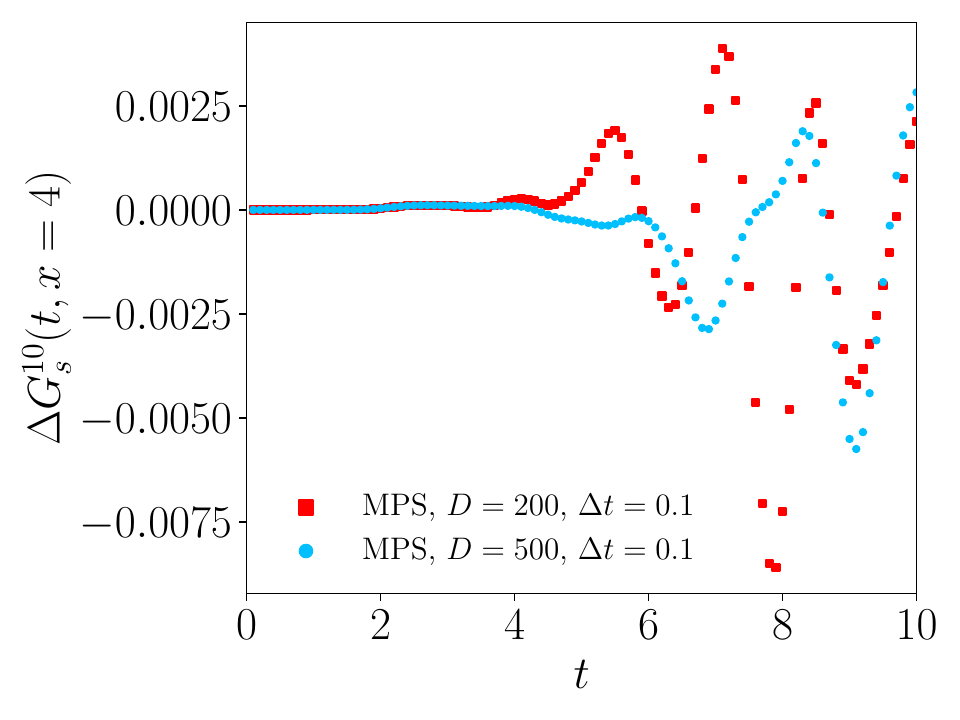}%
}

\subfloat[$x=6$.\label{fig:Delta_Gs_Px_6_bond}]{%
  \includegraphics[width=0.3\linewidth]{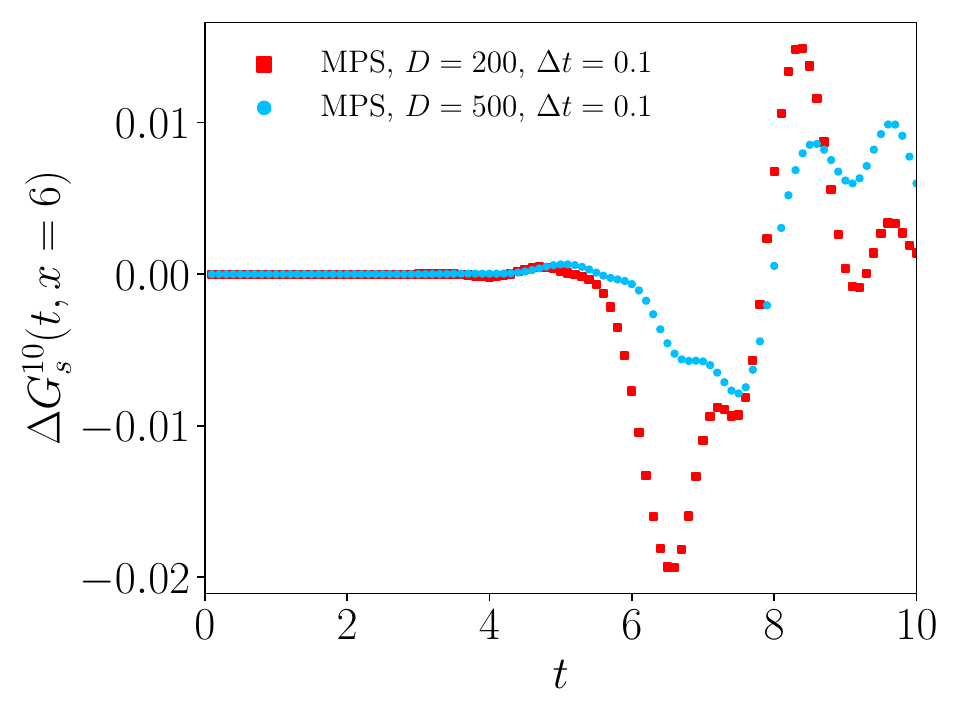}%
}~
\subfloat[$x=8$.\label{fig:Delta_Gs_Px_8_bond}]{%
  \includegraphics[width=0.3\linewidth]{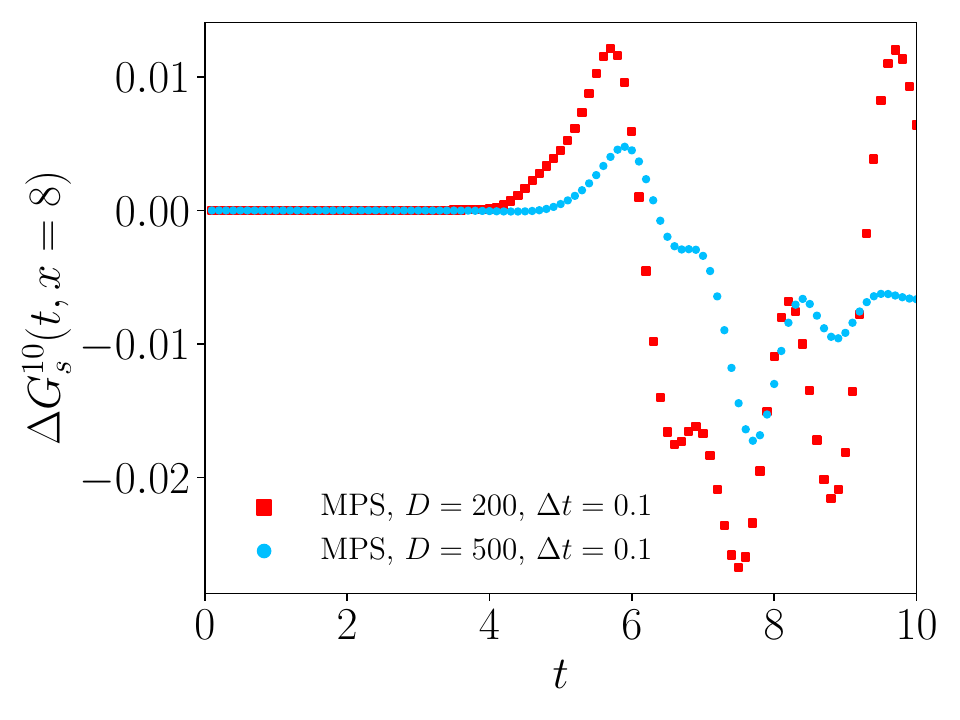}%
}
\caption{MPS calculation results of $G_s^{10}(t,x)$ (first two rows) and their deviations from exact diagonalization results $\Delta G_s^{10}(t,x)$ (last two rows) on a periodic $L=16$ lattice with $\xi_{\rm lat}=0.01$, $h_z=-0.3$ and $T=10$ using two different bond dimension upper limits $D$. The black solid lines are obtained from exact diagonalization. For the MPS calculation, we use 10 steps in the second-order Trotterized imaginary time evolution and $\Delta =0.01$ to apply the local perturbation at $t=0$.}
\label{fig:Gs_Px_16_bond}
\end{figure*}

\begin{figure}[t]
\centering
\includegraphics[width=0.45\textwidth]{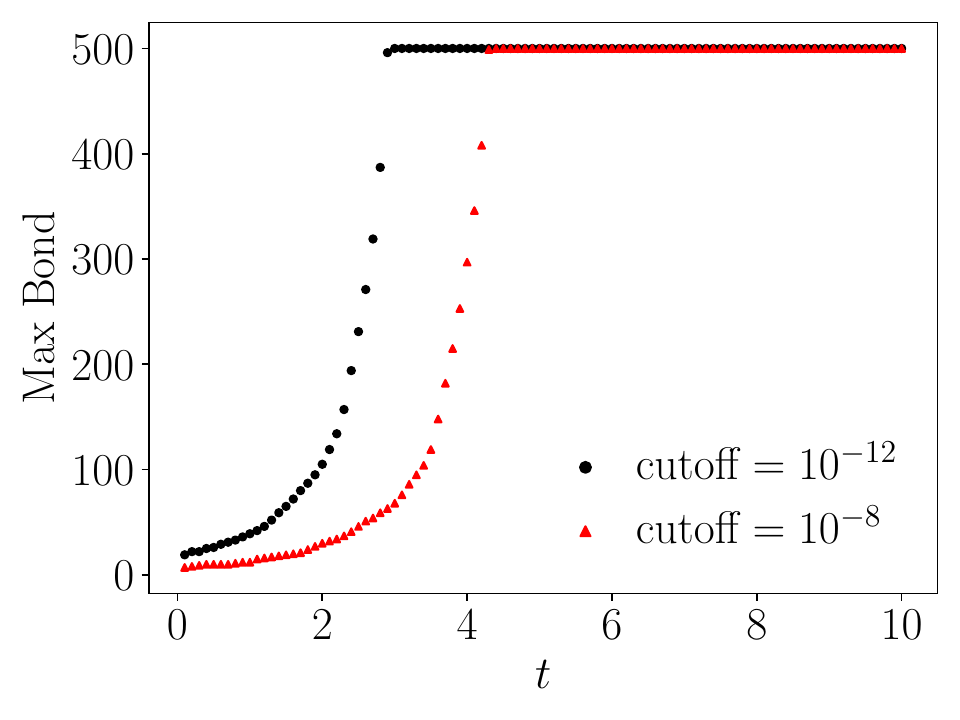}
\caption{Maximum bond dimension across the whole lattice used in the MPS calculation for the $L=16$ lattice as shown in Fig.~\ref{fig:Gs_Px_16_bond} using two different cutoffs $\chi$. Here, the bond dimension upper limit is $D=500$.} 
\label{fig:maxbond_N=16}
\end{figure}

To benchmark the MPS calculations, we compare the MPS-calculated $G_s^{10}(t,x)$ with classical exact diagonalization results on a periodic $L=16$ lattice with $\xi_{\rm lat}=0.01$, $h_z=-0.3$, and $T=10$ in Fig.~\ref{fig:Gs_Px_16_bond}. The deviations of the MPS results from the exact diagonalization ones are also shown. The MPS results are obtained by using 10 steps in the second-order Trotterized imaginary time evolution and $\Delta =0.01$ to apply the local perturbation to the thermal state at $t=0$. We see that the perturbation arrives at the right boundary at $x=8$ at time $t\approx 4$ and the signal there reaches the maximum at time $t\approx 5$. Before $t=5$, the MPS results agree with the exact diagonalization results well. Beyond $t=5$, they start to deviate in a noticeable way. To more systematically understand the precision of the MPS calculation, we plot in Fig.~\ref{fig:maxbond_N=16} the maximum bond dimension used in the MPS calculation for the $L=16$ lattice mentioned above with two different cutoffs: $\chi=10^{-12}$, which is used in the calculations that lead to the bulk viscosity in the main text, and $\chi=10^{-8}$, which is only used here to help us understand the precision. When the ITensor package performs Singular Value Decomposition for tensors, it removes singular values such that the sum of their squared values divided by the total sum of all squared singular values is less than $\chi$. With a finite bond dimension upper limit $D$, it only keeps up to $D$ singular values, independent of $\chi$. Empirically, we find the accuracy for local observables scale as $O(\sqrt{\chi})$ when the maximum bond dimension across the whole lattice is below the upper limit $D$. We want to emphasize that when the bond dimension of a MPS calculation reaches the upper limit, it does not directly mean the calculation results converge poorly. For example, for the time range $3<t<4$ in Fig.~\ref{fig:maxbond_N=16} when the bond dimension saturates to the upper limit $D=500$ in the $\chi=10^{-12}$ case, the MPS results still converge well, as shown in Fig.~\ref{fig:Gs_Px_16_bond}. We know for this time range, the error is bounded by the cutoff as $O(0.0001)$. For the time range $4<t<5$ in Fig.~\ref{fig:maxbond_N=16}, the bond dimension has saturated to the upper limit even in the $\chi=10^{-8}$ case, but we see in Fig.~\ref{fig:Gs_Px_16_bond} that the MPS results with $D=500$ still converge well, with an error smaller than $0.001$.

\begin{figure}[t]
\centering
\includegraphics[width=0.45\textwidth]{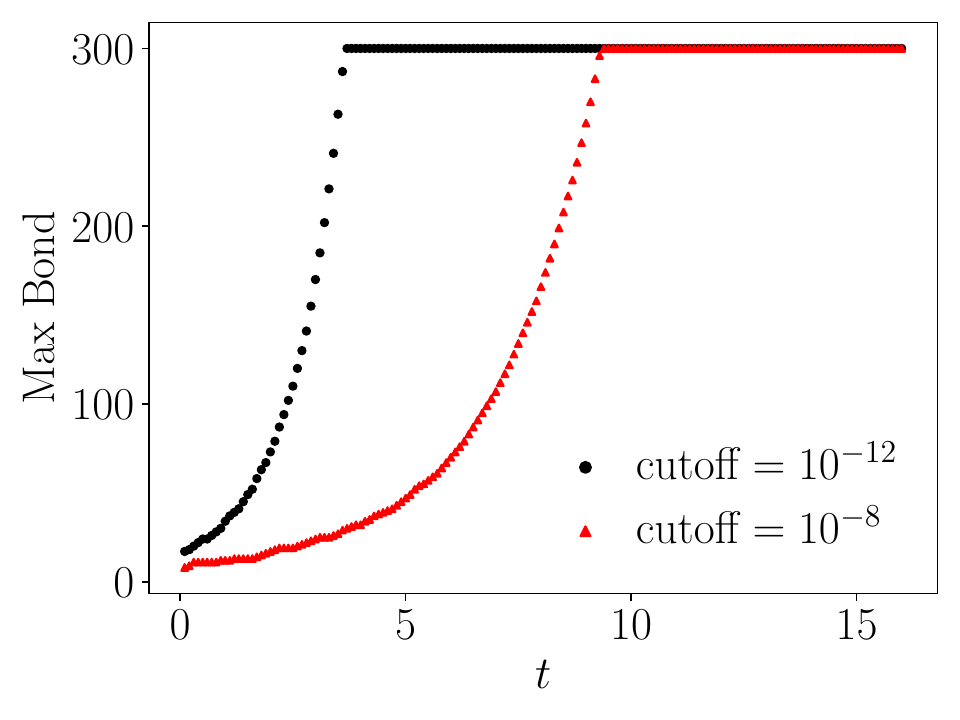}
\caption{Maximum bond dimension across the whole lattice used in the MPS calculation for the $L=64$ lattice as shown in Fig.~\ref{fig:Gs_Px_64_bond} using two different cutoffs $\chi$. Here, the bond dimension upper limit is $D=300$.} 
\label{fig:maxbond_N=64}
\end{figure}

Poor convergence becomes manifest when $t>5.5$. once the perturbation crosses the periodic boundary. This is because in the MPS setup, the local tensors for the first lattice site and the last on the chain are very far away from each other. Thus, the MPS requires a very large bond dimension to maintain the periodic boundary conditions. Using open or fixed boundary conditions can certainly increase the maximum time with good convergence under a given bond dimension upper limit in the MPS calculation. In practice, if we want to demonstrate hydrodynamization before the perturbation reaches the boundaries and extract physical observables with the boundary effects minimized, we choose the maximum time of the simulation to be the time when the perturbation roughly reaches the boundaries. Then we choose the bond dimension upper limit in the MPS calculation for good convergence before the maximum time.

\begin{figure*}[t]
\centering
\subfloat[$x=0$.\label{fig:Gs_Px_0_bond64}]{%
  \includegraphics[width=0.33\linewidth]{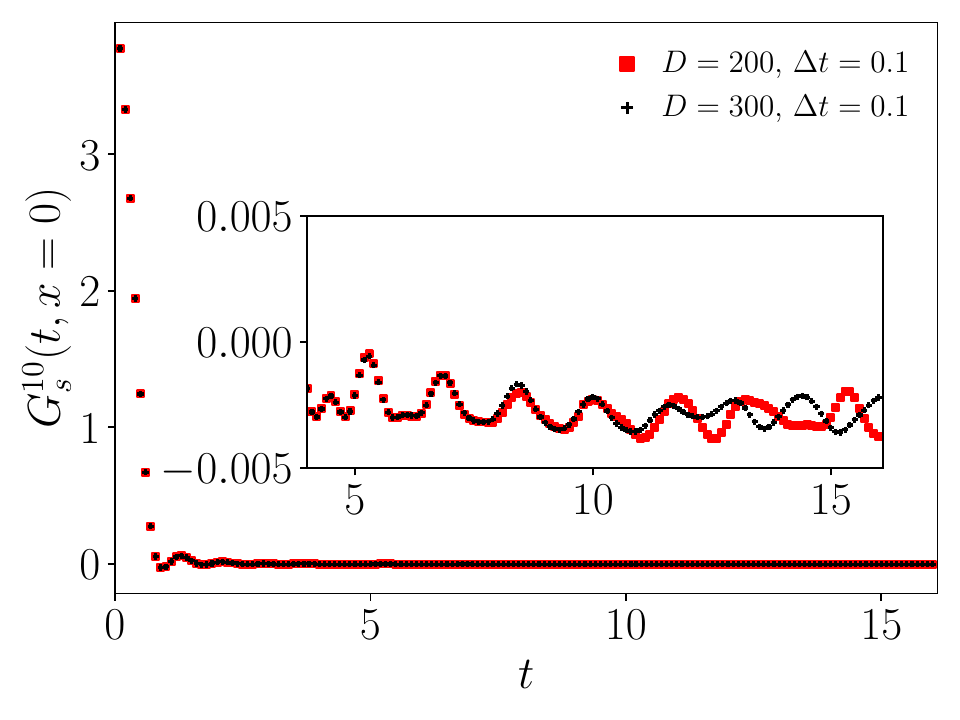}%
}\hfill
\subfloat[$x=4$.\label{fig:Gs_Px_4_bond64}]{%
  \includegraphics[width=0.33\linewidth]{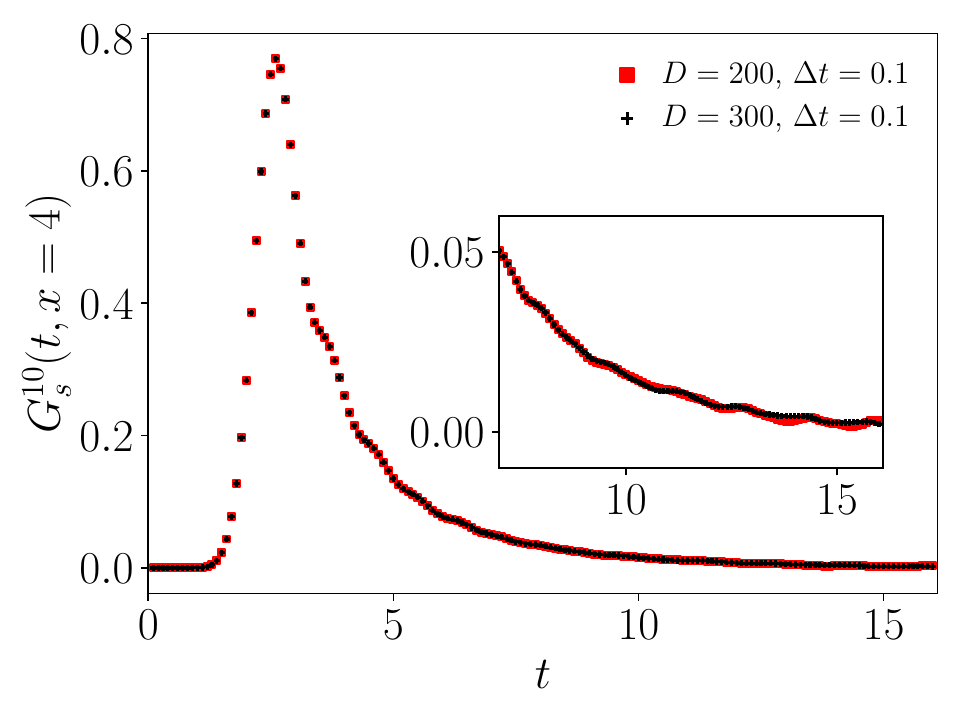}%
}\hfill
\subfloat[$x=8$.\label{fig:Gs_Px_8_bond64}]{%
  \includegraphics[width=0.33\linewidth]{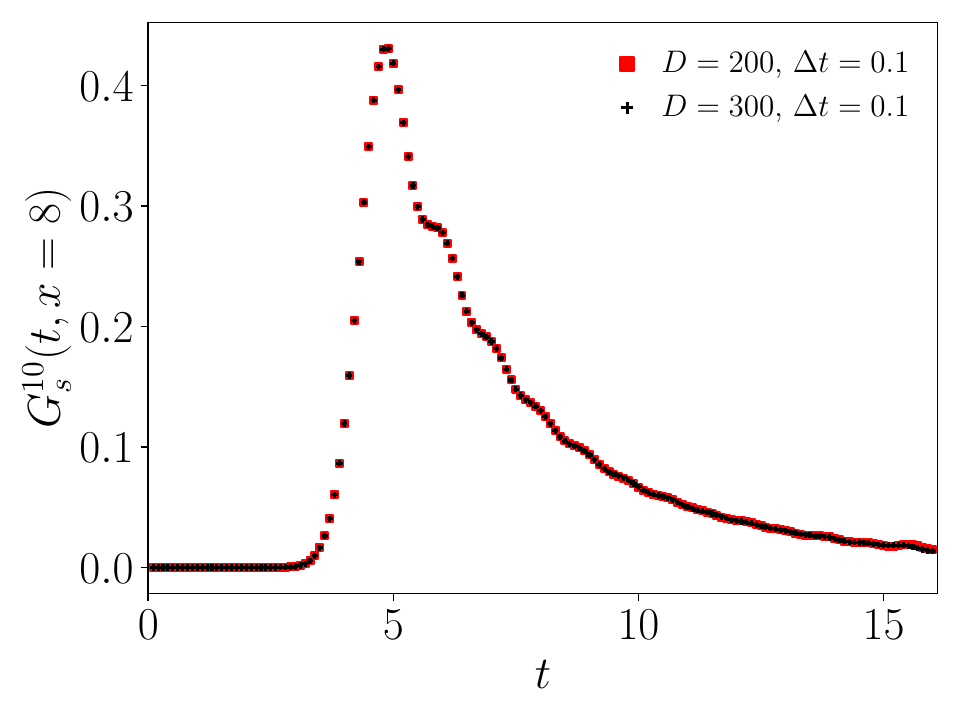}%
}

\subfloat[$x=12$.\label{fig:Gs_Px_12_bond64}]{%
  \includegraphics[width=0.33\linewidth]{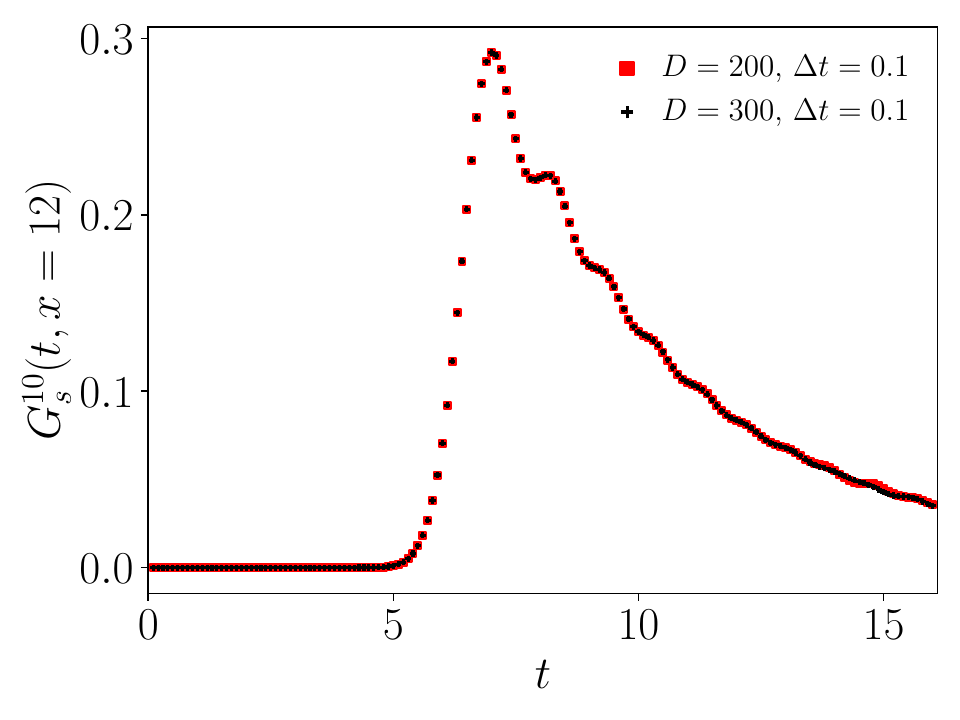}%
}\hfill
\subfloat[$x=16$.\label{fig:Gs_Px_16_bond64}]{%
  \includegraphics[width=0.33\linewidth]{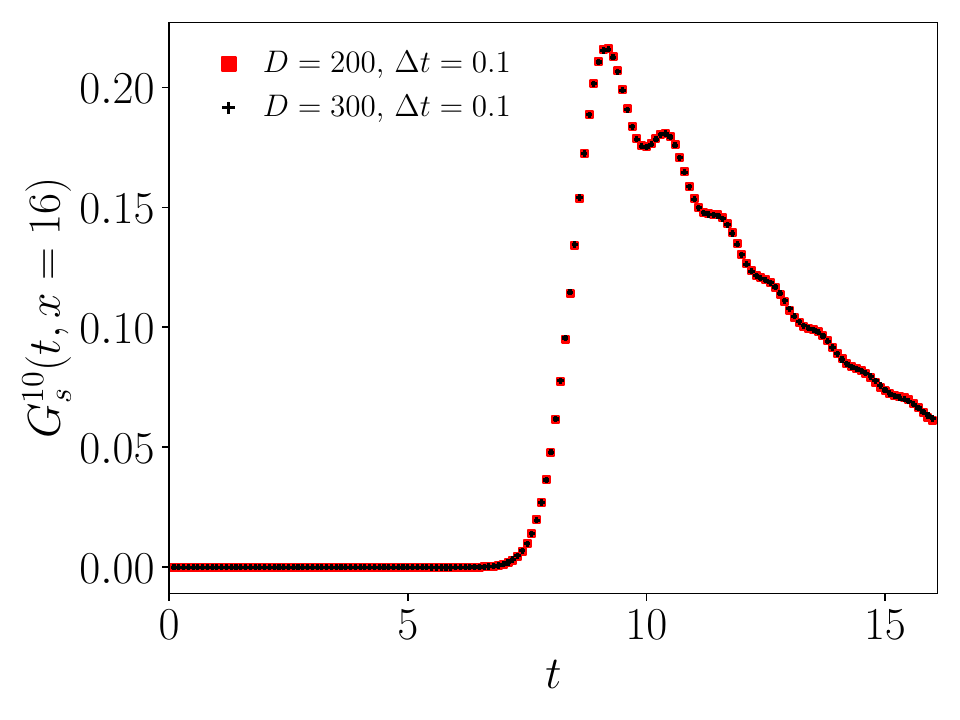}%
}\hfill
\subfloat[$x=20$.\label{fig:Gs_Px_20_bond64}]{%
  \includegraphics[width=0.33\linewidth]{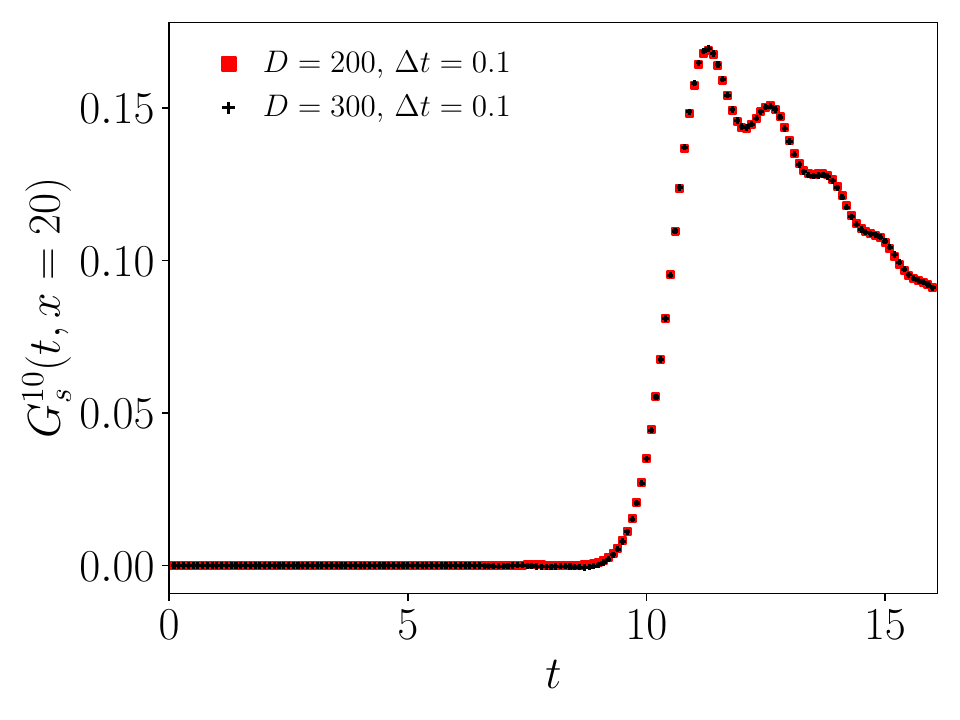}%
}

\subfloat[$x=24$.\label{fig:Gs_Px_24_bond64}]{%
  \includegraphics[width=0.33\linewidth]{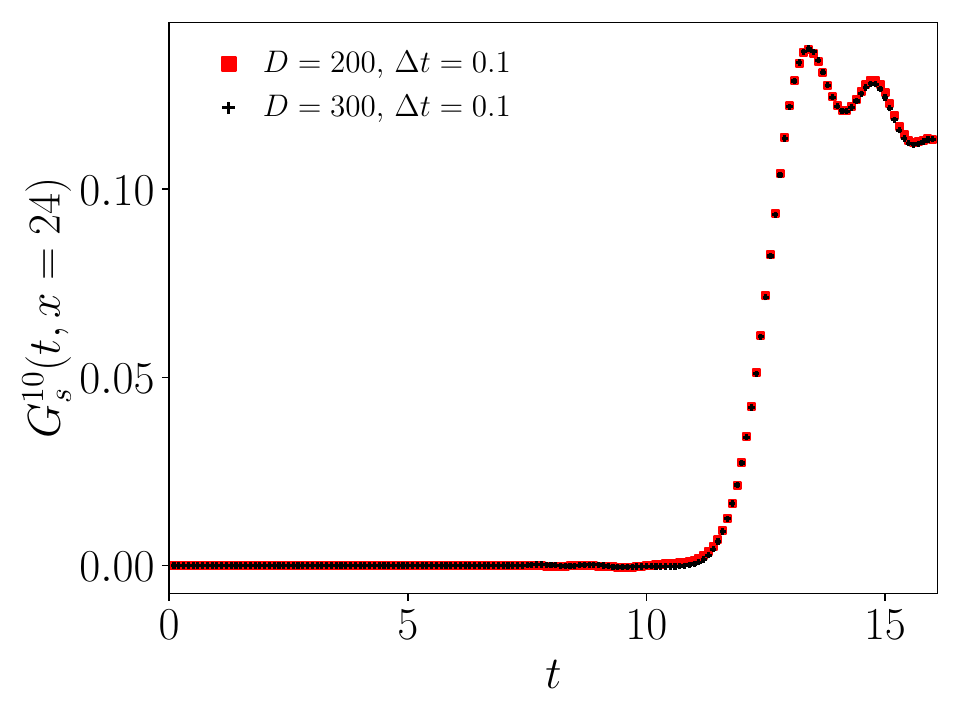}%
}\hfill
\subfloat[$x=28$.\label{fig:Gs_Px_28_bond64}]{%
  \includegraphics[width=0.33\linewidth]{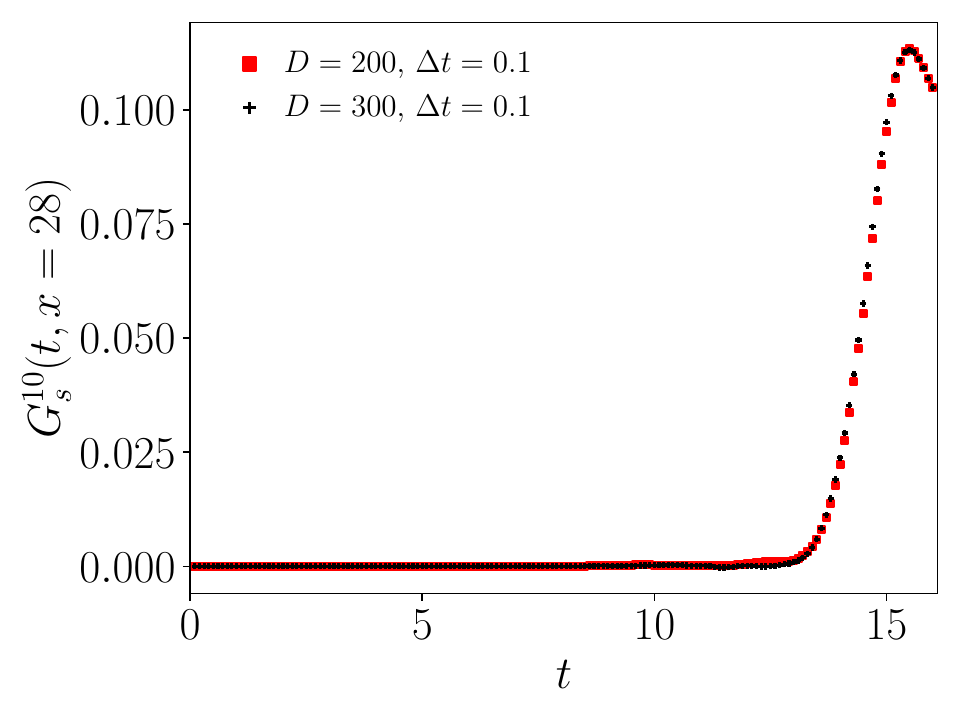}%
}\hfill
\subfloat[$x=32$.\label{fig:Gs_Px_32_bond64}]{%
  \includegraphics[width=0.33\linewidth]{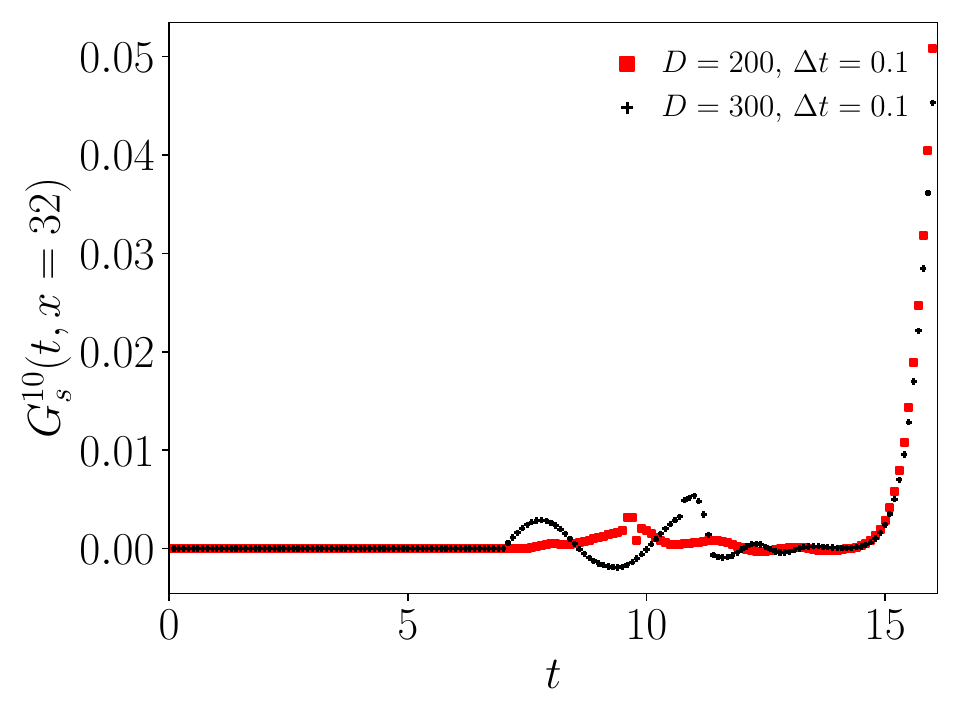}%
}
\caption{MPS calculation results of $G_s^{10}(t,x)$ on a periodic $L=64$ lattice with $\xi_{\rm lat}=0.01$, $h_z=-0.04$ and $T\approx 6.911$ ($\beta=0.1447$) using two different bond dimensions $D$. We use 10 steps in the second-order Trotterized imaginary time evolution and $\Delta =0.01$ to apply the local perturbation at $t=0$.}
\label{fig:Gs_Px_64_bond}
\end{figure*}

Instead of keeping increasing the bond dimension to see complete agreement between the MPS and exact diagonalization results on this small lattice, we move on to bigger lattices and investigate if the MPS calculation converges well within our available computational resources for the time before the boundary effects influence the results. In the main text, we studied the convergence of the bond dimension on a periodic $L=32$ lattice with $\xi_{\rm lat}=0.01$, $h_z=-0.08$, and $T=10$ in Fig.~\ref{fig:Gs_Px_32_bond_dt} and reached the conclusion that a MPS calculation with $D=200$ and $\Delta t=0.1$ can well describe the spreading perturbation, before it crosses the periodic boundary. A similar analysis is performed for a periodic $L=64$ lattice with $\xi_{\rm lat}=0.01$ and $h_z=-0.04$ for the temperature $T\approx 6.911$ used in the viscosity calculation ($T/m_1\approx 7.141$). The maximum bond dimension across the whole lattice used in this MPS calculation is shown in Fig.~\ref{fig:maxbond_N=64} with two different cutoffs and $D=300$. The results of $G_s^{10}(t,x)$ are depicted in Fig.~\ref{fig:Gs_Px_64_bond}. 
Scrutinizing these plots, we conclude that the accuracy of the MPS calculation with $D=200$ used in the main text is better than $O(0.0001)$ (obtained from $\sqrt{\chi}$) for $t<8$ and the uncertainty for $8<t<15$ before the perturbation arrives at the right boundary of the $L=64$ lattice is on the order of $O(0.005)$ in the worst scenario, which occurs at the boundary site $x=32$ as shown in Fig.~\ref{fig:Gs_Px_64_bond}. The entanglement growth is slower on the $L=64$ lattice than the $L=16$ lattice studied previously in this appendix. This is because the effective temperature is higher on the bigger lattice, $T/m_1\approx 7.141$ than that on the smaller one, $T/m_1\approx 3.540$ and higher temperature thermal states have smaller entanglement.

\section{Other Fitting Results}
\label{app:fit}
\begin{figure*}[t]
\centering
\subfloat[$h_z=-0.04$.\label{fig:cs_fit_004}]{%
  \includegraphics[width=0.33\linewidth]{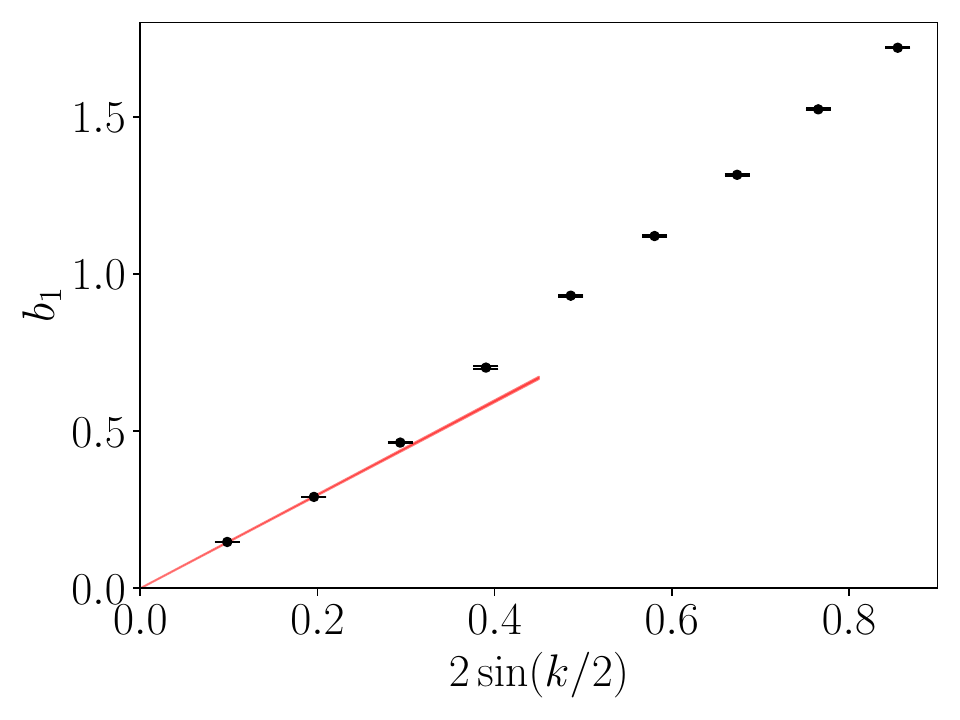}%
}\hfill
\subfloat[$h_z=-0.05$.\label{fig:cs_fit_005}]{%
  \includegraphics[width=0.33\linewidth]{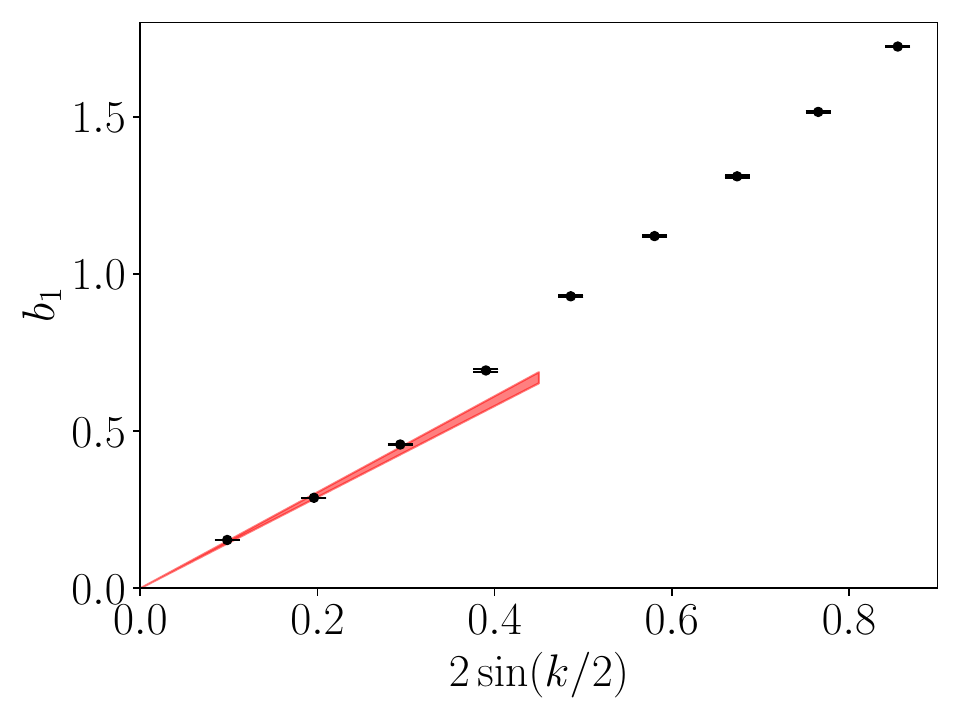}%
}\hfill
\subfloat[$h_z=-0.06$.\label{fig:cs_fit_006}]{%
  \includegraphics[width=0.33\linewidth]{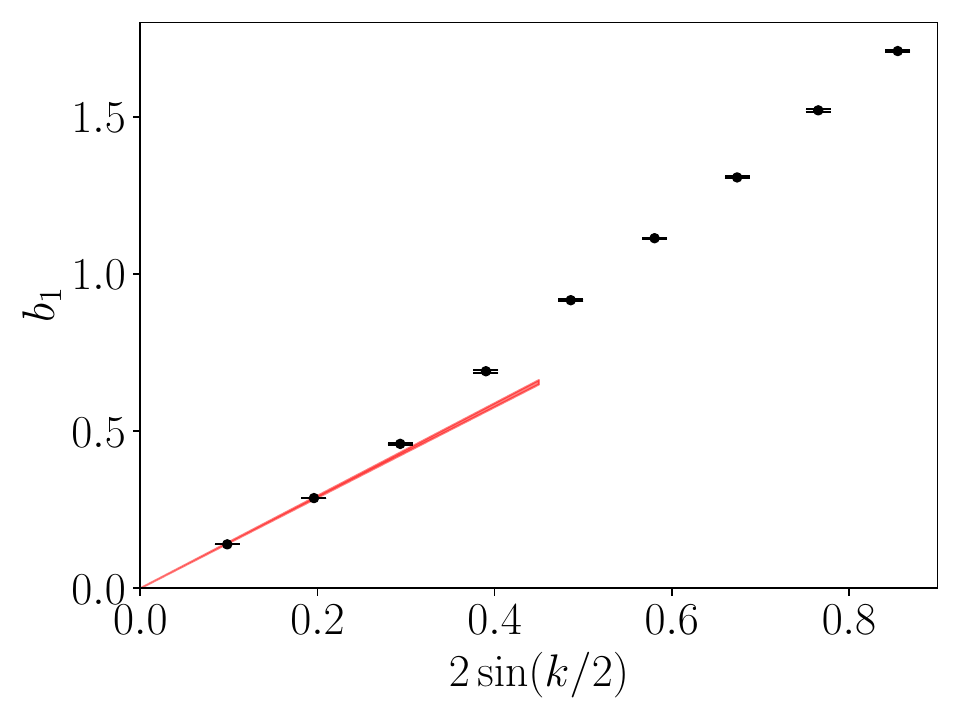}%
}

\subfloat[$h_z=-0.07$.\label{fig:cs_fit_007}]{%
  \includegraphics[width=0.33\linewidth]{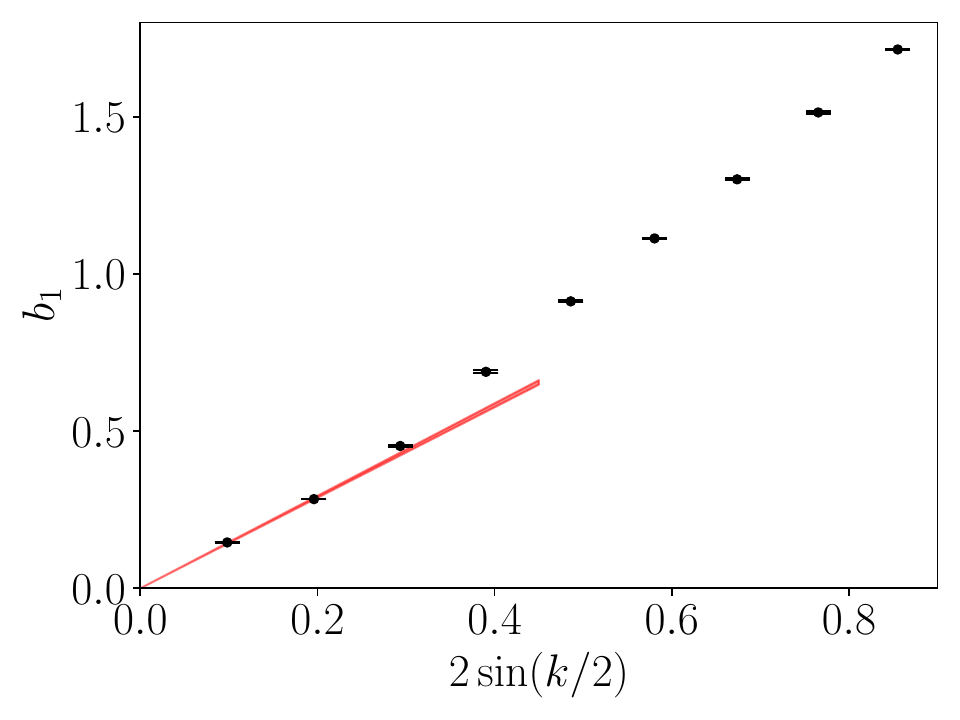}%
}\hfill
\subfloat[$h_z=-0.09$.\label{fig:cs_fit_009}]{%
  \includegraphics[width=0.33\linewidth]{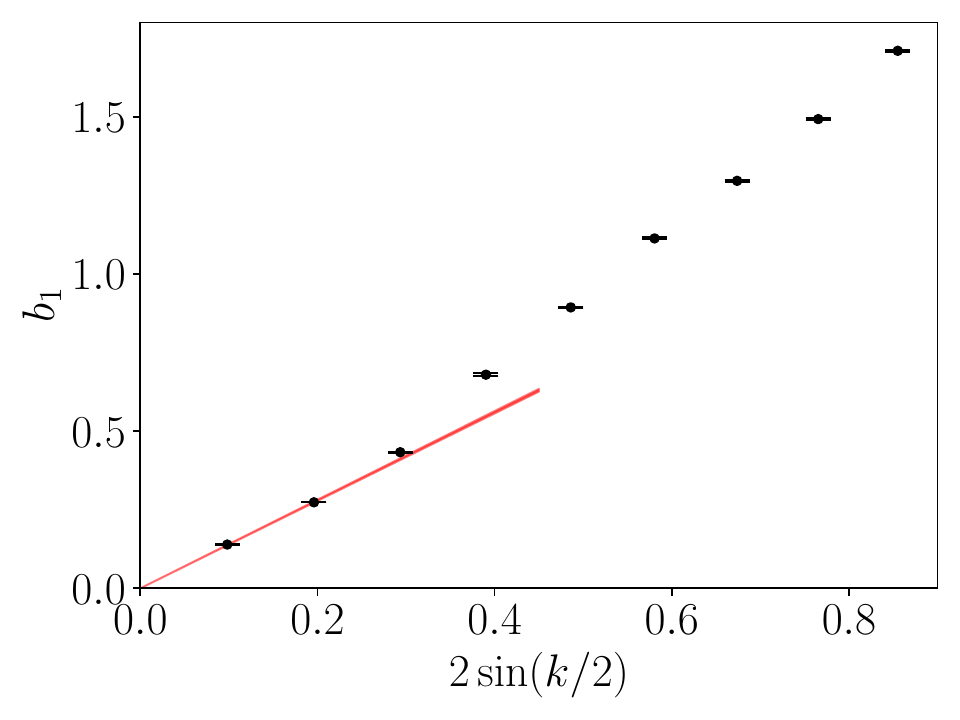}%
}\hfill
\subfloat[$h_z=-0.1$.\label{fig:cs_fit_010}]{%
  \includegraphics[width=0.33\linewidth]{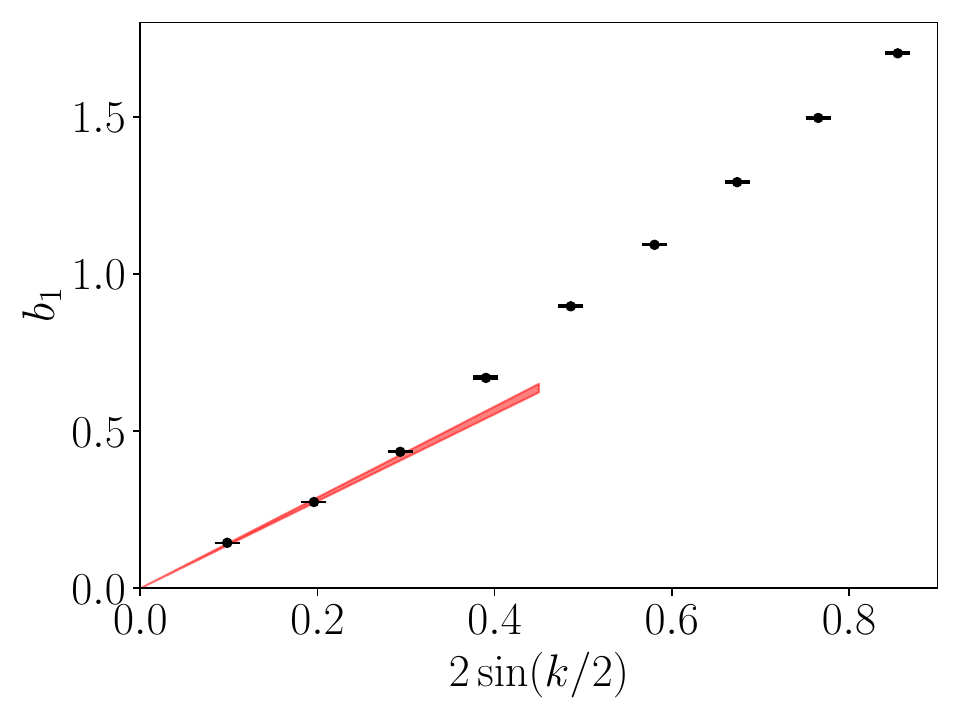}%
}

\caption{Same as Fig.~\ref{fig:cs_k} but for the other couplings used in the continuum extrapolation. The fitted value of the speed of sound for each case is listed in Table~\ref{tab:couplings}.}
\label{fig:cs_fit_all}
\end{figure*}

\begin{figure*}[t]
\centering
\subfloat[$h_z=-0.04$, $\Gamma_0=0.0064\pm 0.0028$.\label{fig:gamma_fit_004}]{%
  \includegraphics[width=0.33\linewidth]{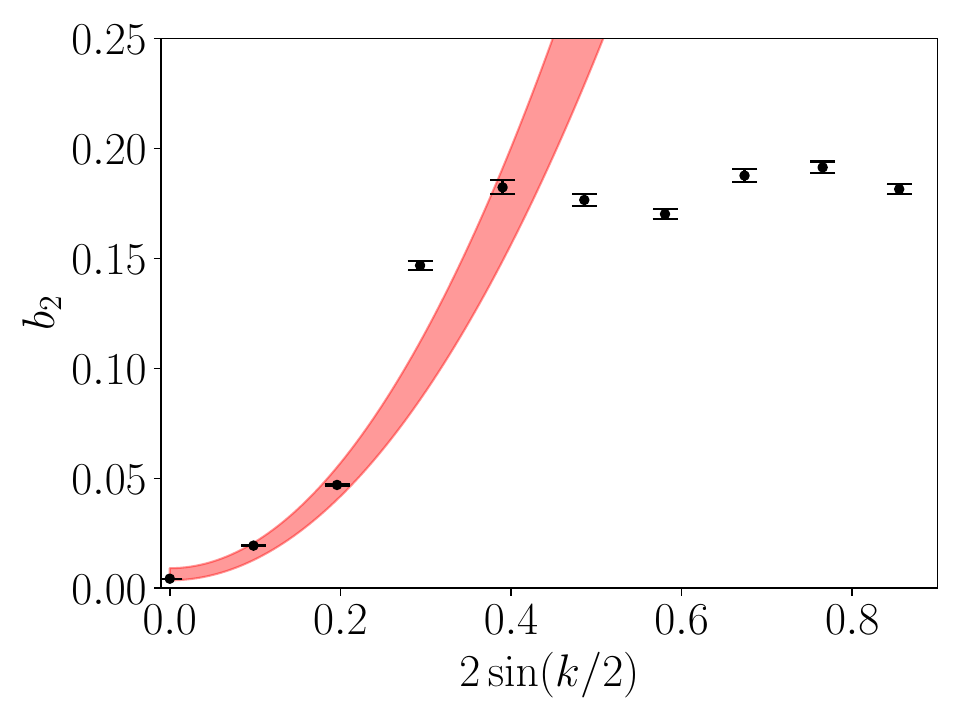}%
}\hfill
\subfloat[$h_z=-0.05$, $\Gamma_0=0.0081\pm 0.0017$.\label{fig:gamma_fit_005}]{%
  \includegraphics[width=0.33\linewidth]{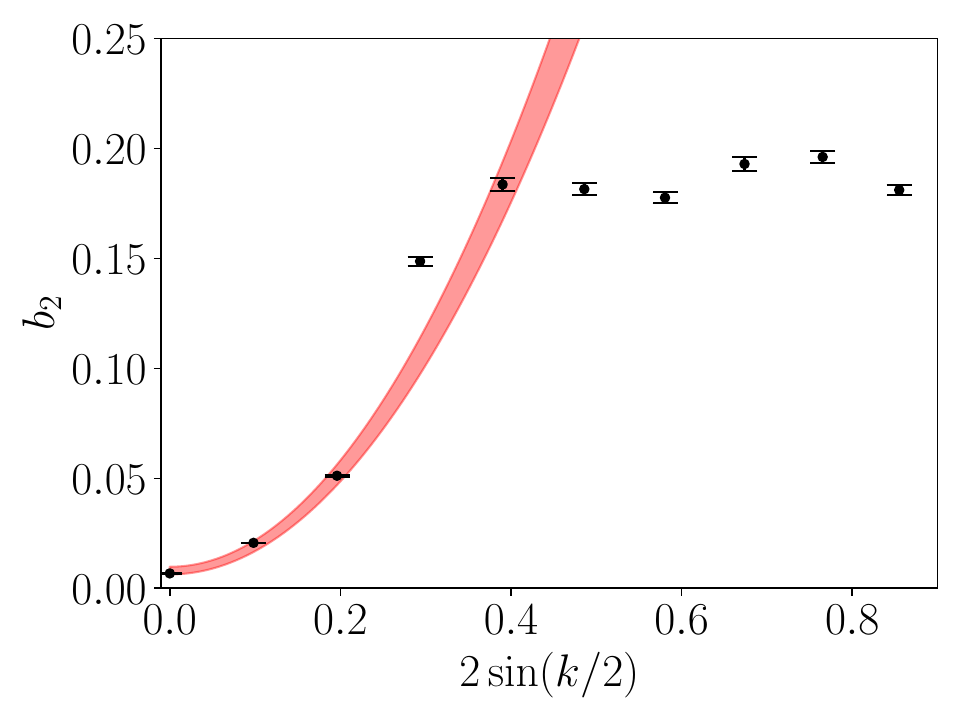}%
}\hfill
\subfloat[$h_z=-0.06$, $\Gamma_0=0.0127\pm 0.0004$.\label{fig:gamma_fit_006}]{%
  \includegraphics[width=0.33\linewidth]{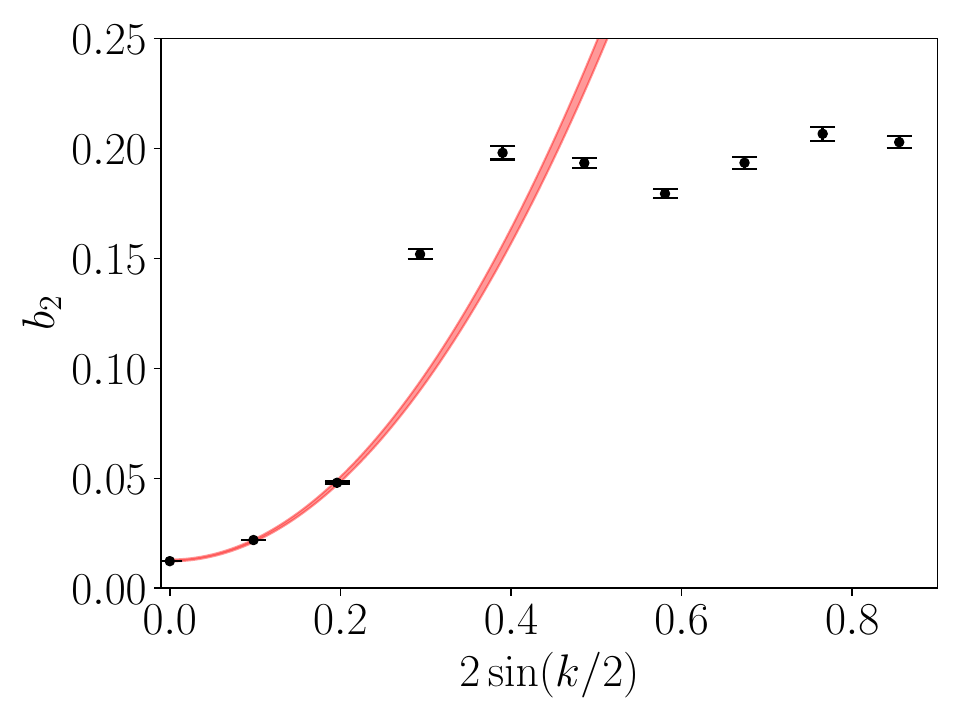}%
}

\subfloat[$h_z=-0.07$, $\Gamma_0=0.0147\pm 0.0006$.\label{fig:gamma_fit_007}]{%
  \includegraphics[width=0.33\linewidth]{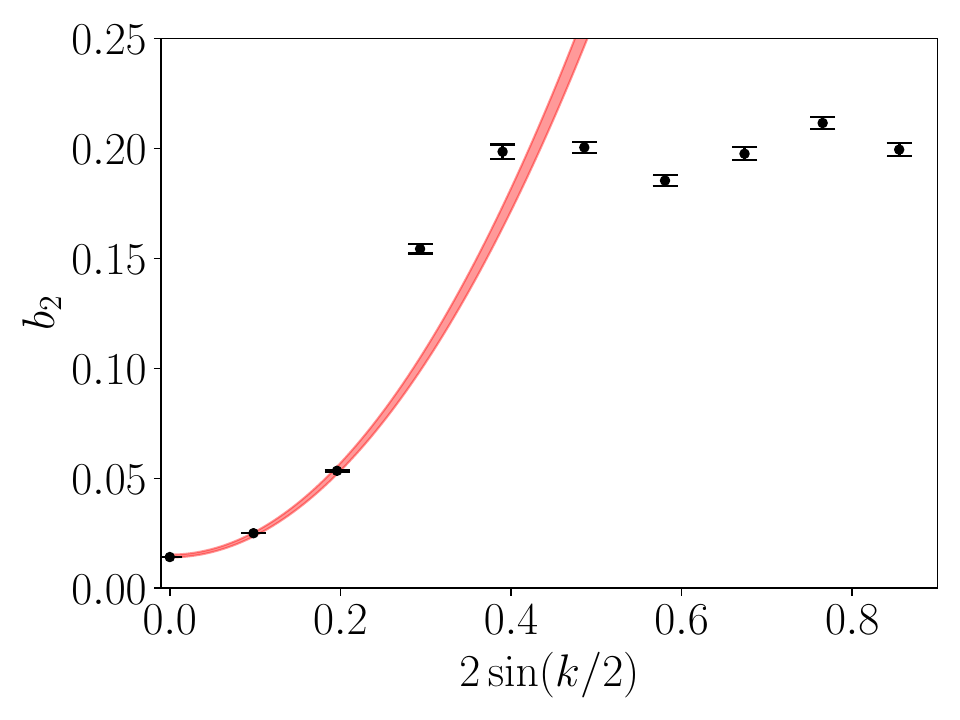}%
}\hfill
\subfloat[$h_z=-0.09$, $\Gamma_0=0.0230\pm 0.0005$.\label{fig:gamma_fit_009}]{%
  \includegraphics[width=0.33\linewidth]{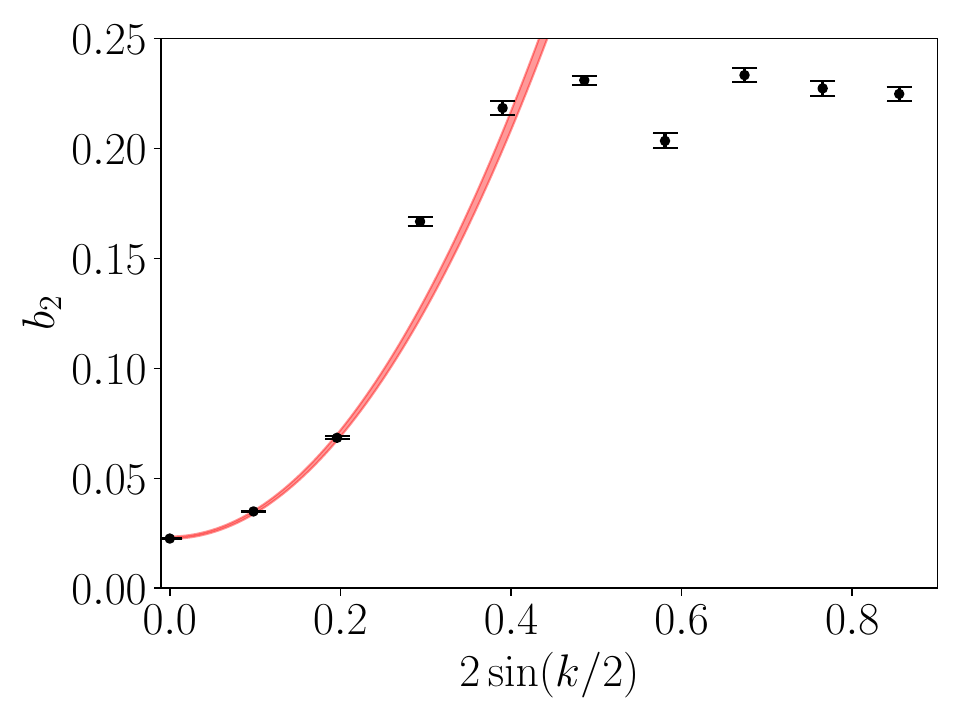}%
}\hfill
\subfloat[$h_z=-0.1$, $\Gamma_0=0.0263\pm 0.0022$.\label{fig:gamma_fit_010}]{%
  \includegraphics[width=0.33\linewidth]{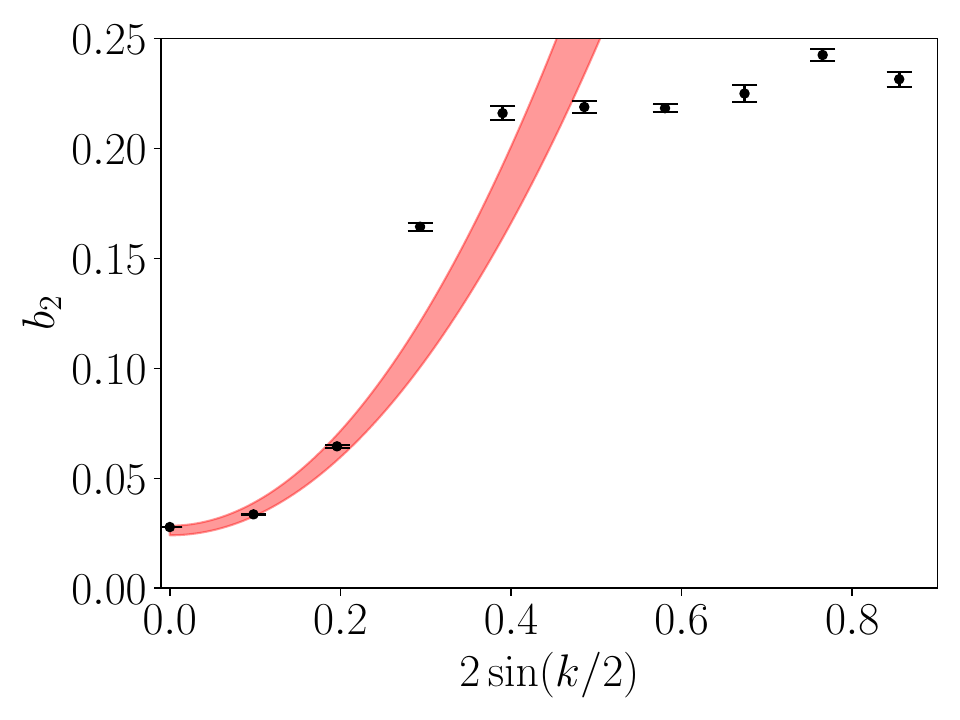}%
}

\caption{Same as Fig.~\ref{fig:gamma} but for the other couplings used in the continuum extrapolation. The fitted value of the bulk viscous damping for each case is listed in Table~\ref{tab:couplings} and that of $\Gamma_0$ is listed in each subcaption here, up to four digits.}
\label{fig:gamma_fit_all}
\end{figure*}

In Figs.~\ref{fig:cs_fit_all} and~\ref{fig:gamma_fit_all} we show the fitting results of the speed of sound $c_s$ and bulk viscous damping $\tilde{\gamma}_\zeta$ for all the couplings used in the continuum extrapolation on the $L=64$ lattice with $\xi_{\rm lat}=0.01$ and $T/m_1\approx 7.141$. The fitted values for the speed of sound and the bulk viscous damping are listed in Table~\ref{tab:couplings}. The fitted values of the zero momentum mode decay rates $\Gamma_0$ are listed in each subcaption of Fig.~\ref{fig:gamma_fit_all}. We see that $\Gamma_0$ approaches zero as coupling decreases, as expected in the continuum limit when momentum conservation is fully restored.

\clearpage
\bibliography{main.bib}
\end{document}